\documentclass[a4paper,twoside,openright]{report}
\usepackage[DIV10,BCOR10mm]{typearea}
\usepackage{amsmath}
\usepackage{amsthm}
\usepackage{amssymb}
\usepackage{graphicx}
\usepackage{stmaryrd}
\usepackage{bbm}
\usepackage{braket}
\usepackage{fancyhdr}
\usepackage{longtable}
\usepackage{alltt}
\usepackage{multicol}
\usepackage{titletoc}
\usepackage{color}
\usepackage{rotating}
\usepackage[british]{babel} 
\usepackage{mparhack}
\usepackage{time}
\usepackage{type1cm} 
\usepackage{calc}
\usepackage{pifont}
\usepackage{url}
\usepackage[format=hang,font=sf,width=0.85\textwidth]{caption}
\newtheorem{theorem}{Theorem}[section]
\newtheorem{thesis}{Hypothesis}[section]
\newtheorem{definition}{Definition}[section]
\newtheorem{observation}{Observation}[section]

\newtheorem{remark}{Remark}[section]
\newenvironment{rationale}{\noindent\emph{Rationale:} }{\hfill\checkmark}
\newcommand{\norm}[1]{\ensuremath{\left|\left|#1\right|\right|}}
\newcommand{\ip}[2]{\ensuremath{\left<#1,#2\right>}}
\newcommand{\normhs}[1]{\ensuremath{\left|\left|#1\right|\right|}_{\text{HS}}}
\newcommand{\iphs}[2]{\ensuremath{\left<#1,#2\right>}_{\text{HS}}}
\newcommand{\trace}{\operatorname{tr}}
\newcommand{\ptrace}[1]{\operatorname{tr}_{#1}}
\newcommand{\BH}{\ensuremath{\mathcal{B}(\mathcal{H})}}
\newcommand{\SA}{\ensuremath{\mathcal{S}(\mathcal{A})}}
\newcommand{\EA}{\ensuremath{\mathcal{E}(\mathcal{A})}}
\newcommand{\CX}{\ensuremath{\mathcal{C}(X)}}
\newcommand{\fix}[1]{\ensuremath{\operatorname{fix}\left(#1\right)}}
\newcommand{\op}[1]{\ensuremath{\hat{#1}}}

\newcommand{\stdsembrack}[1]{\ensuremath{\llbracket #1\rrbracket}}
\newcommand{\sembrack}[1]{\ensuremath{\llbracket #1\rrbracket}(K,T,E)}
\newcommand{\lat}[1]{\textit{#1}} 
\newcommand{\ie}{\lat{i.e.}}
\newcommand{\eg}{\lat{e.g.}}
\newcommand{\cf}{\lat{cf.}}
\newcommand{\Cf}{\lat{Cf.}}
\definecolor{grey}{rgb}{.85,.85,.85}
\newdimen\summarydimen
\newbox\summarybox
\setbox\summarybox=\hbox{\textbf{\textsf{Summary}}}
\summarydimen=\ht\summarybox
\advance\summarydimen by\dp\summarybox
\newbox\summaryrefrot
\setbox\summaryrefrot=\hbox{\reflectbox{\scalebox{1}[-1]{\usebox\summarybox}}}
\newbox\oddrulerbox
\newbox\evenrulerbox
\newdimen\rulethickness
\setbox\oddrulerbox=\hbox{\hspace{0.1\textwidth}%
                     \rule{0.9\textwidth}{\rulethickness}}
\setbox\evenrulerbox=\hbox{\rule{0.9\textwidth}{\rulethickness}%
                     \hspace{0.1\textwidth}}
\newcommand{\summaryodd}[1]{%
  \begin{samepage}    
    \noindent\usebox\oddrulerbox\\[-2pt] 
       {\hspace*{0.1\textwidth}\colorbox{grey}%
     {\parbox[t]{0.9\textwidth-2\fboxsep}{#1}}}%
   \marginpar{\rotatebox[x=0mm,y=\summarydimen]{270}%
                                   {\usebox\summarybox}}\\[-1\rulethickness]%
   \noindent\usebox\oddrulerbox\par
 \end{samepage}}
\newcommand{\summaryeven}[1]{%
  \begin{samepage}
    \noindent\usebox\evenrulerbox\\[-2pt]
     {\raggedright\colorbox{grey}%
     {\parbox[t]{0.9\textwidth-2\fboxsep}{#1}}}%
     \marginpar{\flushright\rotatebox[x=0mm,y=\summarydimen]{270}%
       {\usebox\summaryrefrot}}\\[-\rulethickness]%
   \noindent\usebox\evenrulerbox
 \end{samepage}}
\newcommand{\summary}[1]{\ifodd\thepage\summaryodd{#1}\else\summaryeven{#1}\fi}
\newcommand{\ketbra}[2]{\ensuremath{\ket{#1}\bra{#2}}}
\newcommand{\itext}[1]{\ensuremath{\thickspace\text{#1}\thickspace}}
\newcommand{\sym}{\ensuremath{\operatorname{Sym}}} 
\newcommand{\card}{\ensuremath{\operatorname{card}}}
\newcommand{\dtype}[1]{\ensuremath{\text{\textbf{#1}}}}
\newcommand{\bit}{\dtype{bit}}
\newcommand{\qbit}{\dtype{qbit}}
\newcommand{\cint}{\dtype{int}} 
\newcommand{\qint}{\dtype{qint}}
\newcommand{\short}{\dtype{short}}
\newcommand{\qshort}{\dtype{qshort}}
\newcommand{\void}{\dtype{void}}
\newcommand{\qtype}{\dtype{qtype}}
\newcommand{\yes}{\ding{51}}
\newcommand{\no}{\ding{55}}
\newcommand{\range}{\operatorname{range}}
\newcommand{\mapval}[1]{\ensuremath{#1 \xrightarrow{\pi_{#1}} \range(#1)}}
\newcommand{\mapvalmeas}[2]{\ensuremath{#1 \xrightarrow{\pi_{#1}:\PROJ(#2,T)} \range(#1)}}
\newcommand{\Ycomb}{\ensuremath{\operatorname{Y}}}
\newcommand{\pos}{\ensuremath{\operatorname{pos}}}

\newcommand{\code}[1]{\texttt{#1}}
\newcommand{\inj}{\operatorname{in}}
\newcommand{\smashp}{\operatorname{smash}}
\newcommand{\domfkt}[2]{\ensuremath{\mathcal{#1}\stdsembrack{#2}}}
\newcommand{\LC}{\ensuremath{\mathfrak{L}_{c}}}
\newcommand{\kteinit}{\ensuremath{(K_{\varnothing}, T_{\varnothing}, E_{\varnothing})}}
\newcommand{\kteinittensor}{\ensuremath{(K_{\varnothing}^{\otimes}, 
    T_{\varnothing}^{\otimes}, E_{\varnothing}^{\otimes})}}
\newcommand{\boxvar}[1]{\ensuremath{\text{\fbox{\hspace{1cm}\(#1\)\hspace{1cm}}}}}

\newcommand{\infer}[3]{\frac{#1 \quad\hfill #2}{#3}}
\newcommand{\infersimple}[2]{\frac{#1}{#2}}

\newcommand{\inferdots}[3]{\frac{#1 \cdots #2}{#3}}

\newcommand{\EXP}{\ensuremath{\mathcal{EXP}}}
\newcommand{\EQN}{\ensuremath{\mathcal{EQN}}}
\newcommand{\OP}{\ensuremath{\mathcal{OP}}}
\newcommand{\PROJ}{\ensuremath{\mathcal{PROJ}}}
\newcommand{\DO}{\ensuremath{\mathcal{DO}}}
\newcommand{\MO}{\ensuremath{\mathcal{MO}}}
\newcommand{\SR}{\ensuremath{\mathcal{SR}}}
\newcommand{\RS}{\ensuremath{\mathcal{RS}}}
\newcommand{\PROG}{\ensuremath{\mathcal{PROG}}}
\newcommand{\COMM}{\ensuremath{\mathcal{COMM}}}
\newcommand{\VAL}{\ensuremath{\mathcal{VAL}}}

\newcommand{\forgechapter}[1]{\setcounter{chapter}{#1-1}}

\addtocounter{secnumdepth}{1}

\makeatletter
\def\definecolour#1#2#3{\definecolor{#1}{#2}{#3}}
\def\colour#1{\color{#1}}    
\definecolour{lightgray}{gray}{.80}
\definecolour{gray}{gray}{.50}
\newcommand\titanic{\@setfontsize\titanic{68\p@}{30}}
\newcommand{\titanbf}{\titanic\bfseries}
\newcommand{\titangraybf}{\titanbf\colour{gray}}
\renewcommand{\chapter}[1]{\quotechapter{#1}{\hspace{0pt}}{\hspace{0pt}}}
\renewcommand\chapter{\if@openright\cleardoublepage\else\clearpage\fi
                    \thispagestyle{plain}%
                    \global\@topnum\z@
                    \@afterindentfalse
                    \secdef\@chapter\@schapter}
\def\@chapter[#1]#2#3#4{\ifnum \c@secnumdepth >\m@ne
                         \refstepcounter{chapter}%
                         \typeout{\@chapapp\space\thechapter.}%
                         \addcontentsline{toc}{chapter}%
                                   {\protect\numberline{\thechapter}#1}%
                    \else
                      \addcontentsline{toc}{chapter}{#1}%
                    \fi
                    \chaptermark{#1}%
                    \addtocontents{lof}{\protect\addvspace{10\p@}}%
                    \addtocontents{lot}{\protect\addvspace{10\p@}}%
                    \if@twocolumn
                      \@topnewpage[\@makechapterhead{#2}]%
                    \else
                      \@makechapterhead{#2}{#3}{#4}%
                      \@afterheading
                    \fi}
\def\makequotebox#1#2{\vbox{\hspace{0pt}\hfill\parbox[b]{7cm}{{\raggedright
        #1\\[0.3em]\hspace{0.5cm}\parbox{6.5cm}{\noindent\textit{#2}}}}}}
\def\@makechapterhead#1#2#3{\vspace*{0\p@}%
  \setbox23=\makequotebox{#2}{#3}
  \makequotebox{#2}{#3}\vspace{-\ht23}%
  {\vspace*{50\p@}%
    \parindent\z@ \normalfont
   \ifnum \c@secnumdepth >\m@ne
   \vskip 20\p@
     \settowidth{\@tempdima}{\titanbf\sffamily\thechapter}
     \addtolength{\@tempdima}{15pt}
     \parbox[b]{\@tempdima}{\sffamily\titangraybf\thechapter\hfil}%
   \else \@tempdima=0pt \fi 
   \setlength{\@tempdimb}{\textwidth}%
   \addtolength{\@tempdimb}{-\@tempdima}%
   \parbox[b]{\@tempdimb}{\raggedright\Huge\sffamily\bfseries #1}
   \par\nobreak \vspace{10\p@}
}}
\def\@makeschapterhead#1{%
  \vspace*{50\p@}%
  {\parindent \z@ \raggedright
    \sffamily
    \interlinepenalty\@M
    \Huge \bfseries  #1\par\nobreak
    \vskip 40\p@
  }}
\renewcommand\section{\@startsection {section}{1}{\z@}%
                                   {-3.5ex \@plus -1ex \@minus -.2ex}%
                                   {2.3ex \@plus.2ex}%
                                   {\normalfont\sffamily\Large\bfseries}}
\renewcommand\subsection{\@startsection{subsection}{2}{\z@}%
                                    {-3.25ex\@plus -1ex \@minus -.2ex}%
                                    {1.5ex \@plus .2ex}%
                                    {\normalfont\sffamily\large\bfseries}}
\renewcommand\subsubsection{\@startsection{subsubsection}{3}{\z@}%
                                    {-3.25ex\@plus -1ex \@minus -.2ex}%
                                    {1.5ex \@plus .2ex}%
                                    {\normalfont\sffamily\normalsize\bfseries}}
\renewcommand\paragraph{\@startsection{paragraph}{4}{\z@}%
                                    {3.25ex \@plus1ex \@minus.2ex}%
                                    {-1em}%
                                    {\normalfont\sffamily\normalsize\bfseries}}
\renewcommand\subparagraph{\@startsection{subparagraph}{5}{\parindent}%
                                    {3.25ex \@plus1ex \@minus .2ex}%
                                    {-1em}%
                                    {\normalfont\sffamily\normalsize\bfseries}}
\makeatother

\newcommand{\xquad}{\hspace{0.5em plus.2em minus.2em}}
\titlecontents*{subsubsection}[2.465cm]{\filright\footnotesize}{}{}{}%
[\xquad\textbullet\xquad][.]
\addtocounter{tocdepth}{1}

\begin{document}
\makeatletter
\def\myhrulefill{\leavevmode\leaders\hrule height 0.5mm\hfill\kern\z@}
\makeatother
\newcommand{\revision}{Revision 1.1 (15.~September~2005)}
\thispagestyle{empty}
\vspace*{1cm}
\begin{flushright}
  \huge{\textsf{\textbf{Semantics and Simulation\\ 
        of Communication in\\ Quantum Programming}}}\\
\end{flushright}\vspace{-4mm}
{\hspace*{\fill}\hbox to 0.75\textwidth{\myhrulefill}}\par\vspace{-2.5mm}
{\hspace*{\fill}\hbox to 0.85\textwidth{\myhrulefill}}\par\vspace{2.5cm}

\hbox to\textwidth{\hspace{+5mm}\parbox{\textwidth}{
  \begin{center}\begin{Large}\begin{sffamily}
        Diploma Thesis\\[0.5em]
        Wolfgang Mauerer\(^{1}\)\\
        Quantum Information Theory Group\\
        Institute for Theoretical Physics I and\\
        \hbox to \textwidth{\hss Max Planck Research Group for Optics, 
        Information and Photonics\hss}\hspace{0pt}\\[-1mm]
        University Erlangen-Nuremberg, May 2005
      \end{sffamily}\end{Large}\end{center}}\hss}\vspace{1.5cm}
\hbox to\textwidth{\hspace{+5mm}\parbox{\textwidth}{
\noindent\hfill\textbf{Abstract}\hfill\hspace{0pt}\par 
\noindent We present the quantum programming language cQPL which is an
extended version of QPL~\cite{selinger_qpl}. It is capable of quantum
communication and it can be used to formulate all possible quantum
algorithms. Additionally, it possesses a denotational semantics based
on a partial order of superoperators and uses fixed points on a
generalised Hilbert space to formalise (in addition to all standard
features expected from a quantum programming language) the exchange
of classical and quantum data between an arbitrary number of
participants.  Additionally, we present the implementation of a cQPL
compiler which generates code for a quantum simulator.

\vspace{1ex}\noindent PACS numbers: 03.67.-a, 03.67.Hk, 03.67.Lx, 89.20.Ff\par
\noindent{\footnotesize\sffamily\revision}
}\hss}

\vfill
\noindent{\small\(^{1}\) eMail: wmauerer@optik.uni-erlangen.de}
\newpage\setcounter{page}{1}\pagenumbering{roman}
\thispagestyle{empty}
\tableofcontents
\newpage\thispagestyle{empty}\hspace{1pt}
\newpage\setcounter{page}{1}\pagenumbering{arabic}
\pagestyle{fancy}

\chapter{Introduction}{Dich sah ich, und die milde Freude\\
Flo{\ss} von dem s{\"u}{\ss}en Blick auf mich;\\
Ganz war mein Herz an deiner Seite\\
Und jeder Atemzug f\"ur dich.}%
{Johann Wolfgang von Goethe, Willkommen und Abschied} Although there
is no formal proof that quantum computers offer greater computational
power than classical ones, there are a few quantum computer algorithms
which provide efficient solutions for problems which are up to now
believed to be classically NP-hard, \ie, they cannot be solved in
polynomial time.  One of these problems -- computing discrete
logarithms -- is the cornerstone of basically every modern classical
cryptographic algorithm, so the increased interest in quantum
computing is obvious, both from a fundamental and a practical point of
view.

How can quantum algorithms be described in an efficient, readable 
and precise manner? Usually, this is done with \emph{quantum circuits}
which are combined into a quantum network, but especially for larger programs,
this is a very cumbersome and error-prone method. Much research has
been performed on the construction, implementation and conception of 
classical programming languages; therefore, a great amount of
alternatives are available. The situation is totally different
in quantum computing: Only a handful of languages have been proposed so
far \cite{altenkirch_grattage,betelli,oemer_msc,sanders_zuliani,
selinger_qpl,tonder},  and only two working 
implementations based on quantum computer simulators are available 
\cite{oemer_msc,betelli} (recently, a third, but 
still rough implementation was presented in \cite{altenkirch}). Both 
are based on imperative/object oriented languages (C, Pascal, C++, \dots),
with quantum features as extensions, where\-as QPL \cite{selinger_qpl} models 
a basically functional language.\footnote{We need to note, though, that
  the term \emph{functional} should not be overestimated in this context.
  Important features like higher order functions, which are considered 
  to be key elements of classical functional languages, are missing in QPL.
  The most interesting part of the language from a physicist's point of view
  is the ability to guarantee freedom against runtime errors already at compile
  time, no matter how this goal is achieved.}

In this work, we present an extension of QPL with abilities to cover
\emph{quantum communication}, \ie, the transmission of quantum
mechanical states and exploitation of their highly non-classical
properties like superpositions and entanglement. To distinguish our
approach from QPL, we call it cQPL for \emph{communication capable}
QPL.  In contrast to quantum computers of which only some very
elementary parts have been experimentally realised until now,
implementations of quantum communication are not only available in
several laboratories around the world, but can even be obtained
commercially.

To provide some orientation where our approach can be located in
contrast to work done by other contributors to the field, 
Table~\ref{intro:comparison} presents a comparison of quantum programming
languages and their features.

\begin{table}[htb]
\begin{minipage}{\textwidth}\centering
\begin{tabular}{l|cccccc}
    & QCL & Q Language & qGCL & QML & QPL & cQPL\\\hline
        Reference     & \cite{oemer_msc} & \cite{betelli} 
                      & \cite{sanders_zuliani} & \cite{altenkirch_grattage} 
                      & \cite{selinger_qpl} & \\
        New language  & \no & \no & \yes & \yes & \yes & \yes\\
        Respects physics\footnote{\emph{Respecting physics} is meant in the 
          sense that it is not possible
          to syntactically specify programs which would create unphysical 
          situations; of course, such a state will force the simulation to 
          abort with an error and needs thus to be avoided if possible.}    
                         & \no & \no & \no & \yes & \yes & \yes\\
        Implemented   & \yes & \yes & \no & \yes\footnote{A partially 
          finished implementation was available at the time of writing.} 
               & \yes\footnote{If we consider our cQPL compiler to be a 
                              QPL compiler as well.}  & \yes\\
        Formal semantics & \no & \no & \yes & \yes & \yes & \yes\\
        Communication & \no & \no & \yes & \no & \no & \yes\\
        Universal        & \yes & \yes & \yes & \yes & \yes & \yes\\
\end{tabular}\end{minipage}
\caption{Comparison of quantum programming languages defined in
  other approaches to the problem, their features and their
  shortcomings.}\label{intro:comparison}
\end{table}

In a nutshell, the contribution of our work to the field of
quantum programming languages is twofold:

\begin{itemize}
  \item We provide a compiler which can serve as a testbed for new ideas
    in quantum programming, to teach concepts of quantum algorithms
    or act as an aid to the intuition of users who want to 
    experiment (in the sense of goal oriented playing, not laboratory)
    with quantum protocols.
  \item We present an alternate approach to the compositional semantics of QPL and
    provide the possibility to include and formalise quantum
    communication as part of the language. This is a necessary step
    towards the formalisation of open-world programs, but may also
    prove itself useful in fields like quantum process calculi,
    automated protocol analysis or similar -- cf.
    Chapter~\ref{prospect} for further prospects.
\end{itemize}

The layout of this thesis is as follows: In Chapters~\ref{qprog} and
\ref{cqpl_compiler}, we provide an overview about quantum programming,
present the language cQPL and describe the compiler and its
implementation.  Chapter~\ref{math:chapter} presents the mathematical
tools and requisites necessary for a denotational semantics of cQPL,
and Chapter~\ref{chap:formal_semantics} develops the semantical
description.  Chapter~\ref{prospect} finally provides some short
remarks on possible further directions that may be pursued based on
the results of this work.  All chapters are interlaced with short
introductions to topics, tools and techniques which are uncommon in
physics, but necessary for our work; Appendices~\ref{symbols} and
\ref{glossary} provide a list of symbols and a glossary, respectively.
The formal syntax is presented in Appendix~\ref{comp:formal_syntax}.
To aid the reader in staying on track, we have provided short summaries in
grey boxes at some points along the way.

Since the topic dealt with does not only contain problems of physical
nature, but also touches the fields of computer science and
mathematics, we have tried to make the text as self-contained as
possible for readers with any of these backgrounds.\footnote{In this
  context, it is interesting to note that \(48\%\) of all entries in
  the bibliography are of physical nature and \(36\%\) can be counted
  to computer science; the remaining \(16\%\) belong to mathematics or
  are of general interest.}  Obviously it was not possible to present
everything in as much detail as required without repeating the
introductory textbooks, but numerous references to the literature are
provided which hopefully alleviates any arising problems.


\chapter{Quantum programming with QPL and cQPL}%
{When someone says, ``I want a programming language in which I need only %
say what I wish done,'' give him a lollipop.}%
{\hfill Alan Perlis, Epigrams on Programming}\label{qprog}
This chapter presents an overview about classical and quantum computers,
the underlying models of computation and some principal remarks on
quantum programming languages. Afterwards, a short introduction to
QPL and cQPL is given. Note that this chapter is intentionally kept
as terse as possible to allow a more detailed coverage of other topics
dealt with in this thesis. Therefore, no attempt is made to to provide special
rigour in this chapter.

\section{Programming and quantum physics}
\subsection{Computability}
Classical computers can be described by several models
which are apt for different purposes (for a more detailed
description, cf., \eg,~\cite{asteroth_baier,schoening}):

\begin{itemize}
  \item Turing machines
  \item General recursive functions
  \item Register machines
  \item Lambda calculus
  \item Logical gates
  \item Universal programming languages
\end{itemize}

They can all be brought to a common denominator by showing that they
are able to compute respectively solve the same class of problems.
They are computationally equivalent because one model can be used to
efficiently simulate any other model; a Turing machine is normally
taken to be the normative instance among them. The fact that a turing
machine can compute everything which is computable in principle is
captured in the Church-Turing thesis which is one of the fundamental
axioms of computer science:\footnote{To be precise, one would have to
  note that Church's thesis states that ``a function of positive
  integers is effectively calculable only if recursive''. This is
  equivalent to Turing's thesis, though, and the name Church-Turing
  thesis is conventionally used in the literature.}

\begin{thesis}[Church-Turing]
Every function which would naturally be regarded as computable can be 
computed by a Turing machine.
\end{thesis}

This definition places computability in a purely abstract setting without
regard to the laws of physics. Deutsch \cite{deutsch85quantum} realised
that if the laws of (quantum) physics are used as basis for computation,
an improved version of the Turing machine might lead to greater
computational power. This requires an extended version of the
Church-Turing thesis as well:

\begin{thesis}[Church-Turing-Deutsch]
Every physical process can be simulated by a universal computing device.
\end{thesis}

Greater computational power is meant in the sense that there are some
problems which can be efficiently solved by a universal computing
device according to Church-Turing-Deutsch but which cannot be
\emph{efficiently} solved by a Turing machine, \eg, the simluation
cost for a universal computing device on a normal turing machine would
be at least exponential for the most general case.\footnote{Note that
  there is a very prominent problem which illuminates this field:
  Factorising integers is possible at polynomial cost on a quantum
  computer, but the currently best known algorithms require
  exponential cost on a classical computer.  This does not prove,
  though, that quantum computers are \emph{per se} more powerful than
  classical ones because there is, for example, no proof that there is
  no classical polynomial-time factorisation algorithm.}

Until now, it has not been proven if quantum Turing machines have greater 
computational power than Turing machines or not; cf. 
Refs.~\cite{cleve,aharonov} for further information on this and related 
topics.

\subsection{Characteristics of quantum computers}
The basic building block for quantum computers are \emph{quantum bits}
(qbits), \ie, quantum mechanical objects which can be represented by
two different basis states, usually denoted by \(\ket{0}\) and
\(\ket{1}\). This can, \eg, be realised by two-spin systems such as
electrons, with the two different polarisations of a photon etc. In
contrast to classical machines and models of computation, quantum
computing can draw from two additional resources:

\begin{itemize}
\item Superpositions: The state of a quantum bit can be in a
  superposition \(\alpha\ket{0} + \beta\ket{1}\) with \(\alpha, \beta
  \in \mathbbm{C}\) and \(|\alpha|^{2} + |\beta|^{2} = 1\).  If every
  qbit of a quantum register consisting of \(n\) elements is brought
  into a symmetric superposition with \(\alpha = \beta = 2^{-1/2}\),
  the register contains \emph{all} numbers from \(0\) to \(2^{n}-1\)
  at the same time. Manipulations of the register thus manipulate all
  these numbers in one step, whereas classically, the number of
  required manipulations would grow exponentially with the register
  size. This feature is conventionally referred to as \emph{quantum
    parallelism}.
  \item Entanglement: Two parts of a quantum system can be in an entangled 
    state such that manipulations on one part of the system 
    influence the other system although they may be spatially separated.
\end{itemize}

While superpositions are useful for computational problems,
entanglement is suitable for tasks like secret key growing, but
is for example also necessary to connect input with output 
registers for quantum function application.

In general, there are several equivalent models for quantum computers
which are abstracted from their physical realisation:

\begin{itemize}
  \item Quantum Turing machines
  \item Quantum gates
  \item Universal quantum programming languages
\end{itemize}

\noindent All these models are equivalent, cf., 
\eg,~Refs.\cite{nielsen_chuang,aharonov,preskill} for more detailed
explanations.

\subsection{Static typing, functionality and runtime errors}\label{qprog:funct}
Selinger proves in \cite{selinger_qpl} that QPL has the ability to
avoid runtime errors by detecting them at compile time (and can thus
reject the program) which is not possible in other language proposals
presented at the time of writing. He argues that this must be
attributed to the static type system of the language. Although the
proof for this is solely based on properties of the static type
system, the functional style of QPL does have its merits in this
respect in our opinion, too: It restricts the language to elements
which allow to express universal programs, but do not allow
constructions that can produce runtime errors that cannot be detected
at compile time.\footnote{A static type system is in general not
  enough to ensure this property: Just think of the many possible ways
  to generate runtime-errors in C, which is statically typed as well.
  This is partially caused by the fact that typing in C is weak;
  nevertheless, even strongly typed languages as, \eg, Java and C\#
  still cannot prevail runtime errors. Thus, static typing alone is
  not sufficient to avoid all possible runtime errors.}

When work on this thesis was started, one of the points we
wanted to investigate was if and how principles of functional
languages could be advantageous for quantum languages.  This did not
reach fruition, though, because we could not find a simple way to
transfer any advanced functional method like higher-order functions,
recursively defined data types etc. to the quantum case.\footnote{To
  our knowledge, no fully satisfying mechanism for any of these
  problems has been found until now although the topic is addressed in
  several papers. }

From a physical perspective, the absence of runtime-errors is much
more interesting than any computer science related question like the
type system or functionality. In this thesis, we thus concentrated on
preserving the possibilities of static checking provided by QPL as
far as possible while enhancing usability of the language and
providing means to handle commuication.

\section{Introduction to QPL and cQPL}
This section presents a very short introduction to cQPL; some of the
things stated in the following also hold for a (regularised version)
of block QPL\footnote{The variant of QPL with a textual structure that
  includes blocks; alternatively, there is a textual variant without
  blocks and a flow diagram representation.} as presented in
\cite{selinger_qpl} which is used as basis for cQPL. Note that cQPL is
explicitely meant to be an \emph{experimental} compiler. Only very
modest effort was made to make the compiler easy to use (the error
messages provided can, for example, often only be understood if the
user is familiar with the inner working of the compiler), and only
very few classical operations which are considered standard components
of programming languages (\eg, numerical operations like \(\sin,
\cos\) etc.) were implemented to save time since this is purely
routine work.  Nevertheless, care has been taken to design the
compiler for easy extensibility.\footnote{For example, integration of
  the \(n\)-qbit Fourier gate could be accomplished by adding only
  \(8\) lines of code to the compiler and \(10\) to the runtime
  library. }


\subsection{Model of computation}\label{qprog:comm_model_section}
\begin{figure}[htb]
  \centering\includegraphics[width=0.75\textwidth]{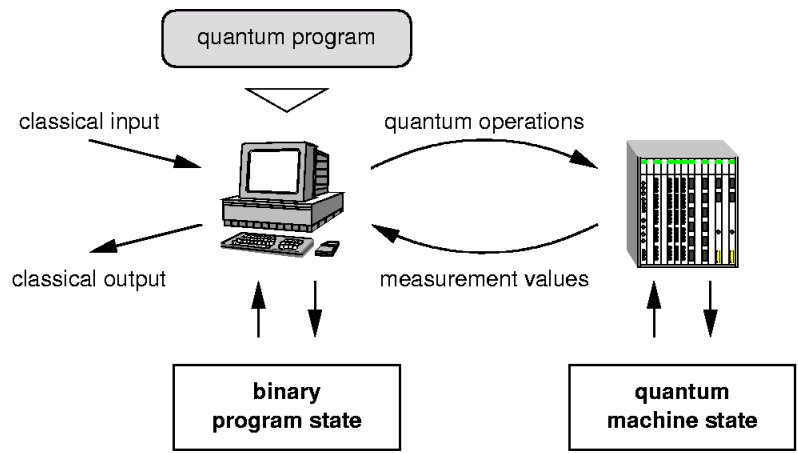}
  \caption{Hybrid architecture for a quantum computer which consists of
    a classical computer and a quantum memory with the ability to 
    apply unitary operators and perform measurements at the disposal
    of the classical system (image taken from \cite{oemer_phd}).}%
  \label{qprog:knill_model}
\end{figure}

Although quantum gates are the most widespread model in the literature
on quantum algorithms, the QRAM model suggested by Knill~\cite{knill}
is more apt as basis for quantum programming languages. The model
consists of two components: A \emph{classical computer} for conventional 
tasks like program flow control, classical calculations etc.,
and a \emph{quantum memory} controlled by the classical computer that
can not only store quantum states, but also apply any unitary operator
and perform measurements. Figure~\ref{qprog:knill_model} visualises
the approach.

cQPL is centred on the assumption that the control flow of a program
can be described by classical means. Although this is considered to
be a loss of generality by some authors (most notably \cite{oemer_phd}),
it does not represent a real restriction because all known quantum
algorithms can be expressed in this framework. Data is, of course, 
quantum mechanical, and can be modified by unitary operators which are 
equivalently called gates in analogy to the gate model of quantum computation.

\subsection{Language elements}
The following remarks provide an introduction to cQPL for ordinary users;
a formal version of the syntax will be presented in 
Appendix~\ref{comp:formal_syntax}. The compiler distribution provides some
example programs which demonstrate all features.

\subsubsection{Identifiers and variables}
Identifiers for variables can be denoted by strings consisting of the 
characters \texttt{A}--\texttt{Z}, \texttt{a}--\texttt{z}, 
\texttt{\_} and \texttt{0}--\texttt{9}, where no digit must be
at the beginning. Allocation of new variables is done with the
operator \texttt{new}:

\begin{verbatim}
new int loop := 10;
new qbit b1 := 1;
\end{verbatim}

Note that it is mandatory to provide an initial value both for 
classical and quantum data types. By default, the data types 
\texttt{bit}, \texttt{int}, \texttt{float}, \texttt{qbit} and
\texttt{qint} are available.

Assigning values to classical variables is possible using the
operator \texttt{:=}:

\begin{verbatim}
loop := loop - 1;
\end{verbatim}

\subsubsection{Arithmetic and logical expressions}
Arithmetic expressions can be given as in most programming languages
(\eg, \texttt{7+3*x-5}). The operators \texttt{+}, \texttt{-}, \texttt{*}
and \texttt{/} are available, they are overloaded to work with any
classical numerical data type.

Operators resulting in a logical value (\texttt{true} or \texttt{false})
are \verb/</, \verb/>/, \verb/<=/, \verb/>=/, \texttt{=} and 
\texttt{!=}; logical negation is
available using \texttt{!}, and predicates can be combined using
\texttt{\&} (and) and \texttt{|} (or). Operator priority 
is defined as usual in arithmetic and logic; parentheses can be
used to explicitely modify this.

\subsubsection{Procedures}
Procedures can take an arbitrary number of input parameters (including
none) which are specified as \texttt{name:type} tuples. Declarations
take the following form:

\begin{verbatim}
proc test: a:int, b:bit, c:float, d:qbit {
...
}
\end{verbatim}

\noindent Procedure calls are achieved with the keyword \texttt{call}:

\begin{verbatim}
(eins, zwei, drei) := call test(a0, a1, a2, b1);
\end{verbatim}

Note that the parameters are passed by value so that modifications of
the classical variables given to the procedure (in this case
\texttt{a0}, \texttt{a1} and \texttt{a2}) are not visible in the
caller's scope.  This means that no matter what \texttt{test} does,
these variables have the same values before and after the procedure is
called. This is not the case with \texttt{b1} because it is a quantum
variable and cannot be cloned to implement call-by-value; any
modifications performed by \texttt{test} on the state of\texttt{b1}
are visible for the caller after \texttt{test} returns.

The result of a procedure is a tuple containing the values the
classical parameters had at the end of procedure execution; the
example stores these in the local variables \texttt{eins},
\texttt{zwei} and \texttt{drei}. Note that the result of a procedure
call may be ignored as well by the caller, so the following variant is
also possible:

\begin{verbatim}
call test(a0, a1, a2, b1);
\end{verbatim}


The example program \texttt{proc\_test.qpl} which accompanies the
compiler further illustrates the described behaviour, so we refer the
reader to it.

\subsubsection{Gates}
Gate application is performed using the operator \texttt{*=} as in the 
following example:

\begin{verbatim}
q *= Not;
\end{verbatim}

\noindent A small number of elementary gates is built into the language 
core (remember that additional ones can be added with really little
effort as we already mentioned before):

\begin{itemize}
  \item \texttt{H} Hadamard transformation on a qbit.
  \item \(\text{\texttt{FT}}(n)\) Fourier transformation on \(n\)
    qbits, \ie, the \(n\)-fold tensor product of Hadamard transforms.
  \item \texttt{NOT} Logical negation on one qbit.
  \item \texttt{CNot} Controlled Not on two qbits.
  \item \texttt{Phase} Phase shift gate on one qbit; the desired shift
    is given as parameter.
\end{itemize}

The dimension of the gate and the destination must match. Variable tuples
where identifiers are combined with commas (\texttt{,}) can be used to 
combine several quantum variables as in the following example:

\begin{verbatim}
new qbit test1 := 0;
new qbit test2 := 1;
test1, test2 *= CNot; 
test1 *= Phase 0.5;
\end{verbatim}

User-defined gates can be defined by enclosing a list of (complex)
numbers in \texttt{[[} and \texttt{]]} as in the following example:

\begin{verbatim}
test1,test2 *= [[0.5,  0.5,   0.5,  0.5,
                 0.5,  0.5i, -0.5, -0.5i,
                 0.5, -0.5,   0.5, -0.5, 
                 0.5, -0.5i, -0.5,  0.5i]];
\end{verbatim}

\subsubsection{Control flow}
If-then-else and While are available for directing the control flow. 

\begin{verbatim}
new int loop := 10;
while (loop > 5) do {
    print loop;
    loop := loop - 1;
};

if (loop = 3) then {
    print "3";
}
else {
    print "Nicht 3";
}
\end{verbatim}

\noindent The meaning of these operations is the same as in classical 
languages.

\subsubsection{Other features}
Several more features which do not fit into any of the above
categories are available:

\begin{itemize}
  \item \texttt{dump} takes one or more quantum variable identifiers as
    argument and provides a dump of the current probability spectrum in
    the canonical basis \(\ket{0}, \ket{1}\):

\begin{verbatim}
dump eins, zwei;
\end{verbatim}

    \noindent To demonstrate the effect of this command, consider the
    following example:

\begin{verbatim}
new qbit a := 0;
new qbit b := 0;
print "State before FT:"; dump a, b;
a, b *= FT(2);
print "State after FT:";  dump a, b;
\end{verbatim}
    
    \noindent The program fragment produces the following output when
    run:

\begin{verbatim}
State before FT: 1 |00>
State after FT: 0.25 |00>, 0.25 |01>, 0.25 |10>, 0.25 |11>
\end{verbatim}
    
    \noindent Note that there is a fundamental difference between dumping
    the state of quantum variables and measuring the state and printing
    the result, as the following example shows:

\noindent\begin{verbatim}
measure a then { print "a is |0>"; } else { print "a is |1>"; };
print "State of b:";  dump b;
print "State of (a,b):"; dump a,b;
\end{verbatim}

    \noindent If the fragment above is supplemented by these lines, running
    the program either yields this

\noindent\begin{verbatim}
a is |0>
State of b: 0.5 |0>, 0.5 |1>
State of (a,b): 0.5 |00>, 0.5 |01>
\end{verbatim}
    
   \noindent or that output (the output of the statements before is omitted):

\noindent\begin{verbatim}
a is |1>
State of b: 0.5 |0>, 0.5 |1>
State of (a,b): 0.5 |10>, 0.5 |11>
\end{verbatim}    
    
    The result of the command \texttt{dump b} never changes, no
    matter how often the program is run. The result of the
    \texttt{measure} command, however, will change so that every
    output appears with probability \(0.5\) for a large number
    of executions.
  \item \texttt{skip} does nothing, but can be used to fulfil syntactic
    requirements if, for example, one branch of an if-then-else-statement
    is supposed to do nothing:

\begin{verbatim}
if (condition) then {
    skip;
}
else {
    ...
}
\end{verbatim}
  \item \texttt{print} prints the value of a variable or an arithmetic 
    expression (\eg, \texttt{print 5+7; print a;}), but can also be
    used to output text strings enclosed in quotation marks to the
    console (\eg, \texttt{print "Hello, world!";}).
  \item \texttt{measure} measures a quantum variable and returns 
    a classical result. The result is governed by a probability distribution
    according to the state the quantum variable is in; thus, successive
    program runs will in general return different results when the
    function is called. Note that the measurement is always performed
    in the standard basis \(\ket{0}, \ket{1}\) for each contributing
    qbit.
\end{itemize}

\subsection{Modelling quantum communication with cQPL}\label{qprog:qcom}
Although quantum communication has already been implemented with
several physical schemes at the time of writing, cQPL does not
consider any of these solutions. Instead, the model presented in
Section~\ref{qprog:comm_model_section} is extended in such a way that
the peculiarities of communication can be replaced by reasonable
simulation alternatives. In a real physical setting, communication
between two parties can be implemented by using whatever kind of
quantum channel which allows to transfer quantum states. Since this is
not quite easy to implement experimentally (quantum states are very
fragile objects), diverse effects like channel loss, decoherence etc.
need to be taken into account in a real-world setting. This is
accounted for by a \emph{channel model} which describes these effects
mathematically.

\subsubsection{Quantum channels}\label{qprog:quantum_channels}
\begin{figure}[htb]
  \centering\includegraphics{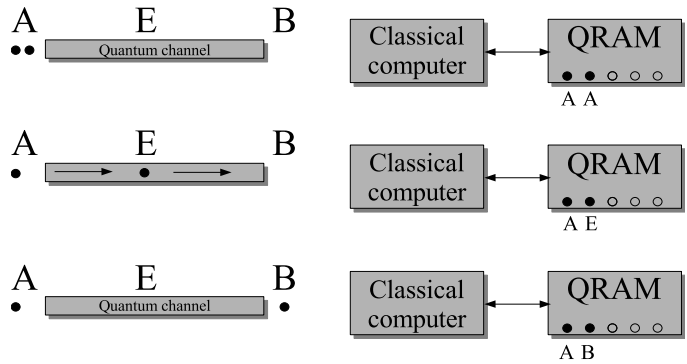}
  \caption{Model of communication used in cQPL. Transmission
    of quantum data is replaced by a quantum heap shared by the
    communicating parties; all qbits have a unique label that identifies
    to whom they belong at the moment. Sending and receiving can be
    modelled by changing the label, the channel itself is modelled by
    a third party.}\label{qprog:comm_model}
\end{figure}

Obviously, no spatial transmission of quantum states is performed in
the simulation.\footnote{It would be possible, for example, to
  consider networked computers between which simulated quantum states
  can be transferred. This would add no new physical insights to the
  problem, but only add technical difficulties, so we did not
  implement such a scheme.}  A quantum channel is thus replaced with a
\emph{label} on each qbit present in the simulation which denotes the
respective owner. Sending a qbit is thus equivalent to changing the
label of it.  Note that this definition differs from the definition of
a channel which is used in some contributions to the literature, \eg{}
\cite{keyl}.  Here, a channel is seen as something that modifies the
state, for example by decoherence, loss or the influence of
eavesdroppers, whereas our definition captures the notion of a channel
as a means of unambiguous quantum state transfer. Inclusion of
eavesdroppers (or any modification of the quantum state) is possible
if a quantum channel is replaced by two quantum channels which are
connected by a third module (which we \texttt{E} for Eve following the
usual convention) which receives quantum states from Alice, performs
appropriate modifications and resends the states to Bob. Obviously,
Eve can collect multiple qbits that pass the channel and manipulate
them collectively to pursue different attack strategies, so this is
in no way a restriction to intercept-resend attacks.
Figure~\ref{qprog:comm_model} provides a visualisation of the
communication model.

\subsubsection{Modules and communication primitives}
Communicating systems are specified in terms of \emph{modules} which
contain the code the participants execute. Every module is identified
by a unique label which identifies it among the participants. Note
that module definitions must only occur at the top-level; modules
must not contain other module definitions. Two parties which are
named Alice and Bob can be implemented by this structure:

\begin{verbatim}
module Alice {
   ...
};

module Bob {
   ...
};
\end{verbatim}

Channels for sending and receiving are implicitly opened between all
participants. The \texttt{send} command is provided for sending
variables.  Two parameters need to be supplied: A list of variables
(which may be classical or quantum) and the identifier of the
receiver. For example, the following code can be placed in module
\texttt{Alice}:

\begin{verbatim}
new qbit q1 := 0;
new qbit q2 := 1;
...
send q1,q2 to Bob;
\end{verbatim}

Receiving works similar; consider for example the following code which
might be located in module \texttt{Bob}:

\begin{verbatim}
...
receive var1:qbit, var2:qbit from Alice;
...
// Do something with var1, var2
\end{verbatim}

Note the a receive commands implicitly introduces new variables into
the present frame, in this case \texttt{var1} and \texttt{var2}. The
data type of the received quantities must be specified after the
variable name where a colon is used as separator.

\section{Example programs}
We present three small, but complete examples together with their output
generated by executing them to allow a look at the cQPL syntax without
resorting to program fragments.

\subsection{Random coin tossing}
This simple algorithm puts a quantum bit into a symmetric superposition
and measures it to simulate the effect of tossing a perfect coin:

\begin{verbatim}
new qbit q := 0;
q *= H;
measure q then { print "Tossed head"; } else { print "Tossed tail"; };
\end{verbatim}

\noindent The output is with equal probability either \texttt{Tossed head}
or \texttt{Tossed tail}, obviously.

\subsection{Distribution of an EPR pair}
Distributing EPR pairs is one possible method to establish a secret
key between Alice and Bob that can, \eg, be used for a Vernam cipher.
The following cQPL program shows how to do this for a single EPR
pair:

\begin{verbatim}
module Alice {
  proc createEPR: a:qbit, b:qbit {
       b *= Not;
       b *= H;
       a,b *= CNot;
  } in {
    new qbit first := 0;
    new qbit second := 0;
    call createEPR(first, second);
    send second to Bob;
    measure first then { print "Alice's qbit is |1>"; } 
                  else { print "Alice's qbit is |0>"; };
  };
};

module Bob {
   receive q:qbit from Alice;
   measure q then { print "Bob's qbit is |1>"; } 
             else { print "Bob's qbit is |0>"; };
};
\end{verbatim}

\noindent Running the program creates one of the following two outputs
with equal probability:

\noindent\begin{tabular}{p{\textwidth/2-5mm}p{\textwidth/2-5mm}}
\noindent\begin{verbatim}
Alice's qbit is |0>
Bob's qbit is |0>
\end{verbatim} &
\noindent\begin{verbatim}
Alice's qbit is |1>
Bob's qbit is |1>
\end{verbatim}
\end{tabular}

Note that the order of Alice and Bob's output may be inverted as well
because after and before the send/receive synchronisation, the
execution order of the threads representing Alice and Bob is
indeterministic.\footnote{To be precise: The execution order depends
  on how the computer used for the simulation implements threads and
  their parallel execution, so it is effectively indeterminate. On
  machines with at least two CPUs and assuming that each thread runs
  on one of them, the indeterminism is real.}

\subsection{Quantum teleportation}
Quantum teleportation is an algorithm that enables to transfer an
unknown quantum state between two parties if both of them share one
part of an EPR state (in this case, \(\ket{\beta_{00}}\)) and can
communicate classically. An easy calculation as is, for example, given in
\cite[p. 27]{nielsen_chuang} or any other quantum information text
shows how this works, so we will not repeat it here. The
implementation in cQPL is as follows:

\begin{verbatim}
module Alice {
   proc createEPR: a:qbit, b:qbit {
      a *= H;
      b,a *= CNot;  /* b: Control, a: Target */
   } in {
     new qbit teleport := 0;  /* Apply unitary operations to set the qbit 
                                 to any other desired state */     
     new qbit epr1 := 0;
     new qbit epr2 := 0; 
  
     call createEPR(epr1, epr2); 
     send epr2 to Bob;        

     teleport, epr1 *= CNot;  /* teleport: Control, epr1: Target */

     new bit m1 := 0;
     new bit m2 := 0;
     m1 := measure teleport;
     m2 := measure epr1;

     /* Transmit the classical measurement results to Bob */
     send m1, m2 to Bob; 
   };
};


module Bob {
   receive q:qbit from Alice;
   receive m1:bit, m2:bit from Bob;

   if (m1 = 1) then {
      q *= [[ 0,1,1,0 ]];  /* Apply sigma_x */
   };

   if (m2 = 1) then {
      q *= [[ 1,0,0,-1 ]];  /* Apply sigma_z */
   };

   /* The state is now teleported */
   print "Teleported state:";
   dump q;
};
\end{verbatim}

\newpage\thispagestyle{plain}

\chapter{A compiler for cQPL}{Denn was wir tun m{\"u}ssen, nachdem wir es 
gelernt haben, das lernen wir, indem wir es tun.}%
{\hfill Aristoteles, Nikomachische Ethik}\label{cqpl_compiler} A
compiler for cQPL was implemented as part of this thesis; since no
quantum computers are available yet, it is obviously targeted at
simulators for such. This chapter presents a very short overview about
the compiler's implementation and the limitations which arise from the
experimental nature of it. Note that this chapter is intentionally
kept as terse as possible; it is not supposed to be a detailed
description of the techniques used in implementing the compiler nor is
its intention to go down to the source code level.  

\section{Structure and implementation}
Since the initial intention when work on the thesis started was to
examine the aptness of functional methods in programming languages for
quantum computers, we decided to implement the compiler for cQPL in a
functional language as well (observe the remarks in
Section~\ref{qprog:funct} why functionality was realised not to be the
important factor). The choice after testing several alternatives fell
on Objective Caml (OCaml) which is, for example, described on
\url{http://caml.inria.fr}. One of the reasons for this choice was
that very good automated generators for lexers and parsers are
available as part of the compiler distribution which considerably
reduce routine work. As simulation backend, we
used the routines supplied with \"Omer's QCL compiler \cite{oemer_msc}
because this library was the most advanced one at that time.  In the
meanwhile, a number of other libraries appeared and existing ones
matured, so most likely, we would have chosen a different simulation
backend now because the QCL library comes effectively without any
documentation for the programmer (only the QCL compiler built on top
of the library is documented) which resulted in quite a few technical
obstacles.

A simple compiler for a simple classical language was implemented as
part of this thesis to provide an example for explaining
compiler technique and to get used to the OCaml language and the
associated tool-chain.

\subsection{Compiler technique}
\begin{figure}[htb]
  \centering\includegraphics{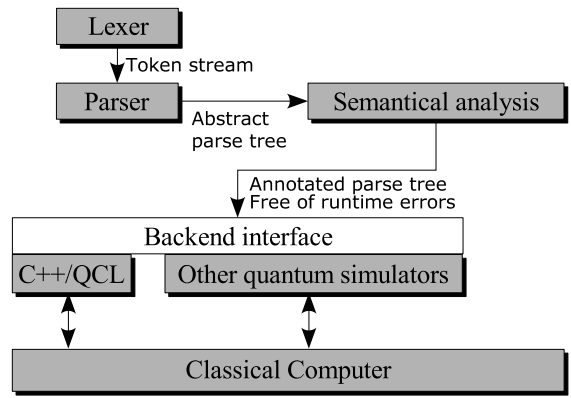}
  \caption{Structure of the cQPL compiler}\label{compiler:structure}
\end{figure}

As usual in compiler technique, the work is separated into several
passes which are shown in Figure~\ref{compiler:structure}.

Lexical analysis is performed by a lexer generated with OCamllex. The
lexer transforms the input stream of characters into a stream of
tokens where tokens are, for example, keywords like \texttt{int},
\texttt{measure} and so on. This simplifies the parsing process
because it does not need to deal with a program representation at the
level of single characters, but can already work with less elementary
units. A more interesting part is the syntactical analysis where the
parser determines if the program is valid according to the rules given
in Section~\ref{comp:formal_syntax}. The syntax for cQPL is specified
as a LALR(1) grammar which is a special kind of an LR(1) grammar
which, in turn, is a variant of a context free grammar as introduced
in Definition~\ref{form:context_free}.  We do not want to go into more
technical details here, but refer to Appendix~\ref{glossary} and the
usual literature on the topic, \eg.
Refs.~\cite{aho_sethi_ullmann,appel,modern_compiler,wilhelm_maurer}
for the exact definitions of such grammars.

After the syntactical correctness of a program has been ensured, the
compiler can analyse the generated parse tree to infer the meaning of
the program and perform compile time checks. These try to find as many
errors in the program as possible to ensure that they will not appear
at runtime and lead to abortion of the program with an error. Such
checks include, for example:

\begin{itemize}
  \item Making sure that all variables were declared before use.
  \item Checking that procedures are called with proper arguments.
  \item Matching the dimension of quantum gates with the dimension
    of variables they are applied to.
  \item Ensuring that tuples of quantum variables are disjoint.
\end{itemize}

Many more checks of this kind can be found in the source code. Some
(\eg, the first mentioned two) appear in classical programming
languages as well, whereas others (\eg, the last mentioned two) are
specific to quantum languages or necessary only for our approach.  The
goal of the analysis phase is to provide a representation of the
program that is augmented with everything necessary so that the
code-generation backend can produce its result directly from this
representation. An additional advantage of this strategy is that the
code generation can be replaced by a variant targeting a different
simulator.\footnote{For a very early version of the compiler, a code
  generation backend for the Fraunhofer simulator \cite{fraunhofer}
  was provided as well, but we dropped support for this to not spend
  too much time on implementation issues.}  The structure of the
annotated parse tree which is passed from the parser to the code
generation backend is documented in the source code.

\subsection{The implementation}
We do not want to go into any implementation details here, but just
provide an overview about the source files which constitute the
compiler. Every detail of the implementation can obviously be found
there.

\begin{itemize}
  \item \texttt{lexer.mll} is an input file for OCamllex which generates
    the lexical analyser for QPL from the rules given in the file.
  \item \texttt{parser.mly} is an input file for OCamlyacc which is
    used to transfer the rules (and code) given in the file into a
    parser for cQPL syntax.
  \item \texttt{parser\_defs.ml} defines the data structures used by
    the parser to build the abstract syntax tree. For every element of
    the grammar, a separate data structure is defined. Most of them
    can be augmented with additional information which is inferred
    during analysis, \eg, lists of type conversions.
  \item \texttt{cqpl.ml} plays a twofold role: On the one hand, it
    implements the user interaction part which handles command line
    processing and directs the different passes of the compiler. On
    the other hand, it implements the semantic analysis where for
    every grammar production, a corresponding function is available to
    perform the appropriate checks.
  \item \texttt{gen\_qcl.ml} implements the code generation backend
    for the QCL library. The generated code depends on the C++ code in
    \verb/qpl_runtime.{cc,h}/ which provides the runtime
    environment for programs. This handles, for example, quantum
    memory management and provides implementations of standard
    operators. \texttt{qpl\_run\-time\_temp\-late.h} contain the
    implementation of runtime routines which use templates and thus
    cannot be compiled to a static object; the routines to implement
    the communication operations (\ie, thread handling, data exchange
    and locking) can be found in \verb/qpl\_runtime\_comm.{cc,h}/.
  \item \texttt{type.ml} contains the type checker and routines
    to perform lossy and non-lossy conversion between different
    data types. Especially interesting here is the conversion between
    classical and quantum variables. The routines provided by this file
    are much more general than required by the compiler and would in
    principle be able to handle mixed data types as well.
  \item \texttt{stacked\_env.ml} provides a stackable environment for
    the semantic analysis.
\end{itemize}

Since the compiler is only one part of this thesis, we deliberately
accepted some limitations which would have to be removed for a
production quality compiler. None of them would present a problem in
principle, but would require some tedious effort that does not justify
the gain:

\begin{itemize}
  \item Error messages are not detailed, and no effort is made to continue
    translation as far as possible or perform error recovery after
    a mistake was spotted.
  \item No specific error productions are provided. Syntactical errors
    are therefore in general reported by the lexer and not properly
    by the parser.
  \item There is absolutely no optimisation for execution speed.
  \item Communication was not too intensively tested because we concentrated
    more on the formal aspects of this. Additionally, it uses big
    locks instead of finer-grained solutions which reduces performance.
  \item There is no set of standard routines or a standard library of
    any kind which would be required for real-world applications.
\end{itemize}

\subsection{Implementation of communication}
Since we do not use a real network of quantum and classical computers
to simulate communicating parties, but only a single system, it is
necessary to find an approach that emulates the characteristic
properties of real-world systems in the simulation environment. It has
already been shown that the QRAM model can -- together with an
appropriate labelling mechanism for the qbits -- provide a suitable
replacement for a quantum channel.  The remaining issue that needs to
be addressed is how the model the independent execution of the parties
together with the synchronisation when send/receive operations take
place is to be implemented. This is, obviously, a problem of the
backend, so the solution presented is specific to the QCL backend.

For every module defined in a program, a separate thread\footnote{A
  \emph{thread} is -- depending on the implementation by the
  underlying library and the operating system kernel -- a lightweight
  process that shares most ressources he has with other threads that
  execute in parallel, but has its own control flow.} is created which
has access to the global memory management routines and communication
channels. Quantum memory management is performed by the main process;
appropriate locking techniques ensure that no race conditions can
occur when quantum bits are allocated by the modules. Sending quantum
data can in this scheme be performed by exchanging pointers to the
respective positions on the quantum heap; the static analysis of the
compiler guarantees that no more than one process is in possession of
a given qbit at a time. Again, appropriate mechanisms in the form of
mutexes are used to provide the required synchronisation between the
modules which is required to implement send and receive operations.
For every pair of modules, a separate bi-directional queue is provided
to handle the exchange of quantum and classical data.

\section{Using the compiler}
As all well behaved programs, the compiler can provide a list of its options:

\begin{verbatim}
wolfgang@meitner> ./cqpl --help

Quantum Programming Language v1.0

Usage: qpl [<input>] [<output>] [--debug] [--nonative] [--norun] 
           [--backend qcl] [--qheap size]
   <input>: Input filename
   <output>: Output filename
  --debug Print debug messages
  --backend Simulation backend (Only qcl is supported at the moment)
  --nonative Generate only backend code, don't create a native executable
  --norun Do not execute the generated native code
  --qheap Size of quantum heap (default: 200 qbits)
  -help  Display this list of options
  --help  Display this list of options
\end{verbatim}

\begin{itemize}
  \item \verb/<input>/ and \verb/<output>/ provide the names of the
    input and output file. If no input file is specified, the compiler
    reads from the stdin channel. If no output if specified, the
    filename of the input file together with some appropriate 
    extension according to the output mode is used.
  \item \texttt{--debug} reports lots of debug messages. Console output
    is really noisy with this option enabled, but provides some
    insight into what the compiler does. Obviously mainly useful 
    for debugging the compiler itself.
  \item \texttt{--nonative} specifies that no native code is generated,
    \ie, the program is not transformed into machine code by the backend.
    For the QCL backend, this means that only a C++ version of the
    quantum program is generated, but the C++ compiler is not called
    to generate a native executable.
  \item \texttt{--norun} does not execute the generated program, but
    only leaves the executable file which can be run later.
  \item \texttt{--qheap size} sets the size of the quantum heap to
    \texttt{size} qbits.
\end{itemize}

The compiler source code will be made available in the near future on 
\url{http://kerr.physik.uni-erlangen.de/qit/qpl.html} once the remaining
changes to make the source code ready for public distribution have been
performed.
\newpage\thispagestyle{plain}
\forgechapter{4}
\chapter{Mathematical structures}{A mathematician is a device for turning 
coffee into theorems.}{\hfill P\'al Erd\H os}%
\label{math:chapter}
Because our work involves the theoretical parts of physics and
computer science, many different notations and conventions enter the
game. To eschew obfuscation, it behoves to define a consistent
notation to be used in this work, which we shall do in the following
sections. Additionally, this chapter introduces some mathematical
structures and their properties which are required for the description
of the denotational semantics of cQPL.

\section{Algebraic structures}
\subsection{Fundamentals}\label{math:algebra}
The concept of an algebra is convenient to cover the properties of
quantum mechanics in an abstract setting, so we remind the reader of
two elementary definitions for terms that are often used sloppily
in physics.

\begin{definition}[Algebra]
Let \(\mathbbm{K}\) be a field. An associative \(\mathbbm{K}\)-algebra 
\(\mathcal{A}\) over \(\mathbbm{K}\) is a nonempty set \(A\) 
together with three operations called addition~\(+\), 
multiplication~\(\times\) and scalar multiplication~\(\cdot\) (the last 
two operations are usually denoted by juxtaposition of symbols) for which 
the following properties hold:

\begin{itemize}
  \item \(\mathcal{A}\) is a linear space under addition and scalar 
    multiplication.
  \item \(\mathcal{A}\) is a ring under addition and multiplication.
  \item If \(r \in \mathbbm{K}\) and \(a,b \in A\), then 
    \(r\cdot (a\times b) = (r\cdot a)\times b = a\times (r\cdot b)\).
\end{itemize}
\end{definition}

\begin{definition}[Subalgebra]
\(S \subseteq A\) is subalgebra of an algebra 
\(\mathcal{A}\) if it has the properties of an algebra and is closed 
under operations of \(\mathcal{A}\).
\end{definition}

\section{Linear operators}
\subsection{General}
Much of our work is based on the grounds of linear operators, so we
present some fundamental definitions and cite a theorem from the
literature which will be useful for us later on.

\begin{remark}
Note that we use only linear operators in this work, so \emph{operator}
is used as a synonym for \emph{linear operator} without mentioning this 
explicitely in the following chapters.
\end{remark}

\begin{definition}[Linear operator]
A \emph{linear operator} \(T\) from a normed space \(X\) to another
normed space \(Y\) is a linear map from \(D(T) \subseteq X\) (the
\emph{domain} of \(T\)) to \(Y\) with the following property for
\(x,y \in D(T)\), \(\alpha, \beta \in \mathbb{K}\):
\begin{equation}
T(\alpha x + \beta y) = \alpha T(x) + \beta T(y).
\end{equation}
\end{definition}

\begin{definition}[Bounded operator]
An operator is called \emph{bounded} if \(\exists C\geq 0, C\in \mathbbm{R}\) 
so that
\begin{equation}
\norm{Tx} \leq C\cdot \norm{x}
\end{equation}
for all \(x \in D(T)\) with \(\norm{x} \leq 1\).
\end{definition}

\begin{theorem}\label{math:props_lin_operators}
Let \(X,Y\) be normed spaces. For a linear operator \(T: X\rightarrow Y\), the
following properties are equivalent:

\begin{itemize}
  \item \(T\) is continuous in every point of \(D(T)\).
  \item \(T\) is continuous at \(0\).
  \item \(T\) is bounded.
\end{itemize}
\end{theorem}

\begin{proof}
Cf. Ref.~\cite[Theorem 2.1]{weidmann}.
\end{proof}

\subsection{Hilbert-Schmidt operators}\label{math:hilbert_schmidt}
Let \(X,Y\) be Hilbert spaces. An operator \(K \in \mathcal{B}(X,Y)\) is called
\emph{Hilbert-Schmidt-operator} if there exists an orthonormal basis
\(\{e_{\alpha}: \alpha \in A\}\) (where \(A\) is some index set) with 
\(\sum_{\alpha\in A}\norm{Ke_{\alpha}}^{2} < \infty\). In a more physical
notation, this means that \(\trace{K^{\dagger}K} < \infty\). This
is obviously fulfilled if \(K \in \BH\) and 
\(\dim(\mathcal{H}) < \infty\).

\begin{theorem}[Hilbert space of Hilbert-Schmidt operators]\label{hs}
For Hilbert-Schmidt operators \(K,L\) of a Hilbert space \(X\) to a Hilbert
space \(Y\), \(\normhs{\cdot}\) is a norm 
on this space induced by the scalar product
\begin{equation}
\iphs{K}{L} := \sum_{\alpha}\ip{Ke_{\alpha}}{Le_{\alpha}}.
\end{equation}
Physicists generally write this as:
\begin{equation}
\iphs{K}{L} = \trace{K^{\dagger}L}.
\end{equation}
\end{theorem}

\begin{proof}
If \(K\) is a Hilbert-Schmidt operator, \(aK\) is a Hilbert-Schmidt operator 
as well for every 
\(a \in \mathbb{K}\). If \(K, L\) are HS operators, 
then for every orthonormal basis \(\{e_{\alpha}\}\), the following 
equation holds:
\begin{equation}
\sum_{\alpha}\norm{(K + L)e_{\alpha}}^{2} \leq 
   2\cdot\sum_{\alpha}\left(\norm{K e_{\alpha}}^{2} + 
   \norm{L e_{\alpha}}^{2}\right) < \infty,
\end{equation}
\ie, \(K + L\) is a Hilbert-Schmidt operator as well. 
By \(\ip{\cdot}{\cdot}\), we denote the scalar product in the space of 
Hilbert-Schmidt operators, and  \(\normhs{K} = \iphs{K}{K}^{1/2}\) 
(as usual, the scalar product induces a metric).
\end{proof}

A comparison of both spaces can be found in 
Table~\ref{math:h_hs_comp}.\footnote{Note that there would be many
  other different choices how the norm for Hilbert spaces could
  be defined which all fulfil the required properties of a norm that
  can, \eg, be found in \cite{achieser_glasmann}.}

\begin{table}[htb]
\newlength{\fr}\setlength{\fr}{1cm}
\centering
\noindent\begin{tabular}%
  {p{\fr+0.5cm}|p{\textwidth*\real{0.3}-\fr/2}|p{\textwidth*\real{0.5}-\fr/2}}
 & Hilbert space & Hilbert space of Hilbert-Schmidt operators \\\hline
State & \(\ket{f} \in \mathcal{H}\) 
      & \(\op{D}: \mathcal{H} \rightarrow \mathcal{H}\)\\
Operator & \(\mathcal{H} \rightarrow \mathcal{H}\): 
                                            \(\op{D}\ket{x} = \ket{x'}\)
         & \(\Lambda: \op{D}\rightarrow\op{D} \equiv 
           \left(\mathcal{H} \rightarrow \mathcal{H}\right) \rightarrow 
           \left(\mathcal{H} \rightarrow \mathcal{H}\right)\)\\
Norm & \(\left|\ket{f}\right| = \sqrt{\braket{f|f}}\) 
     & \(||\op{D}||_{\text{HS}} = \sqrt{\trace{D^{\dagger}D}}\) \\
\parbox[t]{\fr+0.5cm}{\raggedright Operator norm} & 
              \(||\op{D}||_{\text{sup}} = 
              \sup\limits_{\substack{\ket{f}\in\mathcal{H}\\|f|\leq 1}}|
              \op{D}\ket{f}|\)
              & \(\norm{\Lambda} = 
  \sup\limits_{\substack{\op{D}\in\mathcal{H}\\||\op{D}||\leq 1}}
  \Lambda(\op{D}) =
  \sup\limits_{\substack{\op{D}\in\mathcal{H}\\||\op{D}||\leq 1}}
            \trace{\Lambda(\op{D})^{\dagger}\Lambda(\op{D})}\)\\
\end{tabular}
\caption{A comparison between general Hilbert spaces and Hilbert spaces
with Hilbert-Schmidt operators as basis.}\label{math:h_hs_comp}
\end{table}

\section{Connection with quantum mechanics}\label{sec:math:qm}
Quantum mechanics is one of the most advanced theories in physics; most
research performed today is centred around it. There are many ways
to describe the theory mathematically; for our purposes, an abstract
algebraic formulation based on the principles and structures introduced
in the previous part seems to be most apt. In the following, we present
a concise overview about the central elements of the theory, following
the structure of the review given in Ref. \cite{keyl}. More elaborate
introductions can be found in the usual textbooks on the topic,
\eg,~\cite{sakurai}. For those especially interested on the
impact of quantum mechanics on information theory, 
Refs.~\cite{nielsen_chuang, preskill,keyl} can be especially recommended.

Probabilistic processes lie at the very heart of quantum mechanics: 
Predictions made by the theory hold only in the sense that 
\emph{probabilities} for outcomes of measurements can be provided. A 
large number of repeated experiments with the same preparation parameters 
results in a distribution that states how many times a certain outcome will
appear. Exactly this distribution can be predicted by theory. The outcome 
of a single measurement can in general never be forecast with certainty. 

\subsection{States and effects}\label{sec:math:states_and_effects}
Ideally, an experiment resulting in a probability distribution can be
carried out by repeating the following two processes until a sufficient
amount of statistics has been gathered:

\begin{itemize}
  \item \emph{Preparation} of a quantum mechanical state according 
    to some fixed procedure that can be repeated a sufficient 
    number of times.
  \item \emph{Measurement} of some \emph{observable} quantity, \eg,
    spin, energy, etc. \emph{Effects} are a special class of measurements
    which can result in either the answer ``yes'' or ``no'' according
    to some probability distribution.
\end{itemize}

Note that we do not only deal with purely quantum mechanical states,
but may also encounter a mixture between classical and quantum mechanical
properties (which are usually termed \emph{hybrid} systems) that our formalism 
must be able to account for. Measurement results fall into
the classical category since gauges in the macroscopic world are used
to infer them from the quantum system.\footnote{The problem of how 
  measurements of a quantum system are to be interpreted (or even 
  how the whole process can be described consistently) has been and still
  is one of the fundamental philosophical problems of 
  quantum mechanics. We take a pragmatic point of view here and do not 
  consider the problem in greater detail, but refer to the literature,
  \eg, \cite{auletta}.}

Every quantum system can be completely characterised by its observable
quantities which in turn are characterised by self-adjoint operators.
These operators form an algebra \(\mathcal{A}\) as introduced in
Section~\ref{math:algebra}; since we do only deal with
finite-dimensional Hilbert spaces here, we can restrict ourselves to
subalgebras of \(\BH\), \ie, \(\mathcal{A} \subset \BH\).
\(\mathcal{A}\) is called the \emph{observable algebra} of the system
and is often identified with the system itself because it is possible
to deduce all properties of the system from its observable algebra.
The \emph{dual algebra} of \(\mathcal{A}\) is denoted by
\(\mathcal{A}^{*}\) and is the algebra defined on the dual space.

To capture the notions of \emph{state} and \emph{effect} mathematically,
we introduce two sets according to the following definition:
\begin{align}
  \mathcal{S}(\mathcal{A}) &= \Set{\varrho \in \mathcal{A}^{*} | \varrho \geq
    0 \wedge \varrho(\mathbbm{1}) = 1}\\
  \mathcal{E}(\mathcal{A}) &= \Set{A \in \mathcal{A} | A \geq 0 \wedge A
    \leq \mathbbm{1}} 
\end{align} 
\(\mathcal{S}\) represents the set of states, while \(\mathcal{E}\)
contains all effects. For every tuple \((\varrho,A) \in
\mathcal{S}\times\mathcal{E}\), there exists a map
\((\varrho,A)\rightarrow \varrho(A)\in [0,1]\) which gives the
probability \(p = \varrho(A)\) that measuring an effect
\(A\) on a (system prepared in the) state \(\varrho\)
results in the answer ``yes''. Accordingly, the probability for the
answer ``no'' is given by \(1-p\). \(\varrho(A\)) is called the
\emph{expectation value} of a state \(A\); states are thus defined as
expectation value functionals from an abstract point of view.  These
expectation value functionals can by uniquely connected with a
normalised trace-class\footnote{For trace-class operators, the trace
  is independent of the basis chosen to evaluate the trace.} operator
\(\varrho\) such that \(\varrho(A) = \trace(\varrho A)\). In
principle, it would be necessary to introduce two different symbols
for the expectation value functional and the operator, but
for simplicity, we omit this complication.

We have to distinguish between two different kinds of states:
\emph{Pure} and \emph{mixed} ones. This is a consequence of the fact that
both \(\mathcal{S}\) and \(\mathcal{E}\) are convex spaces: For two 
states \(\varrho_{1}, \varrho_{2} \in \SA\) and \(\lambda \in \mathbbm{R},
0 \leq \lambda \leq 1\), the convex combination \(\lambda\varrho_{1} + 
(1 - \lambda)\varrho_{2}\) is also an element of \(\SA\). The same statement
holds for the elements of \(\EA\). This decomposition provides a very nice
insight into the structure of both spaces. Extremal points in this space
cannot be written as a proper convex decomposition, \ie, 
\(x = \lambda y + (1-\lambda)z \rightarrow \lambda = 1 \vee \lambda = 0
\vee x = y = z\). These can be interpreted as follows:

\begin{itemize}
  \item For \(\SA\), they are \emph{pure states} with no associated 
    classical uncertainty.
  \item For \(\EA\), they describe measurements which do not allow any
    fuzziness as is, \eg, introduced by a detector which detects some property
    not with certainty, but only up to some finite error (alas, all
    real-world detectors).
\end{itemize}


\subsection{Observables}
Until now, we have only been talking about effects, \ie,
``yes''/``no'' measurements, but not about measurements with a more
complicated result range which are necessary to describe general
\emph{observables}.  Although we would have to consider an infinite
(even uncountable) number of possible outcomes for a general
description of quantum mechanics, it is sufficient to consider only
observables with a finite range for our purposes.\footnote{This is
  justified because quantum computers process states of the type
  \((\ket{0},\ket{1})^{\otimes n}\). Although quantum computers can
  possess an arbitrary number of qbits, it is still a fixed and (which
  is most important) finite number; additionally, we do not care for
  any continuous quantum properties of these objects.}  Such
observables are represented by maps which connect elements \(x\) of a
finite set \(R\) to some effect \(E_{x} \in \EA\); this in turn gives
rise to a probability distribution \(p_{x} = \varrho(E_{x})\).  More
formally, we can put it as in the following:

A family \(E = \{E_{x}\}, x \in R\) of effects \(E_{x} \in \mathcal{A}\)
if called a \emph{positive operator valued measurement} (POVM) on
\(R\) if \(\sum_{x\in R}E_{x} = \mathbbm{1}\). 

Note that the \(E_{x}\) need \emph{not} necessarily be projectors, \ie,
\(E_{x}^{2} = E_{x}\). Should this nevertheless be the case \(\forall x\),
the measurement is called a \emph{projective measurement}.

Observables of this kind can be described by self-adjoint operators
of the underlying Hilbert space \(\mathcal{H}\) which can (without any 
claim of formal correctness or even a proof) be seen as follows:
Every self-adjoint operator \(A\) on a Hilbert space \(\mathcal{H}\) of
finite dimension can (because of the spectral theorem,
cf., \eg, \cite{achieser_glasmann,weidmann}) be decomposed into the form
\(A = \sum_{\lambda\in\sigma(A)}\lambda P_{\lambda}\) where
\(\sigma(A)\) denotes the spectrum of \(A\) and \(P_{\lambda}\) the
projectors onto the corresponding eigenspace. The expectation value
\(\sum_{\lambda}\lambda\varrho(P_{\lambda})\) of \(P\) for a given 
state \(\varrho\) can equivalently be calculated by 
\(\varrho(A) = \trace(\varrho A)\). Since this is the standard way of
formulating the expectation value of an operator, both points of view
coincide. 

\subsection{Classical components}\label{math:classical_components}
Systems consisting solely of quantum components are generally not
to be found: At the latest after a measurement has been performed,
classical probabilities need to be accounted for. Therefore, we need
to pay attention to hybrid systems composed from quantum and classical
parts as well. Obviously, we have to orient ourselves along the lines
of Section~\ref{sec:math:states_and_effects} to provide proper
grounding for both possibilities. Consider a finite set \(X\) of
elementary events, \ie, all possible outcomes of an experiment.
Again, \(\SA\) and \(\EA\) define the set of states and effects,
respectively, but this time, the observable algebra is given by all
complex valued functions from the set \(X\) to \(\mathbbm{C}\) as
defined by
\begin{equation}
\mathcal{A} = \CX = \Set{f: X \rightarrow \mathbbm{C}}.\label{math:class_map}
\end{equation}
By identifying the function \(f\) with the operator \(\hat{f}\) given by
\begin{equation}
  \hat{f} = \sum_{x\in X}f_{x}\ketbra{x}{x}
\end{equation}
where \(\ket{x}\) denotes a fixed orthonormal basis, the probability
distribution can be interpreted as an operator algebra similar 
to the quantum mechanical case because \(\hat{f}\) is obviously an element
of \(\BH\). Thus, \(\CX\) can be used as an observable algebra \(\mathcal{A}\)
along any other quantum mechanical or classical constituent of a
multipartite composite system.

\subsection{Composite and hybrid systems}
Since quantum mechanical and classical systems can be described with very
similar structures, the presented formalism is obviously well suited
for the presentation of composite systems. Let \(\mathcal{A} \subset \BH\)
and \(\mathcal{B} \subset \mathcal{B}(\mathcal{K})\) be systems 
given in terms of their observable algebras; the composite system is then
given by
\begin{equation}
\mathcal{A}\otimes \mathcal{B} \equiv \operatorname{span}
           \Set{A \otimes B | A \in \mathcal{A}, B \in \mathcal{B}}.
\end{equation}
Three cases for the choice of \(\mathcal{H}, \mathcal{K}\) can be
distinguished:

\begin{itemize}
  \item If both systems are quantum, then \(\mathcal{A}\otimes\mathcal{B}
    = \mathcal{B}(\mathcal{H}\otimes\mathcal{K})\).
  \item If both systems are classical, then \(\mathcal{A}\otimes\mathcal{B}
    = \mathcal{C}(X\times Y)\) with \(\mathcal{C}\) as defined by
    Eqn.~\ref{math:class_map}
  \item If \(\mathcal{A}\) is classical and \(\mathcal{B}\) is
    quantum mechanical, we have a \emph{hybrid} system; the composite
    observable algebra is then given by \(\mathcal{C}(X)\otimes\BH\)
    which cannot be simplified any further. Observables are operator-valued 
    functions in this case, as expected.
\end{itemize}

\section{Domain theory}
The definition of a proper and sound mathematical semantics for a 
programming language necessitates apt structures which can
be used as a solid ground underlying the work. The method we use --
denotational semantics -- is conventionally based on 
\emph{semantic domains}, which in turn rely on partial orders and 
recursion theory. The purpose of this section is to introduce
the elements required for the semantic description of cQPL which will
be given in Chapter~\ref{chap:formal_semantics}; more details are available, 
\eg, in \cite{gunther_scott,winskel}.

\subsection{Basic definitions}
\begin{definition}[Partial order]
  A \emph{partial order} \((P, \sqsubseteq)\)is a set \(P\) on which
  there is a binary relation \(\sqsubseteq\) for which the following
  properties hold \(\forall p,q,r\in P\):

\begin{itemize}
  \item \(p \sqsubseteq p\) (reflexive)
  \item \(p \sqsubseteq q\) and 
    \(q \sqsubseteq r \Rightarrow p \sqsubseteq r\) (transitive)
  \item \(p \sqsubseteq q\) and \(q \sqsubseteq p \Rightarrow p = q\) 
    (antisymmetric)
\end{itemize}
\end{definition}

\begin{definition}[Upper bound]
For a partial order (\(P, \sqsubseteq\)) and a subset \(X \subseteq P\),
\(p \in P\) is an \emph{upper bound} of \(X\) if and only if 
\(\forall q \in X: q \sqsubseteq p\).

\noindent The element \(p\) is a \emph{least upper bound} if:

\begin{itemize}
  \item \(p\) is an upper bound of \(X\)
  \item For all upper bounds \(q\) of \(X\), \(p \sqsubseteq q\)
\end{itemize}
\end{definition}

\begin{remark}
Note that it follows from the definition that the least upper bound is
unique.
\end{remark}

\begin{definition}[\(\omega\)-chain]
Let \((D, \sqsubseteq_{D})\) be a partial order. An \(\omega\)-chain of
the partial order is an increasing chain \(d_{0} \sqsubseteq_{D} d_{1} 
\sqsubseteq_{D} \cdots \sqsubseteq_{D} d_{n} \sqsubseteq \cdots\) of elements
of the partial order. Note that \(\omega\) represents the increasing chain of
natural numbers \(\mathbbm{N}_{0}\).
\end{definition}

\begin{definition}[Complete partial order]
The partial order \((D, \sqsubseteq_{D}\)) is a \emph{complete partial
order} (cpo) if it has least upper bounds of all \(\omega\)-chains, \ie,
any increasing chain \(\{d_{n}|n\in \omega\}\) of elements in \(D\) has
a least upper bound \(\sqcup\{d_{n}|n\in \omega\}\), written as
\(\sqcup_{n\in \omega}d_{n}\). \((D, \sqsubseteq_{D})\) is a cpo with
bottom if it is a cpo which has a bottom element (often also
called least element) \(\perp_{D}\) for which 
\(\perp_{D} \sqsubseteq d\ \forall d\in D\) holds.
\end{definition}

\begin{definition}[Directed-complete partial order]
A partial order \((D, \sqsubseteq)\) in which every directed subset has
a supremum is called \emph{directed}-complete partial order (dcpo).
\end{definition}

\begin{definition}[Monotone function]\label{scott_monotone}
A function \(f: D \rightarrow E\) between cpos \(D\) and \(E\) is
\emph{monotonic} if and only if \(\forall d, d' \in D\):
\begin{equation}
d \sqsubseteq d' \Rightarrow f(d) \sqsubseteq f(d').
\end{equation}
\end{definition}

\begin{definition}[Continuous function]\label{scott_cont}
A function \(f: D \rightarrow E\) between cpos \(D\) and \(E\) is
\emph{continuous} if and only if it is monotonic and for all chains
\(d_{o} \sqsubseteq d_{1} \cdots \sqsubseteq d_{n} \sqsubseteq \cdots\)
in \(D\) there holds

\begin{equation}
\bigsqcup_{n\in \omega}f(d_{n}) = f\left(\sqcup_{n\in\omega}d_{n}\right).
\end{equation}
\end{definition}

\subsection{A fixed point theorem}
\begin{definition}[Fixed point]
Let \(f:D\rightarrow D\) be a continuous function on a cpo \(D\) with bottom
\(\bot_{D}\). A \emph{fixed point} of \(f\) is an element \(d \in D\) such that
\(f(d) = d\). A \emph{prefixed point} of \(f\) is an element \(d \in D\)
such that \(f(d)\sqsubseteq d\).
\end{definition}

\begin{theorem}[Fixed-point theorem]\label{math:fixed_point_theorem}
Let \(f: D\rightarrow D\) be a continuous function on a cpo with a bottom
\(D\). Define
\begin{equation}
\fix{f} = \bigsqcup_{n\in \omega}f^{n}(\perp).
\end{equation}
Then \(\fix{f}\) is a fixed point of \(f\) and the least prefixed
point of \(f\), \ie:

\begin{itemize}
  \item \(f(\fix{f}) = \fix{f}\) 
  \item If \(f(d) \sqsubseteq d\) then \(\fix{f}\sqsubseteq d\)
\end{itemize}

\noindent Consequently, \(\fix{f}\) is the least fixed point of \(f\).
\end{theorem}

\begin{proof}

\noindent It follows from continuity of \(f\) that 
\begin{align}
f(\fix{f}) & = f\left(\sqcup_{n\in\omega}f^{n}(\perp)\right)
= \bigsqcup_{n\in\omega} f^{n+1}(\perp) \\
& = \bigsqcup_{n\in\omega}\left\{\perp \cup \left\{f^{n+1}(\perp)|n\in\omega\right\}\right\}
= \bigsqcup_{n\in\omega} f^{n}(\perp)\label{math:fix_penultimate}\\
& = \fix{f}.
\end{align}

Thus \(\fix{f}\) is a fixed point because \(f(\fix{f}) = \fix{f}\) is
exactly the required property of a fixed point (adding \(\perp\) in step 
\(\ref{math:fix_penultimate}\) is justified because the least upper bound
is not influenced by this). Suppose \(d\) is a prefixed point.
Certainly, \(\perp \sqsubseteq d\). By monotonicity, 
\(f(\perp) \sqsubseteq f(d)\). But \(d\) is a prefixed point, \ie, 
\(f(d) \sqsubseteq d\), so \(f(\perp) \sqsubseteq d\), and by induction
\(f^{n}(\perp) \sqsubseteq d\ \forall n\in\omega\). 
Thus, \(\fix{f} = \sqcup_{n\in\omega}f^{n}(\perp)\sqsubseteq d\)
\end{proof}

\begin{remark}
Note that it is customary to define
\begin{equation}
\Ycomb_{D} f \equiv \bigsqcup_{n\in\omega}f^{n}(\bot)
\end{equation}
such that \(\Ycomb_{D}\) is obviously a function 
\((D \rightarrow D) \rightarrow D\) which maps functions to their fixed
points; it is thus termed the \emph{fixed point combinator}.
\end{remark}

\begin{definition}[Scott topology\footnote{This definition
  is taken from \cite{abramsky_jung}.}]\footnote{A topological space
    is a set \(X\) together with a collection \(T\) of subsets where
  the empty set and \(X\) are in \(T\), the union of any collection of 
  sets in \(T\) is in \(T\) and the intersection of any pair of sets 
  in \(T\) is also in \(T\).}
Let \(D\) be a dcpo. A subset \(A\) is called \emph{Scott closed} if it is a
lower set\footnote{A lower set is a finite, non-empty downward-closed subset
of a partial order, \ie, \(\{x | x\sqsubseteq y\}\).} and is closed under 
suprema of directed subsets.\footnote{A subset \(A\) of a poset
  is directed if it is nonempty and each pair of elements has an
  upper bound in \(A\).}
Complements of closed sets are called 
\emph{Scott open}; they are the elements of \(\sigma_{D}\), the 
\emph{Scott topology} on \(D\).
\end{definition}

\begin{theorem}\label{math:top_scott_cont}
A function \(f: D_{n} \rightarrow D_{m}\) is continuous in the 
sense of definition \ref{scott_cont} (\ie, Scott continuous) if
it is topologically continuous with respect to the Scott topology.
\end{theorem}

\begin{proof}
\Cf{} Ref. \cite[Theorem 2.3.4]{abramsky_jung}
\end{proof}

The last definition and theorem are quite technical, but we need them
for the proof of Theorem~\ref{math:ext_loew_complete} later on.

\subsection{Constructions on domains}
We will use the following constructions on dcpos to create new dcpos:

\begin{itemize}
  \item \(D_{1} \times D_{2} \times \cdots \times D_{n}\) denotes
    \(n\)-tuples respectively cartesian domains. The weaker-than
    relation is defined such that
    \begin{equation}
      \left(x_{1}, x_{2}, \ldots, x_{n}\right) 
      \sqsubseteq_{D_{1} \times D_{2} \times \cdots \times D_{n}} 
      \left(y_{1}, y_{2}, \ldots, y_{n}\right) \Leftrightarrow 
      x_{i} \sqsubseteq_{D_{i}} y_{i}
    \end{equation}
    for \(i=1,\ldots, n\) and \(x_{i}\in D_{i}, y_{i} \in D_{i}\).
  \item \(D_{1} \otimes D_{2} \otimes \cdots \otimes D_{n}\) represents
    the \emph{smash product} which identifies all tuples that contain
    one or more \(\bot\)-elements. Example: \((d_{1}, \bot_{2}), 
    (\bot_{1}, d_{2})\) and \((\bot_{1}, \bot_{2})\) are all
    identified with a new bottom element \(\bot_{D_{1}\otimes D_{2}}\)
    for \(d_{i} \in D_{i}\)
    Formally, the new domain
    \(D \otimes E\) is the set
    \begin{equation}
      \{(x,y) \in D \times E | x \neq \bot \itext{and} y\neq\bot\} \cup
      \{\bot_{D\otimes E}\}.
    \end{equation}
  \item \(D_{1} + D_{2} + \cdots + D_{n}\) is the \emph{separated sum} domain 
    which consists of all elements in \(D_{i}\) together with a
    new bottom symbol \(\bot_{D_{1} + D_{2} + \cdots + D_{n}}\) (usually
    abbreviated to \(\bot_{D}\)).
  \item The \emph{coalesced sum} \(D_{1} \oplus D_{2} \oplus \cdots
    \oplus D_{n}\) is similar to the separated sum, but the new bottom
    element \(\bot_{D}\) is gained by identifying all elements 
    \(d_{1} + d_{2} + \cdots + d_{n}\) (\(d_{i} \in D_{i}\)) which contain 
    one or more of \(\bot_{D_{i}}\). 
  \item \emph{Lifting} is the operation that adds bottom element to a
    domain \(D\); the result is denoted by \(D_{\bot}\); continuity is
    not influenced by this.
\end{itemize}

\summaryeven{We have provided the basic framework required to build the
  denotational semantics of cQPL. This framework is composed of two
  parts: On the one hand, we need an abstract representation of
  quantum mechanics to account for the physical properties of the
  programming language.  On the other hand, the concept of partial
  orders builds the basis for defining semantic domains, \ie, the
  space which will be used to place the equations describing the
  semantics of cQPL in. The choice of partial orders for that is
  especially justified by the fact that fixed points can be
  constructively obtained in them. These in turn are required to solve
  recursive domain equations that will be needed to give a denotation
  for several language constructs of cQPL, most important the
  communication features.}

\section{cp-Maps and their representation}
In quantum mechanics, time evolution is described by transformations of
density matrices with an operator \(\Lambda\) that is called
a \emph{superoperator} \cite{preskill,nielsen_chuang,keyl}.

\begin{definition}[Superoperator]\label{superop_def}
A superoperator \(\Lambda: \BH\rightarrow\BH\) has the following properties 
for all density operators \(\varrho \in \mathcal{D}\) with 
\(\varrho' = \Lambda(\varrho)\):

\begin{itemize}
  \item \(\Lambda\) is linear.
  \item \(\varrho^{\dagger} = \varrho \Rightarrow  \varrho'^{\dagger} 
    = \varrho'\) (hermeticity is preserved).
  \item \(\trace{\varrho'} = 1\) if \(\trace{\varrho} = 1\) 
    (trace preserving).
  \item \(\Lambda \otimes \mathbbm{1}\) is semidefinite positive
    (\(\forall n\in\mathbbm{N}:\Lambda\otimes \mathbbm{1}_{n} \geq
    0\)), \ie, \(\Lambda\) is a completely positive map. In other
    words, this means that \(\Lambda\) is not only semidefinite
    positive (\(\varrho'\) is nonnegative if \(\varrho\) is
    nonnegative) on \(\mathcal{H}_{A}\), but also on any possible
    extension \(\mathcal{H}_{A}\otimes\mathcal{H}_{B}\).
\end{itemize}
\end{definition}

Note that if dissipative processes (\eg, postselection of observed events) 
are considered, the second condition is loosened to
\(\trace(\varrho') \leq 1\).

\subsection{Operator-sum representation}
Kraus~\cite{kraus} proved a result about the decomposability of
completely positive maps which is ubiquitous in quantum 
information theory:

\begin{theorem}[Kraus representation theorem]
  A superoperator \(\Lambda\) as defined in Def.~\ref{superop_def} can
  be written as a partition of \(\mathbbm{1} =
  \sum_{k=1}^{N}A_{k}^{\dagger}A_{k}\) where \(A_{k}\) are linear
  operators acting on the Hilbert space of the system such that
\begin{equation}
\varrho' = \Lambda(\varrho) = \sum_{k=1}^{N}A_{k}\varrho A_{k}^{\dagger}\ 
 \forall \varrho \in \mathcal{D}
\end{equation}
for any density matrix \(\varrho\) that represents a mixed or a pure state.
\end{theorem}

\begin{proof}
\Cf{} Ref. \cite{nielsen_chuang,preskill,kraus}
\end{proof}

To illustrate this representation, consider the situation that the
system under consideration is in contact with a much larger environment, 
a common situation for physical problems. Together, both systems form
a closed quantum system. State transformations in this combined system
can be described by a unitary transformation \(U \in 
U(\dim(\mathcal{H})\cdot \dim(\mathcal{H}_{\text{env}}))\) where 
\(\mathcal{H}\) denotes the Hilbert space of the system under consideration
and \(\mathcal{H}_{\text{env}}\) the Hilbert space of the environment.
Assume that the environment is in a pure state
\(\ketbra{e_{0}}{e_{0}}\).\footnote{This assumption
  holds without loss of generality because it can be shown that 
  a system can be purified by introducing extra dimensions which do
  not have any physical consequences.}
The density operator of the system under consideration \emph{after} the
unitary operation was applied to the total system can be
recovered by tracing out the environment:
\begin{align}
  \varrho' &= \Lambda(\varrho) = 
  \trace(U\varrho\otimes\ketbra{e_{0}}{e_{0}}U^{\dagger})\\
    &= \sum_{k}\bra{e_{k}}U(\varrho\otimes\ketbra{e_{0}}{e_{0}})
                                               U^{\dagger}\ket{e_{0}}\\
    &= \sum_{k}\bra{e_{k}}U\ket{e_{0}}\varrho\bra{e_{0}}
                                               U^{\dagger}\ket{e_{k}} \\
    &= \sum_{k}A_{k}\varrho A_{k}^{\dagger}.
\end{align}
In the last step, \(A_{k}\) is defined by \(A_{k}\equiv
\bra{e_{k}}U\ket{e_{0}}\).

\begin{remark}
We say that a set of Kraus operators \(\{A_{k}\}\) \emph{implements}
a cp-map \(\Lambda\) if \(\forall \varrho \in \mathcal{D}:
\sum_{k}A_{k}\varrho A_{k}^{\dagger} = \Lambda(\varrho)\).
This simplifies the further description.
\end{remark}

\begin{theorem}
The operation elements of a given superoperator \(\Lambda\) are not
unique: If \(\{E_{j}\}\) is a set of Kraus operators, then a different
set of Kraus operators \(\{F_{k}\}\) describes the same operation if 
and only if there exists a unitary matrix \(U \in U(n)\) with
\(n = \card(\{E_{k}\})\) (where \(\card(X)\) is the cardinality of
the set \(X\)) such that
\begin{align}
F_{k} = \sum_{j}U_{kj}E_{j}\label{kraus_unitary}.
\end{align}
Note that the shorter set may be padded with zero elements until the
cardinality of both matches.
\end{theorem}

\begin{proof}
Cf., \eg, Ref.~\cite[Theorem 8.2]{nielsen_chuang} or Ref.~\cite{preskill}.
\end{proof}

\begin{remark}
  Let \(\{A_{k}\}\) be a set of Kraus operators that represents the
  cp-map \(\Lambda\). Note that if any number of elements \(A_{i}\) is
  taken from \(\{A_{k}\}\), the set still remains a completely
  positive map, but is not trace preserving any more.
\end{remark}

\begin{remark}
Note that superoperators are elements of \(\BH\) which makes it possible
to apply many theorems of linear operator algebra to superoperators.
In fact, superoperators can be used as elements of a Hilbert space
as defined in Section~\ref{math:hilbert_schmidt}. The distinction between 
operators and superoperators in physics is therefore in general superfluous. 
\end{remark}

\begin{remark}
It can be shown that the number of Kraus elements needed to express any
arbitrary completely positive map 
\(T: \mathcal{B}(\mathcal{H}_{1}) \rightarrow \mathcal{B}(\mathcal{H}_{2})\)
is bounded by \(\dim(\mathcal{H}_{1})\cdot\dim(\mathcal{H}_{2})\),
confer, \eg, \cite[p. 102]{preskill}).
\end{remark}

\subsection{Equivalence of Kraus operators}
The unitary connection between two sets of Kraus operators defined
in Equation \ref{kraus_unitary} gives rise to an equivalence relation 
between such sets. Two sets \(\{A_{j}\}\) and \(\{B_{k}\}\) are
members of the same equivalence class if there is a unitary matrix
which connects both representations:
\begin{equation}
A \cong B \Longleftrightarrow \exists U\in U(n): A_{i} = 
\sum_{j=1}^{n}U_{ij}B_{j}\ \text{with } i=1,\ldots,n.
\end{equation}
The set of all sets of Kraus operators inducing the same map \(\Lambda\) is 
defined in the obvious way:
\begin{equation}\label{math:kraus_set_equiv}
\mathcal{K}(\Lambda) \equiv \Big\{\{A_{k}\} \Big| 
\sum_{k}A_{k}\varrho A_{k}^{\dagger} =
\Lambda(\varrho)\ \forall \varrho \in \mathcal{D}\Big\}.
\end{equation}
If we talk about a \emph{set of Kraus operators} or simply 
\emph{Kraus operators} in the following, we always mean an arbitrary
set which is an element of the equivalence class inducing
the same cp-map (\ie, an element of \(\mathcal{K}(\Lambda)\)), but will not 
mention this explicitly every time.

\subsection{A partial order for Kraus operators}
The L\"owner partial order \cite{loewner} for two density operators \(A\), 
\(B:\mathcal{B}(\mathcal{H}_{1}) \rightarrow \mathcal{B}(\mathcal{H}_{2})\)
is given by 
\begin{equation}
A \sqsubseteq B \Longleftrightarrow (B - A) > 0.\label{eqn:kraus_po}
\end{equation}
This partial order can be extended to sets of Kraus operators by defining
\begin{align}\label{extended_loewner_po}
\{A_{i}\} \sqsubseteq \{B_{i}\} \Longleftrightarrow& \forall \varrho \in
\mathcal{D}\ \forall n \in \mathbbm{N}:\nonumber\\ 
&\left(\sum_{i}(B_{i}\otimes \mathbbm{1}_{n})\varrho 
        (B_{i}\otimes \mathbbm{1}_{n})^{\dagger} - 
\sum_{k}(A_{k}\otimes \mathbbm{1}_{n})\varrho 
               (A_{k}\otimes \mathbbm{1}_{n})^{\dagger}\right) > 0.
\end{align}

Partial orders are often interpreted as approximations: If an element \(A\)
is \emph{weaker than} \(B\) (\(A \sqsubseteq B\)), then \(A\) is said
to \emph{approximate} \(B\). This point of view will come handy when
we consider solutions of fixed point equations in the denotational description.

It is necessary for our work to see that Kraus operators form a complete
partial order. For this, observe first the following theorem:

\begin{theorem}\label{loewner_cpo}
The partial order on all density operators \(\varrho \in \mathcal{D}\) given
by the L\"owner partial order \(\sqsubseteq\) is complete.
\end{theorem}

\begin{proof}
Cf. Ref.~\cite[Proposition 3.6]{selinger_qpl}
\end{proof}

\noindent From this, we can deduce the required statement:

\begin{theorem}\label{math:ext_loew_complete}
The partial order for cp-maps defined by the extended L\"owner partial order 
given by Eqn.~\ref{extended_loewner_po} is complete, \ie, it forms a cpo.
\end{theorem}

\begin{proof}
  Let \(\{A_{1}\} \sqsubseteq \{A_{2}\}, \ldots \) be an increasing
  chain of topologically continuous (and therefore monotone because of
  \ref{math:top_scott_cont}, \ref{scott_cont} and
  \ref{scott_monotone}) Kraus operators. Because of
  Definition~\ref{extended_loewner_po}, the relation \(\varrho_{1}
  \sqsubseteq \varrho_{2}\) is preserved by applying \(\{A_{1}\},
  \{A_{2}\}\) with \(\{A_{1}\} \sqsubseteq \{A_{2}\}\) to
  \(\varrho_{1}, \varrho_{2}\). An \(\omega\)-chain of density
  operators is conserved if an increasing chain of Kraus operators is
  applied to it. Because of Theorem~\ref{loewner_cpo}, the fact that
  the previous consideration applies to all density operators in
  \(\mathcal{D}\) and the uniqueness of the least upper bound, the
  extended L\"owner partial order is complete as well.
\end{proof}

\subsection{Kraus aggregations}\label{math:kraus_aggregations}
We mentioned that superoperators applied to density matrices describe 
quantum mechanical processes. Operations performed one after another
can therefore be described by the consecutive application of the
corresponding superoperators:

\begin{equation}
\varrho' = \Lambda_{1}(\varrho), \varrho'' = \Lambda_{2}(\varrho') \Rightarrow
\varrho'' = \Lambda_{2}(\Lambda_{1}(\varrho))
\end{equation}

If the sets \(\{A^{1}_{k}\}\) and \(\{A^{2}_{k}\}\) implement \(\Lambda_{1}\)
and \(\Lambda_{2}\), then the same state transformation is given by

\begin{equation}
\varrho'' = \sum_{k}\sum_{l}A^{2}_{k}A^{1}_{l}\varrho{A^{1}_{l}}^{\dagger}
{A^{2}_{k}}^{\dagger}.
\end{equation}

We call a collection of sets of Kraus operators that are to be applied 
subsequently an \emph{aggregation of (sets of) Kraus operators} or
simply \emph{Kraus aggregation}; the Kraus sets involved
are written as a list of the form
\begin{equation}
\Gamma = \{A^{1}_{k}\}, \{A^{2}_{k}\}, \ldots, \{A^{n}_{k}\}
\end{equation}
The list \(\Gamma\) gives rise to the following quantum mechanical operation:
\begin{equation}
\Gamma(\varrho) = \varrho' = \sum_{k_{1}}\sum_{k_{2}}\cdots\sum_{k_{n}}
    A^{n}_{k_{n}}\cdots A^{2}_{k_{2}}A^{1}_{k_{1}}\varrho
    {A^{1}_{k_{1}}}^{\dagger}{A^{2}_{k_{2}}}^{\dagger}\cdots 
                                                  {A^{n}_{k_{n}}}^{\dagger}
\end{equation}
List concatenation is formally described by the operator \(\circ\):
\begin{align}
&\Gamma_{1} = \{A^{1}_{k}\}, \{A^{2}_{k}\}, \ldots, \{A^{n}_{k}\},\\
&\Gamma_{2} = \{B^{1}_{k}\}, \{B^{2}_{k}\}, \ldots, \{B^{m}_{k}\}\\
&\Rightarrow \Gamma_{1} \circ \Gamma_{2} \equiv
\{A^{1}_{k}\}, \{A^{2}_{k}\}, \ldots, \{A^{n}_{k}\},
\{B^{1}_{k}\}, \{B^{2}_{k}\}, \ldots, \{B^{m}_{k}\}
\end{align}
\ie, the effect of \(\Gamma_{1} \circ \Gamma_{2}\) on a state \(\varrho\)
is the same as if first \(\Gamma_{1}\) and then \(\Gamma_{2}\) would have
been applied. Note (since this is a potential source of confusion) that the 
list is ``processed'' from left to right, \emph{not} from right to left!

A Kraus aggregation can also consist of multiple sub-aggregations
which are prefixed by some scalar. Formally, we use the 
operator \(+\) to denote this:

\begin{equation}
\Gamma' = p_{1}\cdot \Gamma_{1} + \cdots + p_{n}\cdot \Gamma_{n}.
\end{equation}

If the \(p_{i} \in \mathbbm{R}\) are to be interpreted as
probabilities, the normalisation condition\footnote{The sum can be
  smaller than \(1\) to account for the possibility of non-termination
  which will happen with probability \(1-\sum p_{i}\).  It also allows
  to describe non trace-preserving effects.}  
is \(\sum_{n} p_{n} \leq 1\).  \(\Gamma'\) can thus be seen as a formal 
combination of lists. The interpretation of such an aggregation is 
straightforward: With probability \(p_{k}\), the Kraus aggregation 
\(\Lambda_{k}\) is selected whenever \(\Lambda'\) acts on a density 
operator. Obviously, lists of this form are apt to introduce mixed 
states into the Kraus list formalism. Consider, for example, the aggregation
\begin{equation}
  \Delta = \frac{1}{2}\cdot \{\text{NOT}\} + \frac{1}{2}\cdot\{\mathbbm{1}\}.
\end{equation}
The effect of it is to apply the unconditional not-operation
(which maps \(\ket{0}\rightarrow\ket{1}\) and \(\ket{1}\rightarrow\ket{0}\)
and may, for example, be implemented with \(\hat{\sigma}_{x}\)) 
with probability \(0.5\) and to leave the state unchanged with the same
probability. If this aggregation is applied to, \eg, the following (pure) 
density operator
\begin{equation}
  \varrho = \ketbra{0}{0},
\end{equation}
the resulting state is the impure density operator given by
\begin{align}
  \varrho' &= \Delta(\varrho) = \frac{1}{2}\{\text{NOT}\}(\varrho) + 
                                \frac{1}{2}\{\mathbbm{1}\}(\varrho) \\
           &= \frac{1}{2}\ketbra{1}{1} + \frac{1}{2}\ketbra{0}{0} 
            = \frac{1}{2}\{\ket{1}\} + \frac{1}{2}\{\ket{0}\}
\end{align}
which describes an impure mixture between \(\{\ket{0}\}\) and 
\(\{\ket{1}\}\).

\begin{remark}
Note that we will use Kraus lists prefixed with probabilities to
describe different measurement outcomes when we provide the semantics
of cQPL in Chapter~\ref{chap:formal_semantics}. The physical way to
think about such operations is to take a density operator
\(\varrho\) and apply the Kraus elements for the projective
measurements on it; this results in the state
\begin{equation}
\varrho' = \sum_{k}M_{k}\varrho M_{k}^{\dagger} = 
  \sum_{k}\mathcal{E}_{m}(\varrho)
\end{equation}
where \(M_{k}\) are the projection operators and
\(\mathcal{E}_{m}(\varrho) \equiv M_{k}\varrho M_{k}^{\dagger}\).  
The probability to obtain the measurement outcome \(k\) is given by
\begin{equation}
p(m) = \trace(\mathcal{E}_{m}(\varrho)).
\end{equation}
The probability factors in Kraus aggregations can be calculated in
exactly this way; both points of view therefore provide the same
information.
\end{remark}

Note that we allow the pre-factors of the sub-aggregations to 
depend on parameters which make the complete aggregation
dependent on the disjoint union of the set of parameters used for
the sub-aggregations. This is necessary to describe Kraus 
aggregations dependent on probability distributions which are unknown 
before a initial state is given or the
outcomes of some measurements are known. The following example shows a
Kraus aggregation where the first sublist depends on the 
parameters \(a_{1}\) and \(a_{2}\) and the second on 
\(a_{1}\) and \(a_{3}\); the complete aggregation obviously depends on 
\(a_{1}\), \(a_{2}\) and \(a_{3}\):
\begin{equation}
  \Gamma(a_{1}, a_{2}, a_{3}) =
       p(a_{1}, a_{2})\cdot\Gamma_{1} + p(a_{1}, a_{3})\cdot\Gamma_{2}.
\end{equation}
For a Kraus aggregation of the most general form (where \(\{a^{i}\}\) 
denotes the set of parameters for the \(i^{\text{th}}\) sub-aggregation) 
given by 
\begin{equation}
\Gamma(\cup_{i}\{a^{i}\}) = \sum_{i}p_{i}(\{a^{i}\})\Gamma_{i},
\end{equation}
the normalisation condition is obviously still given by
\begin{equation}
  \sum_{i} p_{i}(\{a^{i}\}) \leq 1
\end{equation}
which necessitates that \(0 \leq p_{i} \leq 1\ \forall i\) (this
is supposed to hold for all \(p\) used in the following).

It is possible to contract Kraus (sub-)aggregations which consist of more 
than one element to a shorter form because two Kraus sets \(\{A_{i}\}\) and
\(\{B_{i}\}\) can be contracted to a new set \(\{C_{k}\}\) which describes
the subsequent application of both initial sets, as the following simple
calculation shows:
\begin{align}
(\{A_{k}\}, \{B_{i}\})(\varrho) = \{B_{i}\}(\{A_{k}\}(\varrho)) &= 
                \sum_{k=1}^{N}\sum_{i=1}^{N} \hat{B}_{k}\hat{A}_{i}\varrho 
                                  \hat{A}_{i}^{\dagger}\hat{B}_{k}^{\dagger}\\
           &= \sum_{n=1}^{N^{2}} \hat{C}_{n}\varrho \hat{C}_{n}^{\dagger}%
              \label{math:kraus_std_form_wo_p}
\end{align}
with
\begin{equation}
  \hat{C}_{n} \equiv \hat{B}_{\lceil n/N\rceil}\hat{A}_{n\bmod N}.
\end{equation}
Recall that different set cardinalities can be compensated by adding an
appropriate number of zero operators to the smaller set.
Since the calculation is valid \(\forall\varrho\in\mathcal{D}\), the
new single element aggregation \(\{C_{i}\}\) is a unique replacement for 
the aggregation \(\{A_{i}\},\{B_{i}\}\). 


Based on this contraction, it is possible to define a standard
representation for Kraus aggregations which is easier to handle formally 
when aggregations must, for example, be compared. 

\noindent With 
\begin{equation}
\mathcal{P} \equiv \Big\{ \{p_{i}(\{a^{i}\})\} \Big|
                          a^{i}_{k} \in \mathbbm{R} \forall i,k, 
                          0 \leq p_{i}(\{a^{i}\}) \leq 1, 
                         \sum_{i} p_{i}(\{a^{i}\}) \leq 1\Big\}
\end{equation}
being the set of all possible parametrised probability distributions and
\begin{equation}
\mathcal{K} \equiv \Set{ \Lambda | \Lambda\ \text{is a cp-map}}
\end{equation}
being the set of all unparametrised Kraus aggregations contracted to 
the normal form given by Eqn.~\ref{math:kraus_std_form_wo_p}, we can finally 
define the set of \emph{all} possible Kraus aggregations formally by
\begin{equation}
\mathcal{A} \equiv \Big\{ \sum_{i} p_{i}\Lambda_{i} \Big|
                         \{p_{i}\} \in \mathcal{P} \wedge
                         \Lambda_{k} \in \mathcal{K}\Big\}.
\end{equation}

\subsubsection{A partial order for Kraus aggregations}
For a Kraus aggregation of the contracted normal form \(\Lambda =
\{C_{k}\}\), the definition for a partial order can be directly
transferred from Equation~\ref{eqn:kraus_po}. If the aggregation
contains sub-aggregations, \(\sqsubseteq\) is formally a function
dependent on the parameters of the aggregation: For \(\Gamma_{1} =
\Gamma_{1}(A_{1},\ldots,A_{n})\) and \(\Gamma_{2} =
\Gamma_{2}(B_{1},\ldots,B_{n})\), the partial comparison \(\Gamma_{1}
\sqsubseteq \Gamma_{2}\) becomes a function
\((A_{1},\ldots,A_{n},B_{1},\ldots,B_{n}) \rightarrow \{\text{true},
\text{false}\}\), \ie, the comparison depends on the parameters of
both sets of parameters involved. Note that this does not concern
Kraus aggregations where all coefficients have defined scalar values.
Basically, the parametrised comparision is nothing else than a
comparision of all elements of an unfolded Kraus aggregation as
defined in Section~\ref{form:state_transform} followed by folding
everything back afterwards.

\subsubsection{Equivalence of Kraus aggregations}\label{math:kraus_equiv}
One possible task of denotational semantics is to decide wether two
programs which \emph{look} different perform the same actions, \ie, if
their semantics coincide. This question is in general complicated to
answer constructively. Nevertheless, it is possible for some cases. We
will consider this problem in more detail in
Chapter~\ref{chap:formal_semantics}. At this point, we are interested
in the question when two Kraus aggregations are semantically
equivalent, \ie, induce the same physical operations. The method used
for this is almost identical to the method used for Kraus sets.
Consider two aggregations \(\Gamma_{1}\) and \(\Gamma_{2}\) given in
the contracted normal form, \ie,
\begin{align}
  \Gamma_{1} &= \sum_{k=1}^{N}p_{k}^{1}\Lambda_{k}^{1}\\
  \Gamma_{2} &= \sum_{k=1}^{N}p_{k}^{2}\Lambda_{k}^{2}.
\end{align}

Let \(\sym(M)\) by the symmetric group over the finite set \(M\). 
Both lists are equivalent if (but not only if) the following condition holds:
%
\begin{align}
  \Gamma_{1} \cong \Gamma_{2} \Leftrightarrow & 
                   \exists \varphi\in \sym([1,\ldots,N])
                   \forall k\in [1,N]\forall\varrho\in\mathcal{D}:\\
                 & p^{1}_{k}(A^{1}_{k}) = 
                 p^{2}_{\mathcal{P}(k)}(A^{2}_{\mathcal{P}(k)})
    \wedge\Lambda_{k}^{1}(\varrho) = \Lambda_{\mathcal{P}(k)}^{2}(\varrho) .
 \end{align}
Note that this equivalence requires that the same Kraus operators are
used in both lists; it is nevertheless possible that a different set
of Kraus operators prefixed by another probability distribution induces
the same action. The criterion given here is thus sufficient, but not 
necessary.

The set \(\mathcal{E}\) of all aggregations that are equivalent in
this sense can be defined analogous to Eqn.~\ref{math:kraus_set_equiv}:
\begin{equation}
\mathcal{E}(\Lambda) \equiv \Set{\Lambda_{i} \in \mathcal{A} | 
                     \Lambda_{i} \cong \Lambda}.\label{math:kraus_agg_equiv}
\end{equation}

This definition is not very satisfying from a constructive point of
view: There is no simple way to systematically decide if the effects
of two aggregations coincide. This can be improved by giving an
explicit criterion for the equivalence between two Kraus aggregations.
We consider the special case of two lists which are composed of the
same operators, but are ordered differently.  This happens, for
example, when statements in a program are reordered.  With the method
given below, we can thus get a criterion to decide if such
reorderings preserve the semantics of programs which is a very
important case.

Unparametrised Kraus aggregations can always be written in the
standard form given by Eqn.~\ref{math:kraus_std_form_wo_p} and are
thus equivalent to a Kraus set; this again is equivalent to some
cp-map \(\Lambda\). Because we have seen in
Section~\ref{math:hilbert_schmidt} that such cp-maps form a Hilbert
space, it is reasonable to define a commutator (analogous to the case
of regular operators) for two Hilbert-Schmidt operators \(\Lambda_{1},
\Lambda_{2}\) by setting:
\begin{equation}\label{math:hs_commutator}
  [\Lambda_{1}, \Lambda_{2}] \equiv 
         \Lambda_{1}\Lambda_{2} - \Lambda_{2}\Lambda_{1}.
\end{equation}
The following theorem provides a condition for the
identity between a list of operators and a permutation of it which is
based on elementary commutators of the elements.  Unfortunately, this
is not a general solution since the effect of the theorem might just
be to rephrase the problem in different terms if the structure of the
commutators is not apt.


\begin{theorem}
  Let \(A_{1}, A_{2}, \ldots, A_{n}\) be operators and let \(\varphi
  \in \sym(n)\) be a permutation of the index set. Then the difference
  between the commuted product \(A_{\varphi(1)}\cdot
  A_{\varphi(2)}\cdots A_{\varphi(n)}\) and \(A_{1}\cdot A_{2}\cdots
  A_{n}\) can be written as\footnote{This representation (which is
    much more elegant than the one derived by the author) was provided
    by Volker Strehl.}
\begin{equation}
A_{\varphi(1)}\cdot A_{\varphi(2)}\cdots A_{\varphi(n)} =
A_{1}\cdot A_{2}\cdots A_{n} + \sum_{(s,t)} X_{s,t}\cdot 
[A_{s}, A_{t}]\cdot Y_{s,t}\cdot Z_{s}\label{math:comm_def}
\end{equation}
where (s,t) runs over all inversions of \(\varphi\), \ie, \(1\leq t <
s \leq n\) and \(\varphi^{-1}(s) < \varphi^{-1}(t)\) and where
\begin{equation}
X_{s,t} = \prod_{\makebox[0.5cm]{
     \(\begin{array}{cc}\scriptstyle 1\leq i \leq n\\
       \scriptstyle\varphi(i) < s,\ i < \varphi^{-1}(t)\end{array}\)}} 
     A_{\varphi(i)},\hspace{5mm}
Y_{s,t} = \prod_{\makebox[0.5cm]{
            \(\begin{array}{cc}\scriptstyle 1\leq i \leq n\\
              \scriptstyle\varphi(i) < s,\ i > \varphi^{-1}(t)\end{array}\)}} 
            A_{\varphi(i)},\hspace{5mm}
Z_{s} = \prod_{s<k\leq n} A_{k}.
\end{equation}
\end{theorem}
\begin{proof}
We prove this statement by induction on the list length. The 
cases \(n = 0\) and \(n = 1\) are trivial. The induction step
\(n \rightarrow n+1\) can be seen as follows. Let \(j \in [0,\ldots, n+1]\)
such that \(\varphi(j) = n+1\). Then,
\begin{align}
&A_{\varphi(1)}\cdot A_{\varphi(2)}\cdots 
     A_{\varphi(j)}\cdots A_{\varphi(n+1)} =\nonumber\\
&\hspace{1cm}A_{\varphi(1)}\cdots A_{\varphi(j-1)}\cdot A_{\varphi(j+1)} \cdot 
   A_{\varphi(j)}\cdot A_{\varphi(j+2)}\cdots A_{\varphi(n+1)} +\nonumber\\
&\hspace{1cm}A_{\varphi(1)}\cdots A_{\varphi(j-1)} \cdot 
   [A_{\varphi(j)}, A_{\varphi(j+1)}] \cdot 
      A_{\varphi(j+2)}\cdots A_{\varphi(n+1)} = ... = \nonumber\\
&\underbrace{A_{\varphi(1)}\cdots A_{\varphi(j-1)}\cdot 
   A_{\varphi(j+1)}\cdots 
   A_{\varphi(n+1)}}_{\text{I.H.}}\underbrace{A_{\varphi(j)}}_{A_{(n+1)}} +
\sum_{n+1,t} X_{n+1,t}\cdot [A_{n+1}, A_{t}]\cdot Y_{n+1,t}\label{math:ind}
\end{align}
where \(1 \leq t < n+1\), \(\varphi(n+1) < \varphi^{-1}(t)\),
\begin{displaymath}
X_{n+1,t} = \prod_{\makebox[0.5cm]{
     \(\begin{array}{cc}\scriptstyle 1 \leq i \leq n+1\\
       \scriptstyle\varphi(i)< n+1,\ i < \varphi^{-1}(t)\end{array}\)}} 
     A_{\varphi(i)},\hspace{5mm}
Y_{n+1,t} = \prod_{\makebox[0.5cm]{
       \(\begin{array}{cc}\scriptstyle 1 \leq i \leq n+1\\
         \scriptstyle\varphi(i)< n+1,\ i > \varphi^{-1}(t)\end{array}\)}} 
       A_{\varphi(i)}
\end{displaymath}
and \(n+1\) is obviously fixed. The final resulting equation 
thus resembles exactly the form given by Eqn.~\ref{math:comm_def}, but we
have \emph{not} used the induction hypothesis yet. Now, by using the 
induction hypothesis, it follows that
\begin{equation}
A_{\varphi(1)}\cdots A_{\varphi(j-1)}\cdot A_{\varphi(j+1)}\cdots 
A_{\varphi(n+1)} = A_{1}\cdots A_{n} + \sum_{(s',t')}X_{s',t'}\cdot 
    [A_{s'}, A_{t'}]\cdot Y_{s',t'}\cdot Z_{s'}
\end{equation}
where the primed identifiers are defined by \(1 \leq t' < s' \leq n\),
and \(\varphi^{-1}(s') < \varphi^{-1}(t')\).  By placing this into
the part of Eqn.~\ref{math:ind} marked by I.H., we see that
\begin{align}
& A_{\varphi(1)}\cdots A_{\varphi(n+1)} = \left(A_{1}\cdots A_{n} + 
\sum_{(s',t')}X_{s',t'}\cdot [A_{s'}, A_{t'}]\cdot Y_{s',t'} 
Z_{s'}\right)\cdot A_{n+1} + \nonumber\\
& \hspace{1cm}\sum_{(n+1,t)} X_{n+1,t}\cdot [A_{n+1}, A_{t}]\cdot Y_{n+1,t} \nonumber\\
&= A_{1}\cdots A_{n} A_{n+1} + \sum_{(s',t')} X_{s',t'}\cdot 
    [A_{s'}, A_{t'}]\cdot Y_{s',t'} \underbrace{Z_{s'}A_{n+1}}_{Z_{s}} + 
  \sum_{\makebox[1mm]{\(\scriptstyle(n+1,t)\)}} X_{n+1,t}\cdot  
                              [A_{n+1}, A_{t}]\cdot Y_{n+1,t}\nonumber\\
&= A_{1}\cdots A_{n+1} + \sum_{(s,t)} X_{s,t} [A_{s}, A_{t}]Y_{s,t}Z_{s}%
\label{math:comm_ind_final}
\end{align}
where the unprimed variables are now given by \(1 \leq t < s \leq 1\)
and \(\varphi^{-1}(s) < \varphi^{-1}(t)\); the condition for \(k\) in
\(Z_{s}\) is now obviously \(s < k \leq n+1\). The resulting
Equation~\ref{math:comm_ind_final} has thus the form for \(n+1\) as
required by the statement.
\end{proof}

To illustrate this theorem (note, additionally, that a little program
to calculate all elements of the commutator sum is available),
consider the permutation given by
\begin{align}
\left(
\begin{array}{ccccc}
1 & 2 & 3 & 4 & 5\\
5 & 2 & 3 & 1 & 4
\end{array}
\right).
\end{align}
The inversions \((s,t)\) are all pairs of elements
in the permuted list where a bigger element is on the left side of a
smaller element, in this case: \((5,2)\), \((5,3)\), \((5,1)\),
\((5,4)\), \((2,1)\), \((3,1)\).  Note that the inversions characterise
the list completely, cf., \eg, \cite[Section~5.1.1]{knuth}. The method defined above
is a variant of insertion sort which is a standard sorting method, 
covered, \eg, in \cite{sedgewick}. This can be seen by inspecting
the conditions imposed by the products defining \(X\), \(Y\) and
\(Z\):

\begin{itemize}
  \item For \(X_{s,t}\), \(\varphi(i) < s, i < \varphi^{-1}(t)\) selects
    all \(i\) such that the corresponding elements in the permuted list 
    are smaller than the element \(s\) of the inversion and 
    are placed on the left hand side of the element \(t\) in the
    permuted list. For \((5,1)\), the condition would select \(i = 2,3\).
  \item The conditions for \(Y\) make sure that again only elements which
    are smaller than \(s\) are selected. This time, they additionally
    have to be on the right hand side of \(t\) in the permuted list.
  \item \(Z\) specifies all elements which are on the right hand side
    of \(s\) in the \emph{unpermuted} list.
\end{itemize}

By applying these rules, we can calculate the following sets for each 
inversion:
\begin{align*}
(5,1) &\rightarrow X: i = 2,3; Y: i = 5\\
(5,2) &\rightarrow Y: i = 3,4,5\\
(5,3) &\rightarrow X: i = 2; Y: i = 4,5\\
(5,4) &\rightarrow X: i = 2,3,4\\
(3,1) &\rightarrow Z: k = 4,5\\
(2,1) &\rightarrow Z: k = 3,4,5
\end{align*}
This leads to the following identity that is provided by 
Eqn.~\ref{math:comm_def} (note that we use \(i\) instead of \(A_{i}\)
to simplify the notation):
\begin{align*}
52314\hspace{2mm} &=&&\hspace{-5mm}12345 + 231[5,4] + 2[5,3]14 + [5,2]314 + 
                                   23[5,1]4 + 2[3,1]45 + 
         [2,1]345 \\
      &=&&\hspace{-5mm}12345 + 23154 - 23145 + 25314 - 23514 + 52314 - 
                       25314 + 23145 -\\
      &  &&\hspace{-5mm} 23145 + 23145 - 21345 + 21345 - 12345\\
      &= &&\hspace{-5mm} 12345 - 12345 + 23154 - 23154 + 25314 - 25314 + 
                         23145 - 23145 +\\
      &  &&\hspace{-5mm} 21345 - 21345 + 23514 - 23514 + 52314 \\
      &=&&\hspace{-5mm}52314
\end{align*}

It it also instructive to observe the following two identies because they
illuminate the induction step:
\begin{align*}
14532 &= 12345 + 14[5,3]2 + 143[5,2] + 1[4,3]25 + 13[4,2]5 + 1[3,2]45\\
1432 &= 1234 + 1[4,3]2 + 13[4,2] + 1[3,2]4
\end{align*}

\begin{remark}
Because the proof has only made use of general properties of permutations
and of the definition of the commutator, it is not only applicable to
Hilbert-Schmidt-operators as we need, but also for any other
objects fulfilling the mentioned properties.
\end{remark}

\summaryodd{We have explained how to represent quantum operations by
  cp-maps and these in turn by a sum of Kraus operators. The L\"owner
  partial order defined for density matrices was generalised to Kraus
  operators; this order is complete and is therefore a cpo as
  introduced in the beginning of this chapter. Since the denotational
  semantics of cQPL will require lists of Kraus operators, we have
  introduced Kraus aggregations to handle this formally. Since it is
  one of the problems of denotational semantics to decide whether two
  given programs are equal or not, we have also derived general and
  specific criteria for the equivalence of Kraus aggregations.}

\newpage\thispagestyle{plain}
\forgechapter{5}
\chapter{Formal denotational semantics}{A map is not the territory.}%
{Alfred Korzybski, Science and Sanity -- An Introduction to Non-Aristotelian 
 Systems and General Semantics}%
\label{chap:formal_semantics}
In this chapter, we are going to define a denotational semantics for cQPL,
the communication capable version of QPL \cite{selinger_qpl}.
Before we get into the details, we will give an overview about
the ideas of denotational semantics in general, present a 
survey of the denotational semantics of QPL (because we reuse some ideas
for the semantics of cQPL) and show why the approach of annotation-based QPL must fail
for communicating programs.

\section{Fundamentals of denotational semantics}
Denotational semantics is a well-understood standard method of theoretical
computer science which is used to assign precise and mathematically sound 
and rigorous semantics to syntactically specified programs; introductions
are, \eg, given in Refs.~\cite{mosses, winskel, reynolds}.
In this section, we will try to present an elementary
introduction to the field. We align our description along the lines
of \cite[Section 1--3]{mosses}.

Computer programs are (usually) specified in the form of a textual
description; this description must follow certain rules defined
by a grammar. Usually, context-free grammars are used for this purpose because
they are the most apt choice for that kind of problem. They are defined as 
follows regarding to \cite{asteroth_baier, schoening}:

\begin{definition}[Context-free grammar]\label{form:context_free}
A context-free grammar \(G\) is a four-tuple \((N,T,P,s_{0})\) where
\(N\) is a finite set of nonterminal symbols, \(T\) is a finite set of 
terminal symbols with \(T \cap N = \varnothing\), 
\(P\subseteq N\times (N\cup T)^{*}\) is a finite set of productions and 
\(s_{0}\in N\) is the start symbol.
\end{definition}

As a very simple example, consider a grammar for binary strings of the
form \(0\), \(01\), \(100110\), \dots\ which is recursively given
by\footnote{In general, one has to distinguish between abstract and
  concrete syntax respectively grammars defining these. The latter is
  used to specify a representation of programs that can be processed
  with parsers; for that, some syntactical elements for disambiguation
  of certain constructions needs to be introduced.  Additionally, the
  capabilities and, especially, limitations of different parsing
  techniques need to be considered when specifying a concrete grammar.
  Abstract syntax, on the other hand, is a representation of a program
  stripped down to the bare minimum that is able to include all
  available information; additionally, the structure of the syntax can
  be chosen such that it is not most suited for parsing, but for
  further analysis and processing of the program. Usually, data
  structures in the form of trees are used to realise abstract
  syntax.}
\begin{equation}
B ::= \text{'0'}\ |\ \text{'1'}\ |\ B \text{'0'}\ |\ B \text{'1'}.%
                                              \label{form:bin_abs_gram}
\end{equation}
The terminal symbols\footnote{A constant symbol which cannot be resolved
  any further, cf. Appendix~\ref{glossary}.} are \(0\) and \(1\), the 
non-terminal\footnote{A symbol whose definition consists of a chain
  of terminal and (at least one) non-terminal symbols and can thus be 
  resolved further.} symbol is \(B\), and the start symbol
is obvious because there is only one non-terminal. The productions 
are defined by Equation~\ref{form:bin_abs_gram}; explicitely, they
are given by\(\{B \times 0, B \times 1, B \times B0, B \times B1\}\).

This grammar defines the syntactical representation of binary numerals.
The really interesting thing, however, is not how numerals \emph{look
like}, but instead \emph{what they mean} -- in other words, the semantics
of numerals. Obviously, the meaning of a binary numeral is some
natural number, so finding semantics for a binary string
is equivalent with constructing a method which assigns the appropriate
natural number to a given syntactical representation of a binary numeral.
The constitutional parts of which the grammar is made up of are called 
\emph{phrases}. In our case, these are given by the strings \(0\) and
\(1\) and the productions \(B0\) and \(B1\). 

Denotational semantics assigns a meaning to sentences constructed
according to a given grammar by assigning a meaning to every
elementary phrase of a grammar. The meaning of phrases 
which are constructed from multiple sub-phrases (\eg, \(B0\) in the
example grammar) is given by the meaning of these sub-phrases.
The meaning of a complete program is thus determined by the meaning
of its constituents. The denotational approach is -- in a nutshell -- 
characterised by the following points:

\begin{itemize}
  \item Denotational semantics assigns some appropriate semantic object
    to every phrase of the grammar; the object is called the
    \emph{denotation} of the phrase.
  \item \emph{Valuation functions} are used to connect syntactical
    objects with their semantical counterparts. For example,
    \(\mathcal{BIN}\) is a valuation function that maps text strings
    consisting of a series of '0' and '1' to a natural number.
  \item The denotation of compound phrases must only depend on the
    denotations of the sub-phrases, \ie, \(\stdsembrack{A_{1}, \ldots, A_{n}} =
    f(\stdsembrack{A_{1}}, \ldots, \stdsembrack{A_{n}})\).
    This is also known as the \emph{compositionality principle}.
\end{itemize}

The valuation functions for binary numerals can be represented by
the following equations:

\begin{center}
  \begin{tabular}{lcl@{\hspace{1cm}}lcl}
    \domfkt{BIN}{0} &=& 0 & \domfkt{BIN}{1} &=& 1\\
    \domfkt{BIN}{B0} &=& \(2\cdot(\domfkt{BIN}{B})\) &
    \domfkt{BIN}{B1} &=& \((2\cdot(\domfkt{BIN}{B})) + 1\)
  \end{tabular}
\end{center}

The double brackets \(\stdsembrack{}\) are used to distinguish between 
the realms of syntax and semantics, while the valuation function
\(\mathcal{BIN}\) is used to map the phrases in these brackets to natural 
numbers, their denotations. Thus, the domain of this function is the 
\emph{semantic domain} \(\mathbbm{N}\). In Chapter~\ref{math:chapter}, the 
required material for the
specification of domains suitable to support the denotational semantics
of cQPL has been presented; it will be put to use in this chapter.
Especially note that the denotations of the composite phrases 
\(B0\) and \(B1\) are defined only in terms of the denotations
of their sub-phrases as required by the compositionality principle.

To clarify the effect of the denotational definitions, consider how
the meaning of the numeral \(101\) is denoted; the abstract syntax generates
the tree shown in Figure~\ref{form:binary_tree} as representation. This
leads to the following denotation (observe that the \(B\)s used in
the equations are not identical):

\begin{figure}[htb]
  \centering\includegraphics{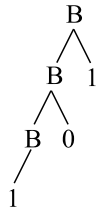}
  \caption{Derivation tree for the binary numeral \(101\) generated
    by the abstract grammar given in Equation~\ref{form:bin_abs_gram}.}%
  \label{form:binary_tree}
\end{figure}

\begin{align*}
  \mathcal{BIN}\stdsembrack{B} &= \mathcal{BIN}\stdsembrack{B1} = 
                                  2\cdot\mathcal{BIN}\stdsembrack{B} + 1 \\
     &= 2\cdot\mathcal{BIN}\stdsembrack{B0} + 1 = 
                     2\cdot 2\cdot\mathcal{BIN}\stdsembrack{B} + 1\\
     &=2\cdot 2\cdot \mathcal{BIN}\stdsembrack{1} + 1 = 2\cdot 2\cdot 1 + 1 = 5
\end{align*}

Since \(B = 101\), the final denotation is given 
by \(\mathcal{BIN}\stdsembrack{101} = 5\). This is precisely the expected 
result.

\section{Survey of QPL}
The semantics of QPL is based on the idea of annotating a flow graph
that represents a quantum program with density matrices for the quantum
mechanical parts and tuples of probabilities covering the classical
components. Additionally, a typing context is used to keep track of
all variables together with their types that are in use at a certain 
stage of a program.

Since our work is based on QPL, it seems appropriate to summarise its
central concepts. The original definition of QPL \cite{selinger_qpl}
provides a more detailed description than given here; an alternative
review can be found in \cite{schroeder}.  We align our summary on both
sources. Note that it is nevertheless useful to have some familiarity
with the paper introducing QPL because we can obviously not repeat
everything here.

\subsection{Notational conventions}
QPL operates on finite-dimensional quantum states represented by
vectors over \(\mathbbm{C}\). The basis states for qbits are defined as 
\(\ket{0} = (1,0)^{\text{t}}\) and \(\ket{1} = (0,1)^{\text{t}}\).
Combination of multiple qbits are as usual represented by tensor products
of these states. Density matrices are used as basis for any manipulations
performed by the language. If a state is defined by some vector 
\(u \in \mathbbm{C}^{2^{n}}\), the corresponding density matrix is 
given by \(\varrho = uu^{\dagger}\) and may also be denoted by
\(\{u\}\). Mixed states are represented by linear combinations
of pure states, \eg, \(\lambda_{1}u_{1}u_{1}^{\dagger} + \cdots + 
\lambda_{n}u_{n}u_{n}^{\dagger}\). Given four matrices 
\(A_{1}, A_{2}, A_{3}, A_{4}\) of identical dimension, they can be 
concatenated horizontally and vertically by 
\begin{equation}
\left(\begin{array}{c|c}A_{1} & A_{2}\\\hline A_{3} & A_{4}\end{array}\right)%
                                             \label{form:qbit_annot}
\end{equation}
which is used to specify composite density matrices. 
This notation is used to specify the semantics of actions possible
in QPL which will be introduced in the following sections.

\subsection{Language elements}
QPL programs are given in terms of \emph{quantum flow charts}\footnote{There
  is also a textual representation for programs, but this is only
  considered as an aside in the definition of QPL.} where
each edge is supplemented with all the information necessary to 
unambiguously specify the meaning of a program. Every edge is
augmented with

\begin{itemize}
  \item a \emph{typing context}, \ie, a mapping from identifiers of variables
    to the types of these. It is written as a list of identifiers
    followed by their type, \eg, \(a,b,c: \bit, d: \cint\).
    Typing contexts encapsulating variables which are not related to
    the present considerations are denoted by \(\Gamma\).
  \item an \emph{annotation}, \ie, a tuple of density matrices 
    which specifies the state of the system.
\end{itemize}

The annotation of a classical bit is given by \((A,B)\) where 
\(A+B = 1\) and \(A\) represents the probability that the value of the 
bit is \(0\), whereas \(B\) is the probability that the value is
\(1\). The annotation for a quantum bit is of the form given by
Eqn.~\ref{form:qbit_annot}.

\begin{figure}[htb]
  \centering\includegraphics[width=\textwidth-1cm]{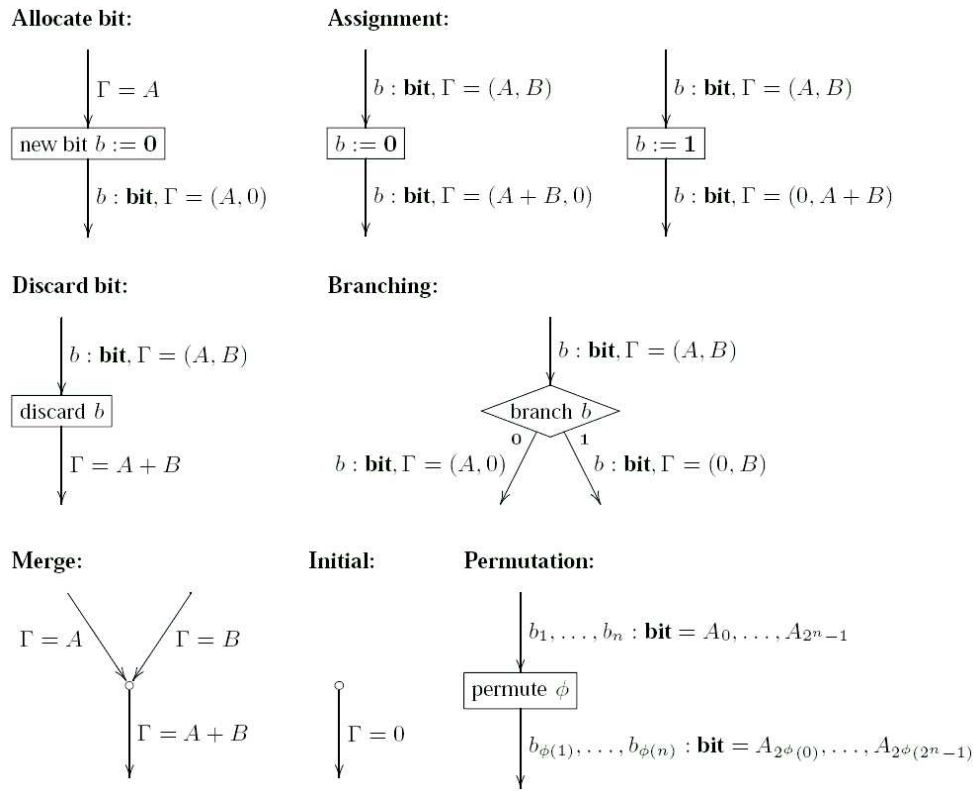}
  \caption{Summary of all classical operations of QPL, taken
    from~\cite{selinger_qpl}. Note that the symbol ``\(=\)''
    is used to separate typing context and the annotation, which
    can be confusing at times because it is \textsl{not} associated
    with \textrm{\(\Gamma\)} alone.}\label{form:qpl_class_fc}
\end{figure}

\begin{figure}[htb]
  \centering\includegraphics[width=\textwidth-1cm]{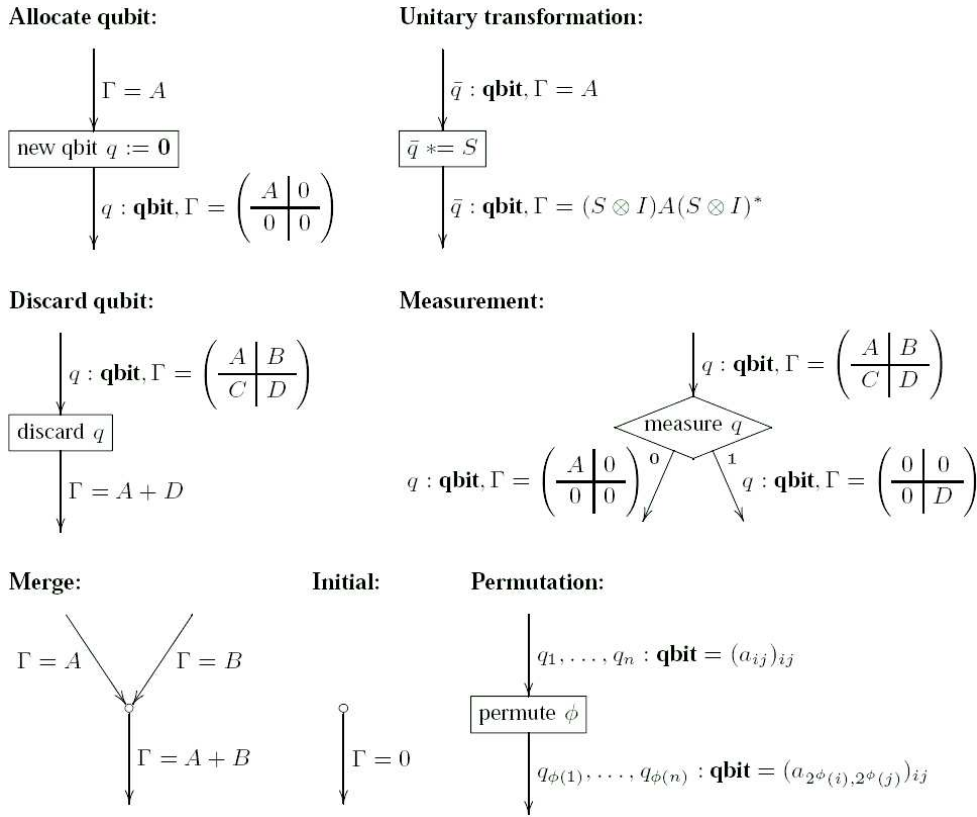}
  \caption{Summary of all quantum mechanical operations of QPL, taken
    from~\cite{selinger_qpl}.}\label{form:qpl_qm_fc}
\end{figure}

All classical operations possible with QPL and their flow graph 
representations are shown in Figure~\ref{form:qpl_class_fc}.
Figure~\ref{form:qpl_qm_fc} depicts the quantum mechanical parts.

\subsection{Semantics}
\begin{figure}[htb]
  \centering\includegraphics[width=\textwidth-1cm]{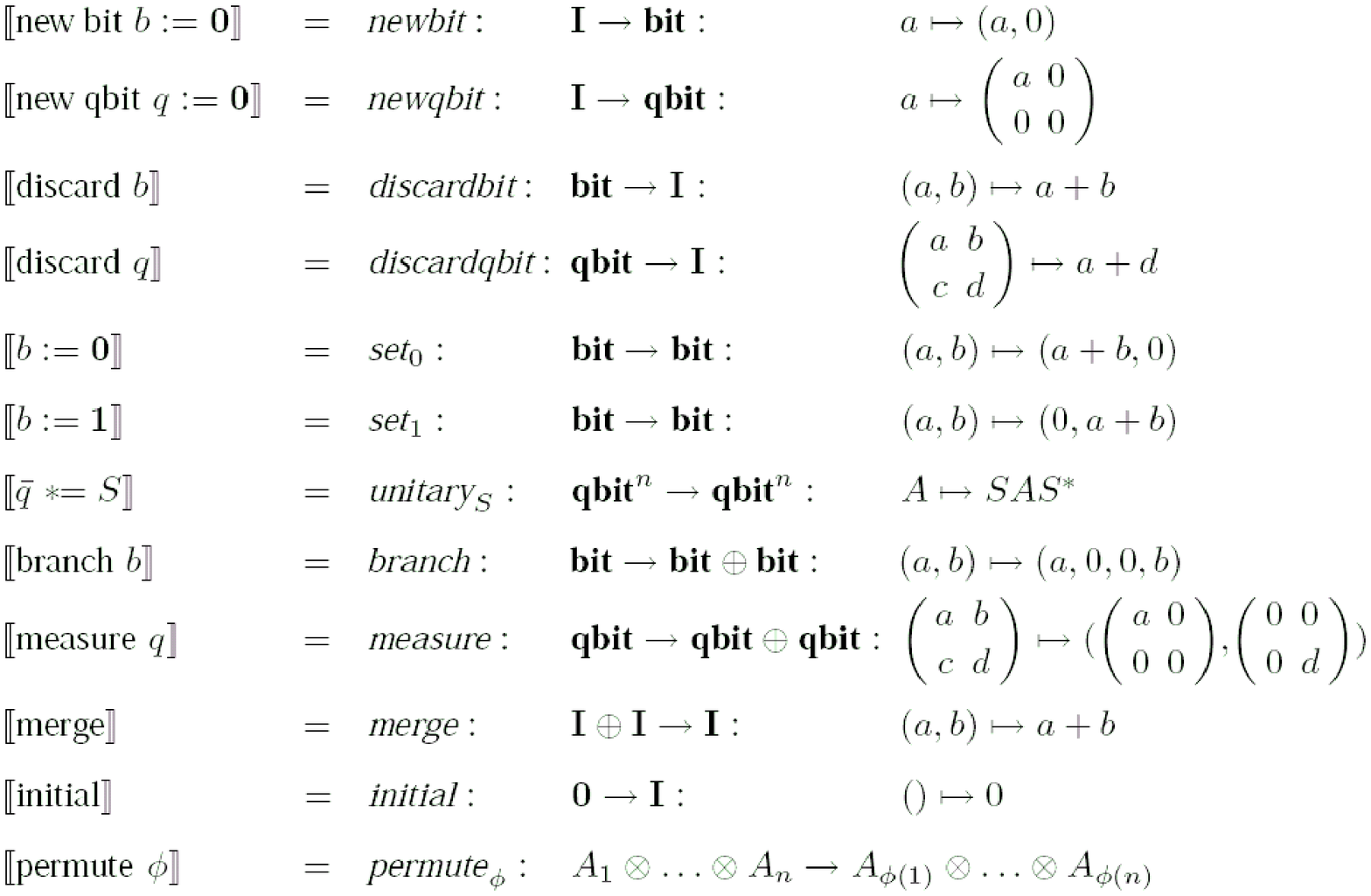}
  \caption{Valuation functions which define the denotational
    semantics of QPL, taken from~\cite{selinger_qpl}.}\label{form:qpl_sem}
\end{figure}

The semantics of a QPL program can be directly inferred from the flow
graph representation. The explicit transformation of a density matrix
given in the annotation of the edges serves as a unique representation
of the meaning of a program. \cite{selinger_qpl} proofs that this
approach is indeed well-defined and also works for recursion and
loops, which can be included into the language. Categorical structures
that allow to accommodate superoperators and morphisms to manipulate
these according to the possibilities of QPL are used as a formal basis
for the definition of the semantics.  This categorical
superstructure\footnote{Note that we are only referring to category
  theory here, \emph{not} to the compositional semantics presented by
  Selinger.} is not too interesting for our purposes. It suffices to
know that the valuation functions for the diverse language elements
are defined as shown in Figure~\ref{form:qpl_sem} and that they indeed
fulfil everything which is necessary for a sound and well-defined
interpretation. Note that the way how the semantics is specified in
Figures~\ref{form:qpl_class_fc} and \ref{form:qpl_qm_fc} is \emph{not}
equivalent to the method of Figure~\ref{form:qpl_sem}: While the first
one relies on explicit transformations of density matrices, the second
one uses a more abstract representation in form of superoperators and
is almost completely identical to the basis of our approach (the
functions computed by both approaches of Ref.~\cite{selinger_qpl} are
nevertheless identical except for loops and recursion). Especially,
the second variant is compositional, which is a necessary condition to
describe multipartite systems in such a way that the description of
one part is independent of other parts.

\subsection{Limitation: Quantum communication}
Before we lay out the denotational semantics of cQPL, it is advisable
to sketch in which sense the different approaches used in QPL do not
work for programs dealing with communication.

First of all, let \(\varrho_{\text{AB}}\) denote the density matrix
of the state shared by Alice and Bob. The information available for 
each party can be inferred by calculating the partial trace:
\(\varrho_{\text{A}} = \ptrace{B}(\varrho_{\text{AB}})\) and
\(\varrho_{\text{B}} = \ptrace{A}(\varrho_{\text{AB}})\).
The bipartite density matrix can never be 
recovered from these partial density matrices because
there are many bipartite density matrices which give rise to the same
partial density matrices.

One of the goals of denotational semantics is to assign sufficient information
to \emph{every} edge of a quantum flow graph such that the complete
semantics of a program can be reconstructed by combining only the
information given by the edges constituting the program. The denotation
of a statement composed of several sub-statements must be
completely determined only by a function of the denotations of 
the sub-statements.






This is impossible in the annotation-based semantics of QPL because
transformations between explicit density matrices are considered.
Since a combination of the partial density matrices
\(\varrho_{\text{A}}\), \(\varrho_{\text{B}}\) which were manipulated
by Alice and Bob does not restore the total bipartite state
\(\varrho_{\text{AB}}\), the QPL annotation would obviously not
comply with the physical state afterwards.

A possible solution is the annotation of the \emph{complete} flow
graph, \ie, of both paths representing the control flow for Alice and
Bob. In this case, the operations performed by Alice and Bob would be
written as tensor products of the type
\(\hat{A}\otimes\mathbbm{1}_{\text{B}}\) and
\(\mathbbm{1}_{\text{A}}\otimes \hat{B}\) which act on the complete
density matrix \(\varrho_{\text{AB}}\). This way, we could assign
semantics to the program as a whole, but would loose the ability to
construct the denotation of a phrase from the denotations of its
subphrases. This means that the semantics of the complete program
could not be constructed from the denotation of Alice's and Bob's
program alone which is in contrast to the key idea of denotational
semantics.

Therefore, we need to seek a solution that does not characterise
quantum operations by showing transformations of explicit density
matrices, but uses something that captures the notion of a
transformation in a more abstract sense. Completely positive maps
represented by a set of Kraus operators fulfil this need as we will
show in Section~\ref{form:formal_definitions}; this is basically the
same approach as used for the compositional semantics of QPL.
Nevertheless, QPL does not provide any means of parallel composition,
communication and other details which are necessary to describe
quantum communication as we will do in the remaining parts of this
thesis.

\summaryodd{We have presented a quick summary of QPL and the associated
  denotational semantics which is based on partial orders of density
  operators.  Additionally, we have shown why this approach is not
  suitable to describe quantum communication respectively the
  interaction of spatially separated systems where the combined
  density matrix is not available as whole in the framework of the
  annotation-based semantics.} 

\section{Denotational semantics of cQPL}
cQPL is an extended variant of QPL with the ability to express
and formalise quantum communication, \ie, the ability to describe
multiple parallel flow graphs that exchange quantum and classical
information at well-defined points, but do otherwise know nothing
about each other. 
To achieve this, we have to base the semantic description on three
components in contrast to the two components (typing context and 
tuples of density matrices) of QPL:

\begin{itemize}
  \item A Kraus aggregation \(K\) as defined in 
    Section~\ref{math:kraus_aggregations}
    which is used to keep track of the quantum operations performed
    on the qbits of the system.
  \item A typing context \(T\) used to specify which quantum and
    classical variables are allocated at a given moment and which
    data type they have. This is also important to describe communication
    because it allows to uniquely determine to which party a variable
    belongs at a given stage of a program.
  \item A probabilistic environment mapping identifiers 
    to values. Since the interaction between quantum and classical
    parts of the system introduces probability, the values of 
    classical variables are subject to such a distribution.
    In general, only the range of possible values together with the
    fact that it is governed by a probability distribution 
    is known in the semantical description.
\end{itemize}

\noindent We refer to these three elements as the three-tuple \((K, T, E)\).

A Kraus aggregation \(K\) specifies a quantum mechanical operation 
which has the same effect \emph{for all} density matrices \(\varrho\) (in the
sense that the application of a Hadamard gate will yield different effects
according to the state it was applied to. Nevertheless, it is still
a Hadamard gate in every case, and this is the really important thing). Thus, 
the operation is completely characterised \emph{without} the need to specify 
any density matrix \emph{at all}. This is exactly what is required when 
spatially separated operations performed by several parties on multipartite
states are to be described, as we will see later in greater detail.

To realise the benefits of this approach, consider how the generation
of a new qbit in state \(\ket{0}\) subsequently followed by the
application of a Hadamard gate is described in QPL (we do not show the
complete flow graph, but only the relevant parts of the annotation):

\begin{displaymath}
\Gamma = A \xrightarrow{\text{create new qbit \(q\)}} q:\text{\textbf{qbit}}, 
\Gamma = \left(\begin{array}{c|c}A & 0 \\\hline 0 & 0\end{array}\right) 
    \xrightarrow{\text{apply \(H\) on \(q\)}} 
    q:\text{\textbf{qbit}}, \Gamma
    = \frac{1}{2}\left(\begin{array}{c|c}A & A \\\hline A & A 
      \end{array}\right)
\end{displaymath}

Although only the newly created qbit is concerned, the state of the remaining
system is still implicitly present in \(A\). This is more than needed: It
suffices to consider the application of two operations given by
the following Kraus sets:
\begin{equation}
\{C_{i}\}_{\#q}; \{H_{i}\}_{\#q}
\end{equation}
where \(\{C_{i}\}_{\#q}\) stands for ``create a new qbit with label \(q\)''
and \(\{H_{i}\}_{\#q}\) for ``apply a Hadamard gate on \(q\)''. With
these, we can describe the same operation \emph{without}
resorting to a density matrix or any other part of the system unconcerned
by the operation at all.

The typing context \(T\) is basically adopted from QPL.
An extension to the framework used by QPL is the \emph{probabilistic
  environment}. For every allocated classical variable in the current
frame, it is used to specify a probability distribution that maps the
variable name to the range of possible values. This distribution is
parametrised by density operators because it depends on the initial
conditions of the program fragment and on the path taken in the flow
graph (an example explaining this will follow in the next section).
The probabilistic environment could in principle be replaced by the
tuples for classical states as used in QPL, but this works only well
for data types with a very low number of bits. Because of this reason,
QPL tries to hide these tuples most of the time, so we eliminate them
completely and replace them by the probabilistic environment.

The probabilistic environment also deals with quantum variables: For
every such variable, the position of the allocated qbits in the global
quantum heap is given by the probabilistic environment.  This is
necessary because quantum variables cannot be characterised by a value
as it is possible for classical variables because they do not have a
state as such. The state is replaced by the series of operations which
have been performed on the variables; since these operations need some
location to act on, every quantum variable needs to have a unique
position on the quantum heap, \ie, where the qbits are stored.

Note that this approach is somewhat contrary to the spirit
of functional programming because it introduces stateful global
variables, but a closer examination reveals that QPL implicitly uses
the same model and that compile-time checking (and thus the
protection against runtime errors) is not affected by this.

In addition to allocated variables, branches in programs are also
present in the probabilistic environment. They are identified by a unique 
ID which is assigned to every branching node.\footnote{To be precise:
  Which is assigned whenever the branching node is transversed because we
  need to account, \eg, for branches in loops where the same
  branch might be traversed multiple times.} This is necessary because the
branching conditions -- being based on comparisons of probability 
distributed quantities -- are in general not represented by some fixed
values, but represented by a probability distribution as well.

\subsection{Formal definitions}\label{form:formal_definitions}
In this section, we will present some methods to characterise and
describe the semantic components of cQPL.  Note that in the following,
we use \(\mathcal{D}_{n}\) to denote the set of all density operators
of dimension \(n\), dropping the subscript if the exact dimension is
not important or can be deduced from the context.

\subsubsection{Typing context}\label{form:type_cont}
Let \(\sigma\) be a list of numbers \(n_{i} \in \mathbbm{N}_{+},
i=1,\ldots,k\) as given by \(\sigma = n_{1}^{\tau}, n_{2}^{\tau},
\ldots, n_{k}^{\tau}\). \(\sigma\) is also called the \emph{signature}
of a data type. An associated Hilbert space \(\mathcal{H}_{\sigma}\)
is given by
\begin{equation}
  \mathcal{H}_{\sigma} \equiv 
           \mathcal{H}_{1}\otimes\cdots\otimes\mathcal{H}_{k}
\end{equation}
where \(\mathcal{H}_{i}\) is either \(\mathcal{B}(\mathcal{H})\) for
quantum or \(\mathcal{C}(X)\) for classical data (cf.
Section~\ref{math:classical_components}) where both are distinguished
by the index \(\tau\): \(\tau = q\) for quantum variables and \(\tau =
c\) for classical variables. The dimension of the \(i\)-th space is
given by \(2^{n_{i}^{q}}\) for quantum mechanical and \(n_{i}^{c}\)
for classical variables. Since we restrict ourselfs to
finite-dimensional Hilbert spaces, this means that we can use
\(\mathcal{H}_{i} = \mathbbm{C}^{n_{i}}\) for quantum mechanical and
\(X = [0,1,\ldots,n_{i}-1]\) for classical data.  To distinguish
between both cases, we define the function \(q: n_{i}^{\tau}
\rightarrow [0,1]\) given by
\begin{equation}
q(n_{i}^{\tau}) = \begin{cases}
                     1 & \text{if \(\tau = q\)}\\
                     0 & \text{otherwise}
                  \end{cases}
\end{equation}
Note that although our formalism allows to define data types which consist 
of both quantum mechanical and classical components, we do not exploit this 
possibility because we could not find any reasonable application for this in 
our work. To keep the formalism as general as possible, we will 
nevertheless still retain the possibility as long as no 
noteworthy effort is necessary to do so.

We can define a function \(t_{q}: \sigma \rightarrow \mathbbm{N}\)
to compute the total number of qbits necessary for a given signature
(\(\card(\sigma)\) denotes the cardinality of \(\sigma\)): 
\begin{align}
t_{q}(\sigma) = \sum_{i=1}^{\card(\sigma)}q(n_{i}^{k})\cdot n_{i}^{k}
\end{align}
The analogous function \(t_{c}\) for the classical components is given by
\begin{equation}
t_{c}(\sigma) = \sum_{i=1}^{\card(\sigma)}(1-q(n_{i}^{k}))\cdot n_{i}^{k}
\end{equation}
Finally, two functions \(q(\sigma)\) and \(c(\sigma)\) to check if a given 
data type is purely classical or purely quantum are 
necessary:\footnote{Note that
  we define a data type consisting of \(0\) quantum and \(0\)
  classical bits, \ie, \void, as classical.}
\begin{align}
q(\sigma) &= \begin{cases}
               1 & t_{c}(\sigma) = 0 \wedge t_{q}(\sigma) > 0\\ 
               0 & \text{otherwise}
             \end{cases}\\
c(\sigma) &= \begin{cases}
               1 & t_{q}(\sigma) = 0 \wedge t_{c}(\sigma) \geq 0\\ 
               0 & \text{otherwise}\end{cases}
\end{align}
For simplicity, we label data types required for practical use with
special mnemonics; some examples can be found in
Table~\ref{data_types}. Note that the type \(\void\) can, for example,
be used to formally describe statements which return no value and thus
have no type.

\begin{table}[htb]
  \begin{center}
    \begin{tabular}{c|c}
      Mnemonic & Signature \\\hline
      \bit     & \(2^{c}\) \\
      \qbit    & \(2^{q}\) \\
      \short   & \(8^{c}\) \\
      \qshort  & \(8^{q}\) \\
      \cint    & \(16^{c}\) \\
      \qint    & \(16^{q}\) \\
      \void    & \(0^{(c)}\) 
    \end{tabular}
  \end{center}
  \caption{Signatures and the mnemonics commonly
    used in programming languages for data types supported by cQPL.}
  \label{data_types}
\end{table}

Since two finite-dimensional Hilbert spaces \(\mathcal{H}_{1}\) and 
\(\mathcal{H}_{2}\) are isomorphic if the sum over the 
dimensions of their subsystems is equal (this is, \eg, proved 
in~\cite[Theorem 2.62]{weidmann}), \ie, 
\begin{equation}
  \bigotimes_{k}\mathcal{H}_{1k} = \mathcal{H}_{1} \cong 
  \mathcal{H}_{2} = \bigotimes_{l}\mathcal{H}_{2l} \Longleftrightarrow 
  \sum_{k}\dim(\mathcal{H}_{1k}) = \sum_{l}\dim(\mathcal{H}_{2l}),
\end{equation}
the description of types is not unique. For example, the types
given by \((2^{q},2^{q},2^{q},2^{q})\) and \(\qshort = 8^{q}\) are identical 
and provide only different aspects of the same thing. This equivalence
can also be extended to mixed data types:
\begin{align}
  n_{1}^{\tau},\ldots,n_{K}^{\tau} \cong m_{1}^{\tau},\ldots,m_{L}^{\tau}
    \Longleftrightarrow 
  &\sum_{k}q(n_{k}^{\tau})\cdot n_{k}^{\tau} =  
  \sum_{l}q(m_{l}^{\tau})\cdot m_{l}^{\tau} \nonumber\\ \wedge&
  \sum_{k}(1-q(n_{k}^{\tau}))\cdot n_{k}^{\tau} =  
  \sum_{l}(1-q(m_{l}^{\tau}))\cdot m_{l}^{\tau}\label{form:type_equiv}
\end{align}
This creates an equivalence class for data types which will be useful
in Section~\ref{form:type_system}. The class of all data types
equivalent to \(\sigma\) is given by
\begin{align}
  \mathcal{T}(\sigma) &\equiv
  \left\{\bigoplus_{i=1}^{t_{q}(\sigma)}\inj_{\varphi(i)}(\Xi\#i)^{q}
    \oplus
    \bigoplus_{i=1}^{t_{c}(\sigma)}\inj_{\varphi(i+t_{q})}(\Omega\#i)^{c}
  \right\}\label{form:decomp_dtype}\\
  \text{such that } &\Xi\in\mathcal{S}(t_{q}(\sigma)),
  \Omega\in\mathcal{S}(t_{c}(\sigma)),
  \varphi\in\sym([1,\ldots,t_{q}(\sigma)+t_{c}(\sigma)])\nonumber
\end{align}
where \(M\#i\) denotes the \(i^{\text{th}}\) element of the ordered set
\(M\) and \(\mathcal{S}(k)\) is the decomposition of the scalar value
\(k\) into all possible sums given by
\begin{equation}
\mathcal{S}(k) \equiv \Big\{Z\ \text{is a set with members
                      \(\in \mathbbm{N}\)} \Big| 
                    \sum_{i=1}^{\card(Z)}Z\#i = k\Big\}.
\end{equation}
If \(\sigma_{2} \in \mathcal{T}(\sigma_{1})\), we write 
\(\sigma_{1}\cong\sigma_{2}\).

To illustrate the effect of Eqn.~\ref{form:decomp_dtype}, consider a
data type which consists of \(3\) quantum and \(3\) classical bits.
Structurally, it does not make any difference how these components are
ordered, \eg, \((1^{q}, 1^{c}, 2^{q}, 2^{c})\) is identical with
\((3^{q}, 3^{c})\) in this sense.\footnote{Note that there \emph{is} a
  difference between these orderings from the compiler's point of view
  because the different components are located at different locations
  in memory if different orderings are used. The semantics is
  nevertheless unconcerned by this.}  Eqn.~\ref{form:decomp_dtype} is
a generalisation of this idea: The scalar \(3\) can be decomposed as
\(1+1+1\), \(2+1\) and \(3\) as given by \(\mathcal{S}(3)\), so there
is no difference between any of these groupings.  Additionally, it is
not interesting how the components are ordered, \eg, \((2 + 1)\) is
equivalent to \((1 + 2)\).  Finally, the quantum and classical
components can be arbitrarily interchanged, so we have to consider
this as well. The effect of Eqn.~\ref{form:decomp_dtype} is to
construct all equivalent representations of a data types following
these considerations.

Note that QPL uses a Cartesian product of complex vector spaces given
by \(\mathbbm{C}^{n_{1}\times n_{1}}\times\cdots
\times\mathbbm{C}^{n_{k}\times n_{k}}\) for both classical and quantum
mechanical signatures (the set of complex \(d\times d\) matrices is
used to represent the complex Hilbert space of dimension \(d\)). This
does not reflect the relationship between corresponding quantum and
classical objects directly. For example, the data type for bits is
given by \(\bit = (1,1)\), whereas for qbits, the definition is
\(\qbit = 2\)). This leads to appropriate spaces for these objects,
but does not present the relation between them directly.

We thus used a different approach that makes correspondences more
clear which is important when, for example, quantum variables are
measured and the result is stored in a classical variable. Besides, it
fits better into the more abstract description of quantum mechanics as
introduced in Section~\ref{sec:math:qm}.

Let \(\Sigma\) be the set of finite strings over the alphabet
\(\alpha\). With this and the notion of types, we can define the
typing context used in the semantic description of cQPL.

\begin{definition}[Typing context]
  A typing context \(\tau\) is a three-tuple \(\tau = (\iota, \theta,
  \chi)\) where \(\iota\) is a set of identifiers in \(\Sigma\),
  \(\theta\) is a set of types and \(\chi: \iota \rightarrow \theta\)
  is a surjective mapping which assigns a type to every identifier.
  Identifiers starting with \(\#\) must not be used by programs.
\end{definition}

Because cQPL is strongly and statically typed (\ie, the type of an
expression is completely determined by the types of its components and
the type of an elementary component cannot be changed after it has
been declared, cf. Appendix~\ref{glossary}), information contained in
the typing context cannot be modified any more once it has been
introduced. Note that this does not hinder the possibility of
\emph{overshading} entries. This happens when, \eg, a variable
declared in an inner block has the same name as a variable declared in
an outer block.  Although both have identical names, their types do
not need to match because they are otherwise completely unconnected.

Typing contexts are modified when new variables are declared (and thus
added to the context) or when variables are removed from the scope
(and thus have to be removed from the typing context). Since it is
obvious how this influences a given context \(\tau\), we only note
that it is easy to define appropriate morphisms \(\tau \rightarrow
\tau'\) which perform the desired job.

Formally, we use the notation 
\begin{equation}
  \tau \rightarrow \tau' = \tau \oplus (\xi \rightarrow \qbit)
\end{equation}
to introduce some new identifier \(\xi\) with type \qbit{} into the
context \(\tau\). Equivalently, the notation \(\tau \ominus \xi\) is
used to remove \(\xi\) which is needed to describe sending quantum
variables.

\subsubsection{Probabilistic environment}
The probabilistic environment can be defined formally as follows:
    
\begin{definition}[Probabilistic environment]
Let \(\pi\) be a probability distribution on a finite set \(X\) with
probabilities \(p_{i}\) for every element of \(X\) such that
\(\sum_{i}p_{i} = 1\).  Let \(\iota\) be a set of identifiers, 
\(P\) be a set of probability distributions and \(M\) be a surjective 
map \(M: \iota \rightarrow P \cup \bot\). Then \(E = (\iota, P, M)\) is a 
probabilistic environment.
\end{definition}

As usual in denotational semantics, the symbol \(\bot\) is used to
denote an undefined identifier, \ie, a variable which is not present
in the environment. To specify components of a probabilistic environment, 
we use the notation
\begin{equation}
  \mapval{x}
\end{equation}
to denote an element of the probabilistic environment where \(x\) is
the identifier, \(\pi_{x}\) the associated probability distribution
and \(\range(x)\) the set of possible values which obviously depends
on the data type of \(x\). Adding a new binding to a given environment
\(E\) is once more done with the operator \(\oplus\) which is formally
a morphism \(E = (\iota, P, M) \rightarrow (\iota', P', M') = E'\):

\begin{equation}
  E' = E \oplus \mapval{x}
\end{equation}

Note that multiple inclusion of variables overrides the previous 
definition. Thus, the meaning of

\begin{align}
E' =  (E \oplus \mapval{x}) \oplus x \xrightarrow{\pi_{x}'} \range(x)
\end{align}

is to create a probabilistic environment which contains \(\pi_{x}'\)
as probability distribution for the variable \(x\). The previous
distribution \(\pi_{x}\) can then not be recovered any more in \(E'\).

In direct analogy to Kraus aggregations, probabilistic environments can
be combined with \(+\); the summands are prefixed by some constraint
that states which one has to be chosen with which probability:
\begin{equation}
p_{1}(\{c_{1}\})\cdot E_{1} + p_{2}(\{c_{2}\})\cdot E_{2} + \cdots
\end{equation}
where \(\{c_{i}\}\) are the conditions which determine the values of
\(p\) and \(\sum_{i}p_{i} = 1\). This construction is necessary for
the description of, \eg, if-conditions when it is not a priori
determined which path will be selected. If both paths of an
if-condition perform a modification on the same variable that already
existed before the branch, then the variable will have different
values after the merge point. The entries of the probabilistic environment 
which record this assignment are then prefixed by the branching
probability. This is also one of the reasons why the branching
probability needs to be kept even after the branched paths are merged
again.

\begin{definition}[Distributivity of \(\oplus\) over \(+\)]
The operation \(\oplus\) is defined to be distributive over \(+\), \ie, 
\((E_{1} + E_{2}) \oplus \mapval{x} = E_{1} \oplus \mapval{x} + 
E_{2} \oplus \mapval{x}\). This ensures that adding a new binding
to a sum of environments results in adding the binding to all contributing
environments automatically.
\end{definition}

\begin{remark}
This formalism is \emph{not} equivalent with the functionality
introduced by stores (cf. \eg, \cite{mosses,reynolds}). It does still not make 
use of stateful variables per se, but rather updates the binding of a 
variable, \ie, the value it is associated with.
\end{remark}

Observe that the probabilistic environment is only necessary for classical,
but not for quantum variables: The state (or, rather: the history
of all operations performed until the present moment) of the quantum 
mechanical constituents of the computation can be reconstructed with
the aid of the Kraus aggregation. Nevertheless, the probabilistic
environment is necessary to keep track of quantum variables in a different
way which will be introduced in a moment.

Note that the view on quantum variables differs slightly from that on
classical ones: It is not only necessary to keep track of the
structure of a variable (as is done by the typing context), but also
of the \emph{position} within the quantum heap -- this is necessitated
by the underlying model of computation as introduced in
Section~\ref{qprog:qcom}.\footnote{The value of a quantum variable can
  obviously not be directly stored in an environment because the state
  might be in a superposition.  The operations performed on the
  quantum bit are recorded in the Kraus aggregation and unambiguously
  specify the state.}  The environment can be used to provide this
kind of information by supplying a map
\begin{equation}
  \#: \iota \owns v \rightarrow (i_{1}, i_{2})
\end{equation}
where \(i_{k}\) are integer numbers with \(0 \leq i_{k} < Q\) and
\(Q\) is the size of the quantum heap. The tuple \(i_{1}, i_{2}\) 
denotes the interval \([i_{1}, i_{2}]\) which contains 
\(i_{2} - i_{1} + 1\) quantum bits. Obviously, the number of qbits 
allocated in the quantum heap must agree with the number of qbits
necessary for the type of the variable as given by the typing context.

In theory, it is possible to assume that the quantum heap
can always be partitioned into consecutive intervals; we do not need
to take care of issues like fragmentation which does obviously appear
in implementations and simulations. We assume that the hardware of the
quantum memory take care of this issue by acting like a memory
management unit.\footnote{This component of a processor creates a view
  of the available memory such that every application -- roughly --
  thinks that it would have an own linear address space which is as
  big as the the one available for the whole system.}  Note that if
the user is allowed to directly address the components of the quantum
heap, it is possible to cause run-time errors as in QCL.  Therefore,
we do not allow this.

Consider a subset \(M = [0,n]\) of \(\mathbbm{N}\). The set of all 
interval partitions is given by\footnote{An example might illustrate this
 definition: Consider the set \(\{[0,1],[2],[3,4]\}\). This is 
 a proper partition since no elements overlap and the boundaries
 are adjacent. These conditions can be ensured by considering
 the last element of the \(i^{\text{th}}\) set given
 by \((m\#i)\#(\card(m\#n) - 1)\) and the first element of the 
 \((i+1)^{\text{th}}\) set given by \((m\#(i+1))\#0\). If the difference
 between these is \(+1\), then both the adjacency and no overlap conditions
 are fulfilled. If this holds for all subsets, we have a proper partition.}
\begin{align}
I(M) \equiv \{&m \subseteq \mathcal{P}(M) | 
             \forall n\in [1,\ldots,\card(m)-1]:\nonumber\\
             &((m\#n)\#0 -
             (m\#(n-1))\#(\card(m\#(n-1))-1)) = 1\}
\end{align}
where we suppose that the contents of all sets \(m\#i\) is sorted in
ascending order. This can be used to formally define how the
probabilistic environment can be adapted to the requirements for
quantum variables:

\begin{definition}[Quantum part of the probabilistic environment]
Let \((\iota, P, M)\) be a probabilistic environment. It can be extended
to fulfil the requirements for the description of quantum variables by
the following construction:

\begin{itemize}
  \item \(P\) is extended to \(P \oplus (I([0,Q-1]) \cup \Sigma^{*})\) 
    where \(Q\) is the total number 
    of quantum bits present in a system. The set of intervals is used
    to represent quantum variables which reside on the local quantum
    heap, \ie, which were allocated in the module the probabilistic
    environment belongs to. \(\Sigma^{*}\) is used to denote
    the originating module for variables which were received from
    some other party.\footnote{We have to make sure that every
      quantum bit in the system belongs to exactly one place in 
      a quantum heap. This is simple for single-party programs, but
      gets more complicated when communicating programs are considered
      because the case of sending the same quantum bit back and forth
      between participants must be taken into account. }
  \item Let \(\mathcal{Q} = \Set{v\in\iota|q(\chi(v)) = 1}\) be the set of 
    all identifiers for variables with quantum data type. Then, \(M'\) is an 
    injective morphism \(\mathcal{Q} \rightarrow I\) for which 
    \(\bigcap \range(M') = \varnothing\) (this ensures that 
    quantum variables do not overlap on the quantum heap) must hold. Then 
    \(M\) is replaced by \(M \oplus M'\) in the previous definition.
\end{itemize}
\end{definition}

Note that this definition reflects a fundamental difference between
classical and quantum variables: While a classical variable is nothing
else than a mapping between an identifier and a value that can be
governed by a probability distribution, such a mapping is in general
impossible for quantum variables because they do not have a value per
se, but only a certain quantum state. To describe this quantum state
precisely (disregarding the principal impossibility of implementing a
measurement that delivers this information by inspecting a single copy
of a quantum system), one would need an infinite amount of classical
information because even a simple system such as a qbit takes values
in a \emph{continuous} space as a consequence of quantum
superpositions. This makes the classical approach of mapping the
identifier to a probability distribution of values impossible.
Nevertheless, the quantum variable is completely characterised if its
location on the quantum heap together with the operations performed on
its initial state are known.

\begin{remark}
Although we retain the name \emph{probabilistic environment} also for 
the version of the environment extended to quantum variables, there are 
no probabilities involved in the connection between variable names and the 
allocated positions on the quantum heap. The convention just simplifies the 
notation.
\end{remark}

\begin{remark}
Support for mixed quantum/classical types would require a little more
effort compared to the case of full separation because with the introduction
of such types, the direct decomposability of the probabilistic environment
would not be feasible any more. Nevertheless, no fundamental difficulties
would be associated with this.
\end{remark}

Adding a new quantum variable with name \(\nu\) which occupies the quantum 
heap positions given by \((q_{1}, \ldots, q_{n})\) is written as

\begin{equation}
  E \oplus_{q} \nu:(q_{1}, \ldots, q_{n})
\end{equation}

Removing a quantum variable is denoted by \(\ominus_{q}\); this is
required when qbits are transmitted and thus cannot be accessed any
more (we drop the index if there is no danger of confusion).  Note
that there is no need for a corresponding operation for classical
variables because overlays provide the required functionality, as will
be shown later.

\subsubsection{Kraus aggregations}
We can directly adopt the definition of Kraus operators as given
in Section~\ref{math:kraus_aggregations}. Nothing needs to be
modified for our purposes (note that the composition of two
Kraus aggregations was denoted by \(\circ\) instead of 
\(\oplus\) as used for the other elements of the three-tuple
\((K,T,E)\) to avoid confusion with the symbol \(+\) used to
combine sub-aggregations).

It is possible to show that this semantic framework can be used to 
formalise standard QPL, but we omit the details here.

\summaryeven{We have introduced the structures which are necessary to
  specify the denotational semantics of cQPL; they fulfil the required
  properties as described in Chapter~\ref{math:chapter}. The semantic
  framework consists of three components: A Kraus aggregation which is
  used to store all quantum mechanical operations performed by commands
  of cQPL, a probabilistic environment which maps identifiers to
  values (possibly governed by a probability distribution) and
  provides mechanisms to ensure that quantum variables do not
  interfere with each other, and a typing context whose information is
  the basis for compile-time correctness checks of programs. These are
  grouped in a \((K,T,E)\) tuple which will be omnipresent in the
  following.  Additionally, we have derived some criteria for the
  identity of data types.}

\subsection{Some examples}\label{form:informal_examples}
Before we commence to extend the formal definitions for multipartite
systems, we want to demonstrate the introduced concepts with some
examples which should aid the reader to see the rationale behind
their definition.

\subsubsection{Semantics of sequential programs}
\begin{figure}[htb]
  \centering\includegraphics[width=0.7\textwidth]{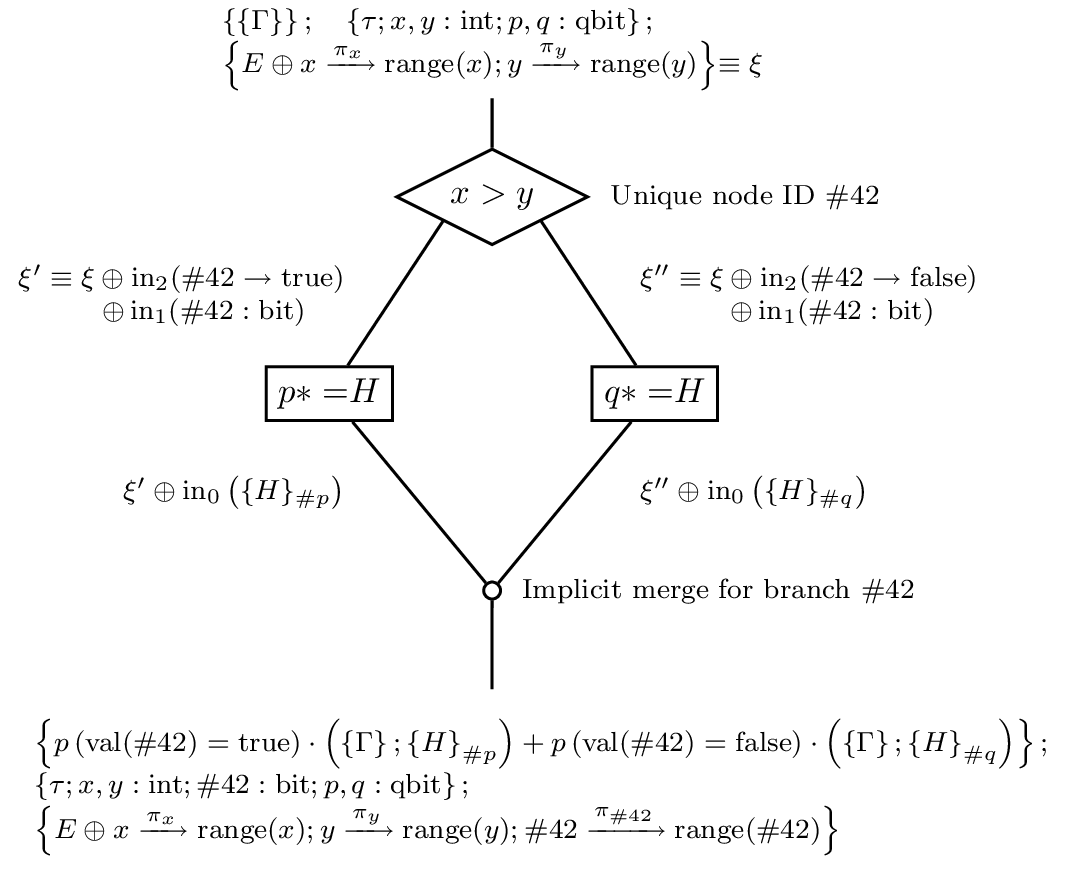}
  \caption{Flow graph of a simple branching operation to demonstrate the
    elements of the semantic framework: A \(3\)-tuple \((K,T,E)\) is used
    to annotate every edge of the graph; \(K\) is a \textsl{list} 
    (or aggregation) of \textsl{Kraus operators}, \(T\) is the 
    \textsl{typing context} and \(E\) the \textsl{probabilistic environment}.
    Note that the annotation of the graph uses several abbreviations as
    defined in the text.}\label{seq_fgraph}
\end{figure}

Consider the following program fragment which applies a
Hadamard matrix to the quantum variable \texttt{p} or \texttt{q}
depending on the result of a branch based on a comparison of
two classical variables \(x\) and \(y\):

\begin{verbatim}
if (x > y) {
   q *= H;
}
else {
   p *= H;
};
\end{verbatim}

\noindent The flow graph representation of the fragment is given in 
Figure~\ref{seq_fgraph}. To shorten the annotation of the edges and to avoid
repeating the same information over and over, we introduce the
injection functions \(\inj_{i}\). \(\inj_{i}\) are \(0\)-based
injections into the \(i^{\text{th}}\) element of an \(n\)-tuple. If we
want to add the contribution \(\{U\}\) to the element \(K\) of the
tuple \((K,T,E) = \xi\), we can write this as \(\xi \oplus
\inj_{0}(\{U\})\). The initial \((K,T,E)\) tuple of the example
flow graphs is abbreviated by \(\xi\); modifications derived from
this are denoted by \(\xi', \xi'', \ldots\).

The initial configuration of the 3-tuple \((K,T,E)\) is given by
\(\Gamma\) (initial list of Kraus operators), \(E\) (probabilistic 
environment) and \(\tau\) (typing context). We don't care about their contents 
in detail, they can represent the semantics of any valid program fragment 
that might be placed before our example. Things which are of 
interest for our code fragment are:

\begin{itemize}
  \item The two classical variables \(x\) and \(y\), both of type
    \texttt{int}.
  \item The probability distributions \(\pi_{x}, \pi_{y}\) which map
    the variables \(x\) and \(y\) to a value contained in
    \({[0,2^{\text{bits\_per\_int}} - 1]}\).
\end{itemize}

The probability distribution for classical variables arises because we
work with Kraus operators describing quantum operations instead of
density matrix transformations. Consider the measurement of a single
qbit whose result is stored in a classical bit variable: We know that
the range of the measurement outcome is \(\{0,1\}\), but we don't know
with which probability the ``\(0\)'' and the ``\(1\)'' will appear
because we do not have an explicit density matrix to describe the
qbit. This piece of information can only be gained when the final
semantics of the program (in form of a total set of Kraus operators)
is ``applied'' to a well-defined initial configuration; only then
quantitative statements about the distribution are feasible. All we
know is that the measurement result \(x\) will be governed by a
probability distribution \(\pi_{x}\) with a certain well-defined
range, so we preserve that information.

Since the values of \(x\) and \(y\) are given by a probability
distribution, the result of the comparison \(x > y\) (with outcome
range \(\{\text{true},\text{false}\}\)) can only be specified by
another probability distribution which can be deduced from \(\pi_{x}\)
and \(\pi_{y}\). Since we need to reference the outcome of the
comparison at a later point in the flow graph (when the two edges of
the branch are merged), a unique identifier for the node is created
(\(\#42\) in this case) and the probabilistic environment is extended
accordingly.

Depending on the outcome of the comparison, a Hadamard gate is applied
on either \(p\) or \(q\). This does not change the probabilistic
environment or the typing context, but is recorded by placing an
appropriate Kraus operator in the Kraus aggregation (\(\#p\) and
\(\#q\) denote the position of the quantum bits in the quantum heap).

After every \texttt{if-then-else} construction, an implicit merge
operation which unites the two branches takes place. The probability
distribution of the branching condition is preserved in the
probabilistic environment under the label assigned to the branch
statement; the Kraus aggregation is transformed into a sum that
formally resembles a mixed state: With probability \(val(\#42) ==
\text{true}\) (which is the probability that \(x > y\)
evaluated to true), the operation \(\{\hat{\Gamma}\}; \{H\}_{\#p}\)
was performed, while with probability \(val(\#42) == \text{false}\),
the operation was \(\{\hat{\Gamma}\}; \{H\}_{\#q}\).

\subsubsection{Communication with EPR pairs}
Consider the following (pseudo-)code which describes how Alice creates
an EPR pair and transmits half of it to Bob:

\begin{verbatim}
module Alice {
   new qbit p := 0;
   new qbit q := 0;
   createEPR(p,q);
   send q to Bob;
   new bit b := measure(p);
   if (b) { ... } else { ... };
};
\end{verbatim}

\begin{verbatim}
module Bob {
   receive m from Alice;
   new bit b := measure(m);
   if (b) { ... } else { ... }; 
};
\end{verbatim}
 
\begin{figure}[htb] 
  \centering\includegraphics[width=0.5\textwidth]{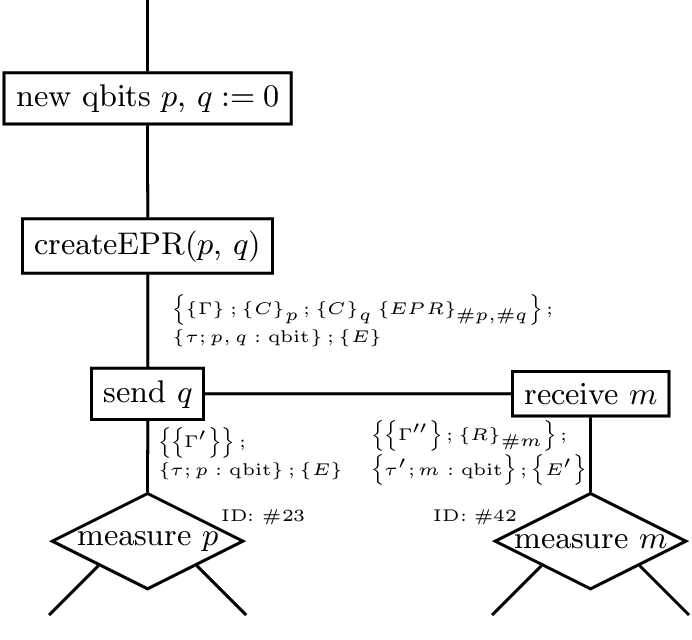} 
  \caption{Flow diagram which describes the creation of an EPR pair
    by Alice; she keeps the first half, while the second half is sent
    to Bob. Afterwards, both of them measure their qbit.
    \(\{\Gamma'\}\) is a shorthand for \(\{\Gamma\}; \{C\}_{p};
    \{C\}_{q}; \{\text{EPR}\}_{\#p, \#q}\), \(\{\Gamma''\}\) is the
    initial Kraus aggregation of Bob, \(\tau\) and \(E\) respectively
    \(\tau'\) and \(E'\) are the initial typing contexts and
    probabilistic environments of Alice and Bob.}\label{comm_graph}
\end{figure}

\noindent The corresponding flow diagram is given in Figure~\ref{comm_graph}.

Note that the labelling formalism is reduced to the basic necessities
in this example in order to emphasise the central elements.  Because
Kraus operators are used to describe the quantum mechanical
operations, it is possible to perform spatially disjoint actions by
parties who do not know the total state. When the edges are merged
together, the operations performed by Alice and Bob can (as was
mentioned before) be factorised as
\(\hat{A}\otimes\mathbbm{1}_{\text{B}}\),
\(\mathbbm{1}_{\text{A}}\otimes\hat{B}\) for the combined semantics of
both branches; the total density matrix is not involved in this,
contrary to annotation-based QPL. The framework to generate the semantics of
the total system from the semantics of the components will be introduced
in the following section.

\subsection{Extension to multipartite systems}%
\label{form:multi_party_components}
Since we want to consider the formal semantics of programs which
deal with communication, we need to extend the previous definitions
to the multi-party case. For this, observe that the number of
participants can naturally be assumed to be finite which makes it possible to 
label each party with a unique identifier of finite length.
For formal simplicity, we assume that the set of labellings for
participants \(\LC\) is disjoint with the standard
labels used for variables, \ie, \(\LC \cap \Sigma^{*} = 
\varnothing\). Let \(\LC(i)\) denote the unique label given by
the \(i^{\text{th}}\) entry of \(\LC\).
 
Assume that we have \(n\) communicating parties which are
labelled with \(\mathfrak{l}_{1},\ldots, \mathfrak{l}_{n} \in \LC\). 
According to the principle of compositionality (we will explain this
for the context of communication in more detail in 
Section~\ref{form:comb_proc} on Page~\pageref{form:comb_proc}),
there is a three-tuple \((K,T,E)\) for every participant, \ie,
\((K_{1}, T_{1}, E_{1}), \ldots, (K_{n}, T_{n}, E_{n})\). The combined
three-tuple for the complete system is then given by

\begin{equation}
\left(\otimes_{i=1}^{n}K_{i}, \otimes_{i=1}^{n}T_{i}, 
      \otimes_{i=1}^{n}E_{i}\right)
\end{equation}

We will consider how the tensor product needs to be defined for each
component to provide a sound basis for our further
needs.\footnote{Note that although formally, different tensor products
  are used for each element of the \((K,T,E)\) tuple, we use the same
  symbol for all of them to simplify notation.}

\subsubsection{Kraus Aggregation}
Consider, for simplicity, two Kraus aggregations 
\begin{align}
\Lambda_{1} &= \{A_{k}^{1}\}, \{A_{k}^{2}\}, \ldots, \{A_{k}^{n}\}\\
\Lambda_{2} &= \{B_{k}^{1}\}, \{B_{k}^{2}\}, \ldots, \{B_{k}^{m}\}
\end{align}
(in case of \(n \neq m\), the shorter list can be padded with zero elements
so that \(m=n\) can be assumed, but note that it is only for
notational convenience). If both
lists operate on a disjoint set of qbits, \ie, no send/receive
(pseu\-do-)op\-er\-a\-tions are contained in the lists, then \(\forall i,j:
[\{A_{k}^{i}\}\otimes\mathbbm{1}_{\text{B}},
\mathbbm{1}_{\text{A}}\otimes\{B_{k}^{j}\}] = 0\) holds. The Kraus
aggregation for the composite system can be written as:
\begin{equation}
\Lambda_{1} \otimes \Lambda_{2} = \{A_{k}^{1}\}\otimes \{B_{k}^{1}\},
\ldots, \{A_{k}^{n}\}\otimes \{B_{k}^{n}\}\label{form:comp_sys_noinfl}
\end{equation}
Note that members of type \(\{A_{k}\}\otimes\{B_{k}\}\) can be written
as \(\{A\} \otimes \mathbbm{1}_{\text{B}} + 
\mathbbm{1}_{\text{A}}\otimes \{B\}\). If this is done for all list 
elements, we see that \emph{all} combinations of \(A \otimes \mathbbm{1}\) 
with \(\mathbbm{1}\otimes B\) commute; this
induces many equivalent orderings of the lists -- in addition to the
normal commutative equivalences as given by 
Eqn.~\ref{math:kraus_agg_equiv} for the sublists -- which 
needs to be considered when multi-party aggregations are tested for semantic 
equality in Section~\ref{form:interp_equiv}.

Formally, the equivalence class of compatible Kraus aggregations for
two independent parallel systems is given by
\begin{align}
  &\left\{\bigoplus_{i=1}^{n}\inj_{\varphi(i)}
    \{A_{k}^{i}\}\otimes\mathbbm{1}_{\text{B}}
    \oplus 
    \bigoplus_{i=1}^{n}\inj_{\varphi(n+i)}
    \mathbbm{1}_{\text{A}}\otimes\{B_{k}^{i}\}
  \right\}\label{form:kraus_two_comm_equiv}\\
  \text{with }
  &\varphi\in\sym(2n):\varphi(i)<\varphi(i+1)
  \forall i\in [1,\ldots,n]\wedge \forall i\in [n+1,\ldots,2n]\nonumber
\end{align}
where the additional constraints on the permutation make sure that
the order of \(\{A_{k}^{i}\}\) and \(\{B_{k}^{i}\}\) is preserved.
We can generalise this approach to \(n\) Kraus aggregations given in
the contracted normal form (padding is applied as usual to compensate
for different cardinalities):

\begin{definition}[Tensor product for Kraus aggregations]
  Let \(\Lambda_{1} =
  \sum_{k_{1}=1}^{N}p_{k_{1}}^{1}\Lambda_{k_{1}}^{1}\), \dots,
  \(\Lambda_{n} = \sum_{k_{n}=1}^{N}p_{k_{n}}^{n}\Lambda_{k_{n}}^{n}\)
  be Kraus aggregations in the contracted normal form. The tensor
  product of these is given by

\begin{equation}
\bigotimes_{i=1}^{n} \Lambda_{i}\equiv 
 \sum_{k_{1}=1}^{N}\cdots\sum_{k_{n}=1}^{N}p_{k_{1}}^{1}\cdots p_{k_{n}}^{n}
 \cdot\Lambda_{k_{1}}^{1}\otimes\cdots\otimes\Lambda_{k_{n}}^{n}
\end{equation}

Note that the naming of qbits changes when multipartite systems are
merged. For Kraus sets which symbolically refer to qbits, the labels
must be updated accordingly (the exact renaming scheme is given in
Definition~\ref{form:def_mp_tc}). Two aggregations are equivalent if
they are member of the same equivalence class as given by
Eqn.~\ref{form:kraus_two_comm_equiv} or member of the equivalence
class given by the same formula, but induced by a compatible ordering
of one or more of \(\Lambda_{i}\) as defined by
Eqn.~\ref{form:comp_sys_noinfl}.
\end{definition}

Note that we will show in Section~\ref{form:comb_proc} how send and receive
operations can be integrated into this formalism. 

\subsubsection{Typing Context}
\begin{definition}[Tensor product for typing contexts]\label{form:def_mp_tc}
The tensor product of \(n\) typing contexts \((\iota_{1},\theta_{1},\chi_{1}), 
\ldots, (\iota_{n},\theta_{n},\chi_{n})\) is given by

\begin{equation}
\bigotimes_{i=1}^{n}(\iota_{i},\theta_{i},\chi_{i}) = 
\left(\bigcup_{i=1}^{n}\LC(i)\iota_{i},
      \bigcup_{i=1}^{n}\theta_{i},
      \bigcup_{i=1}^{n}\LC(i)\chi_{i}\right)
\end{equation}

where \(\LC(i)M\) denotes a set which contains all elements of \(M\)
prefixed by the unique identifier \(\LC(i)\); \(\LC(i)\chi\) denotes
the morphism \(\chi\) adapted to the new naming scheme.
\end{definition}

\subsubsection{Probabilistic Environment}
The probabilistic environment can be adapted to multipartite systems
analogous to the typing context, \ie, by prefixing the variable
names with appropriate labels and adapting the
morphism used to connect the set of identifiers with the set of 
data types. Obviously, the prefix for each subsystem must be the
same as used for the typing context.

The quantum part needs to be modified as well: The local quantum heaps
must be united to a single one; necessarily, the positions of all
variables on the new heap still need to be disjoint. This is simple to
achieve: If \(E_{0}\) uses the range \([0, n_{1}]\) and \(E_{1}\) the
range \([0,n_{2}]\), the new range is given by \([0,n_{1}+n_{2}+1]\)
and the morphism \(M'\) needs to be adapted such that the mapping
remains unchanged for variables originating from \(E_{0}\) and the
constant offset \(n_{1}+1\) is added to its codomain for variables
originating from \(E_{1}\).\footnote{Note that both local quantum
  heaps could have already used the full number of available quantum
  bits (\(Q\)); we do not consider this problem any further because it
  is alway possible to limit the number of quantum bits for \(n\)
  communicating systems to \(n\cdot Q\) because both \(n\) and \(Q\)
  are finite. We are not too concerned about this problem because we
  always assume that there are enough qbits available. The reason
  behind the restriction to a finite, but fixed number of qbits is to
  ensure the boundedness of all Kraus operation as explained in
  Chapter~\ref{math:chapter}.} Likewise, the function \(\#\) which
associates variable names with quantum heap positions needs to be
updated such that the modified variables names are mapped to the
modified positions.

We will not consider the renamings any more in the following parts, but
just take them as given; this simplifies the notation considerably.

\subsubsection{Quantum channels}
Quantum channels are used to exchange information between
processes.\footnote{A user in a real-world implementation is nothing
  else than a process in the simulation. We thus use both terms
  interchangeably.}  Although not only quantum data, but also
classical variables can be sent, we restrict our considerations to the
first case because the second one is not too interesting from a
physical point of view and would only obstruct the view on the central
elements. In particular, classical communication can be seen as a
special case of quantum communication (\cf, \eg,~\cite[Section
6.2.2]{keyl}), hence the generality of the approach does not suffer
from this restriction).  Besides, the topic of classical communication
has been investigated in classical programming language theory for a
long time, so we can refer the reader to the wealth of existing
literature about this topic, \eg, Ref.~\cite{reynolds}.

We have already introduced quantum channels informally in 
Section~\ref{qprog:quantum_channels}; here, we consider the concept formally.

\begin{definition}[Quantum channel]
  A quantum channel is a five-tuple \((O, D, \mathfrak{S},
  \mathfrak{R}, \mathcal{F})\) where \(O\) is the origin and \(D\) the
  destination for a quantum variable (these can, \eg, be represented
  by processes), \(\mathcal{F}\) is a FIFO\footnote{First in, first
    out queue. Informally, this is a queue where objects can be put in
    on one side and taken out on the other side. The object which is
    put in first comes out first, the second one comes out second
    etc.} containing objects of type \((\varrho, \sigma)\),
  \(\mathfrak{S}\) is a morphism to place two-tuples \((\varrho,
  \sigma)\) in \(\mathcal{F}\) and \(\mathfrak{R}\) is a morphism to
  retrieve two-tuples \((\varrho, \sigma)\) from \(\mathcal{F}\). As
  usual, \(\varrho\) represents the density matrix of a quantum
  variable and \(\sigma\) the associated type.
\end{definition}

Thus, the quantum channel can be used to make sure that not only typing
is guaranteed to be preserved along communication (for this, confer
further Section~\ref{sem:runtime_errors}), but also that quantum data
does not appear multiple times in a composite system at the same time
which is necessary to avoid unphysical situations in the simulation.

\begin{remark}
Note that quantum channels are only necessary when the parallel
composition of two or more processes is considered. For single modules,
the functions used to deposit respectively request information from the
channel together with an abstract representation of the channel
(\eg, an identifier) are sufficient.
\end{remark}

\begin{remark}
Also note that most descriptions of quantum communication protocols
do not consider typing of the exchanged data explicitely, it is nevertheless
implicitly implied by the physical realisation of the protocol,
\eg, in the measurement process, by the hardware used to realise the
communication channel or in the way the quantum part is implemented in
general.
\end{remark}

\subsection{Existence of fixed points}
Because fixed points are important for the denotational description,
we need to prove the following theorem which states a condition for
the existence of such. In the following, the condition can shown to be
fulfilled for every element of the semantics.

\begin{theorem}\label{fixed_point_bounded_op}
Let \(T\) be a linear operator \(D\rightarrow D\) acting on a complete partial 
order \(D\) with bottom \(\bot_{D}\). If \(T\) is bounded, then a fixed 
point of \(T\) exists.
\end{theorem}

\begin{proof}
  Since \(T\) is bounded, we can see from
  Theorem~\ref{math:props_lin_operators} that it is continuous as
  well. Topological continuity implies Scott continuity as was shown
  in Theorem~\ref{math:top_scott_cont}.  The existence of a fixed
  point is now given by Theorem~\ref{math:fixed_point_theorem}, as
  required.
\end{proof}

\begin{remark}
  Note that the same proof could have been deduced at a slightly more
  abstract level using the notion of a \emph{pointed dcpo}, \ie, a
  dcpo with a least element. For structures fulfilling this condition,
  Theorem~2.1.19 in \cite{abramsky_jung} ensures that the desired
  least fixed points exist.  Ref.~\cite{selinger_qpl} uses a somewhat
  similar reasoning in the description of recursive procedures where
  the existence of least fixed point solutions for these is explained
  by the fact that for Scott-continuous endofunctions on pointed
  complete partial orders, these always exist.
\end{remark}

\begin{remark}
  For those who want to be extra sure, the classical fixed point
  theorem by Schauder which states that any continuous map with a
  countably compact image on a compact, convex subset of a Banach
  space has a fixed point could also be used to derive the required
  property of cp-maps.
\end{remark}

We will need fixed points to solve recursive equations which occur in
denotations specified by recursive equations. These are required for
loops and the combined semantics of communicating systems.

\subsection{Types of interpretational equivalence}\label{form:interp_equiv}
The term ``equivalence'' is not unambiguous without further specification.
Under which conditions can two programs or, respectively, the denotations
of two programs be considered as equivalent? QPL has to distinguish
between two different types of equivalences as noted  
in~\cite[Section 6.6]{selinger_qpl}; likewise, we can define several
types of equivalence:\footnote{Note that our definitions of
  equivalence do not coincide with the types of equivalence given
  by Selinger.}

\begin{itemize}
  \item Two programs are \emph{textually} equivalent if there is a
    bijective mapping between the set of all variables the programs use
    and for every constituent of program A, there is a constituent of
    program B such that \(A_{i} = B_{i}\), \ie, the programs are
    identical already at the level of the syntax. The ordering of these
    constituents must be identical.
    
    For example, the two program fragments \texttt{new int a; a:=1;}
    and \texttt{new int b; b := 1;} are equivalent because the
    sequence of commands is identical if the replacement
    \(\text{\texttt{a}}\leftrightarrow \text{\texttt{b}}\) is applied
    to the variables.
  \item Two programs are \emph{denotationally} equivalent if
      their denotations are identical.
\end{itemize}

The second definition only shifts the problem because it leaves the
question of how to identify equivalent denotations. This is
problematic for cQPL because we do consider multiple representations
of superoperators which have identical effects; some care needs to be
taken to exactly specify the meaning of ``\emph{identical}'' in this
setting. Denotational equivalence can be refined to the following
cases for cQPL:

\begin{itemize}
  \item \emph{Direct denotational equivalence}: We can distinguish three 
    different scenarios which exhibit
    direct denotational equivalence for \((K_{1}, T_{1}, E_{1})\) and 
    \((K_{2}, T_{2}, E_{2})\) given as denotations of \(A_{1}, A_{2}\):

    \begin{enumerate}
      \item \(\card(K_{1}) = \card(K_{2}), \forall i \in [0,\ldots, 
        \card(K)[:{K_{1}}_{i} = {K_{2}}_{i}\) and \(E_{1} = E_{2}, 
        T_{1} = T_{2}\). This means that both programs have have the same
        number of Kraus sets as denotation, the probabilistic environment
        and the typing context contain the same information and the
        denotations of the statements are pairwise identical.
      \item \(E_{1} = E_{2}\), \(T_{1} = T_{2}\), \(\card(K_{1}) =
        \card(K_{2})\),  \(\exists \varphi\in\sym(n):\forall i \in
        [0,\ldots, \card(K)[:{K_{1}}_{i} = {K_{2}}_{\varphi(i)}\) such
        that the sum of commutator products calculated according to
        the method given in Section~\ref{math:kraus_equiv} vanish
        identically, \ie, only permutations with vanishing commutator
        have been used. This equality holds if only commuting
        statements have been exchanged to match the lists, the total
        denotation is thus identical.
      \item \(\forall \varrho \in \mathcal{D}: (K_{1}, T_{1}, E_{1})(\varrho)
        = (K_{2}, T_{2}, E_{2})(\varrho)\) where \((K,T,E)(\varrho)\) means
        the application of the Kraus set in \(K\) on \(\varrho\) where the
        information contained in \(E\) is utilised to construct the
        proper superoperators because the exact representation of
        \(K\) in general depends on information given in \(T\) and \(E\). Note
        that the initial state resolves any symbolic parametrisations
        which may be present in \(K\).
    \end{enumerate}
    
    The first condition obviously implies the second and third condition;
    the second implies the third, but the other direction is not true
    in general, so equivalences of decreasing strength are defined
    by this enumeration.
  \item \emph{Heap-permutative} respectively \emph{variable-permutative 
      denotational equivalence} is given if there exists a permutation 
    of the quantum heap positions (respectively the variable names) such
    that direct denotational equivalence holds. These permutation
    schemes can be used to align different probabilistic environments
    to each other.
\end{itemize}

Note that textual equivalence implies denotational
equivalence, but the converse statement is not valid in general.

A last form of equivalence that needs to be considered concerns the
parallel execution of programs. If \(\{A_{i}\}\) represents a set of
\(n\) communicating modules, the order in which the subsystems are
given does not make any difference, \ie, \(\stdsembrack{A_{1} || A_{2} ||
  \cdots || A_{n}} \cong \stdsembrack{A_{\varphi(1)} || A_{\varphi(2)}
  || \cdots || A_{\varphi(n)}}\) for any \(\varphi \in \sym(n)\) and
an according update of the references to the other subsystems \(A_{k},
k \neq m\) in \(A_{m}\) for all \(m\).  Likewise, relabelling of
communicating modules does not change the meaning of parallel
execution if the reference names in all participating modules are
updated correspondingly. This type of equivalence can be referred to
as \emph{communicative equivalence}.

The problem of how to detect equivalence between different representations
will emerge several times in the following remarks and is not easy
to solve constructively.

\summaryeven{We have augmented the definitions of the semantic basis
  \((K,T,E)\) with the elements required to represent communicating
  systems, \ie, cQPL programs which are generally independent of each
  other, but can exchange quantum mechanical and classical data in a
  well-defined way. Additionally, we have shown that fixed points
  exist in this framework; they are necessary to assign semantics to
  numerous components of the language as explained in
  Chapter~\ref{math:chapter}. Criteria for the equivalence of cQPL
  programs were specified as well; this allows to check if programs 
  which are specified by different sequences of commands have the 
  same effect.}

\subsection{Semantics of the language components}
Chapter~\ref{qprog} gave an informal\footnote{It should be noted that
  although \emph{informal} may sound a little fuzzy, such a description
  is normally the maximal level of accuracy with which users of 
  programming languages (and in most cases, implementors as well) are 
  confronted.}
introduction the the language elements
of cQPL. In this chapter, we will use the mathematical and semantical
formalism introduced in the preceding sections to give a rigorous 
mathematical meaning to these statements. By the compositionality of
denotational semantics, this means that all cQPL programs (which are,
necessarily, composed of cQPL statements) have a defined semantics.
There are two possible representations 
for cQPL programs in form of textual descriptions and graphical flow charts;
we resort to the particular representation that is more convenient for
the desired purpose in the following remarks.
Establishing a formal correspondence between both possible
representations of cQPL is obvious and follows exactly the argumentation
used in \cite{selinger_qpl}; we will thus not bore the reader with 
details on how to relate both representations uniquely.

Note that we try to keep the purely classical formalism as terse as
possible because most problems related with this are not too interesting
from a physical point of view. More elaborate descriptions or gentle
introductions can be found, \eg, in Refs.~\cite{mosses,reynolds,winskel}.

\subsubsection{Some notational remarks}\label{form:notat_remarks}
Some conventions and notations widespread in semantics are uncommon
in physics, thus we want to make two short remarks before proceeding
further.

\paragraph{Typed lambda calculus}
Computer science literature habitually uses the typed lambda calculus
(cf., \eg, \cite{rechenberg_pomberger}) to formulate the equations for
valuation functions; this is useful to not only clarify which
parameters are used, but also to determine their type. We, in
contrast, use a different notation. Observe, for example, the
denotation of the dyadic operator \(+\) which adds two natural
numbers:
\begin{equation}
\mathcal{DO}\stdsembrack{+}(n_{1}, n_{1}) = \text{sum}(n_{1}, n_{2})
\end{equation}
It is intuitively clear that we are talking about a function which
takes two natural numbers as arguments and computes another natural
number as result. Nevertheless, we did not formally specify the data
types of the arguments nor of the result of the function.

This can be solved by using the typed lambda calculus in which the
function would be written as:
\begin{equation}
\mathcal{DO}\stdsembrack{+} = \lambda n_{1}\in\mathbb{N}. 
      \lambda n_{2}\in\mathbbm{N}.\text{sum}(n_{1}, n_{2})
\end{equation}
This very simple example already demonstrates that the representation
requires a considerable notational effort. Since nearly all valuation
functions for the semantics of cQPL defined in the following require
\((K,T,E)\) tuples in addition to the effective parameters, this would
unduly inflate the length of equations which does not really increase
lucidity. Thus, we stick to a simplified notation which follows the
algebraic convention for functions as presented above.  The domains
where parameters originate from are normally clear from the context;
we will mention it explicitely should this not be the case since
typing is obviously not explicitely part of the simplified
description.

\paragraph{Currying/Sch{\"o}nfinkeln}
Another point we want to mention is the insight that functions of more
than one parameter may always be composed by a number of subsequent
functions which take exactly one parameter. Thus, a function
\(f(x_{1}, x_{2}, x_{3}) = y\) with \(x_{i}, y \in \mathbbm{N}\) which
is an element of \((\mathbbm{N}\times\mathbbm{N}\times\mathbbm{N})
\rightarrow \mathbbm{N}\) can also be seen as a mapping \((\mathbbm{N}
\rightarrow \mathbbm{N} \rightarrow \mathbbm{N}) \rightarrow
\mathbbm{N}\), the parentheses may also be omitted. The technique of
transforming a function with multiple arguments into a function which
takes only one argument, but returns another function which requires
the remaining arguments and returns the result is conventionally
termed \emph{currying}, although it was first introduced by
Sch\"onfinkel \cite{schoenfinkel}. The process can obviously be
repeated so that there are only functions left which take exactly one
argument. In the following, we will use the form which is more apt for
the respective purpose.

\subsubsection{State transformations and fixed points}%
\label{form:state_transform}
Valuation functions for top-level elements of cQPL (\ie, those
for expressions) work on a three-tuple \((K,T,E)\) and produce
a new three-tuple \((K', T', E')\) as we will see in the course
of the following remarks. Thus, these tuples form the domain which
is the basis of semantics. Since we will need fixed points as solution
of several recursive domain equations, we need to show
how to calculate them for \((K,T,E)\) tuples. We have already shown 
that fixed points exist for Kraus aggregations which fulfil certain
conditions. Now, we need to transfer this to \((K,T,E)\) tuples.

For this, note that the typing context is not involved in the
calculation of fixed points: Its purpose is to ensure well-typedness
of programs (which will be explained in more detail in
Section~\ref{form:type_system}) and (as a consequence) to make sure
that ownership of qbits is unique. Otherwise, it has no semantical
meaning.

To consider the contribution of the probabilistic environment, observe
that we could do without it in principle by using a different notation
as is, for example, the case in QPL (we did not adopt this approach
because it quickly leads to very long annotations which are cumbersome
to handle; additionally, our approach has a greater similarity with
the standard notation commonly used in denotational semantics).  For
every possible value of a variable in the probabilistic environment, a
specific Kraus aggregation can be inferred.  Consider a measurement of
two qbits whose result is stored in a classical variable of two bits.
A probability distribution \(x \xrightarrow{\pi_{x}:
  \mathcal{PROJ}(q,T)} [0,1,2,3]\) is stored in the probabilistic
environment where \(\mathcal{PROJ}(q,T)\) represents the information
that the quantum variable \(q\) contained in the typing context \(T\)
was measured (this will be covered in more detail in
Section~\ref{form:measurements} on page~\pageref{form:measurements}).
Since the exact form of the probability distribution depends on the
state of the measured variable about which nothing is known in the
worst case (if, \eg, the variable was received from a remote party
which did not characterise it any further), we have to account for all
possible cases, \ie, for all values the variable can have in
principle. Let \(\{\Gamma\}; \{M\}_{\#q}\) be the contents of the
Kraus list K from the \((K,T,E)\) tuple immediately after the
measurement where \(\{M\}_{\#q}\) denotes the Kraus set for a
projective measurement. This list can with the help of the
probabilistic environment be rewritten into a four-tuple
\begin{equation}
(\{\Gamma\}; \{P_{0}\}, \{\Gamma\}; \{P_{1}\}, \{\Gamma\}; 
 \{P_{2}\}, \{\Gamma\}; \{P_{3}\})
\end{equation}
where \(\{P_{i}\} = \ketbra{i}{i}\) is one of the projection operators
which constitute the POVM elements of the measurement. Note that the
Kraus aggregations contained in tuples of this kind are always
unparametrised.\footnote{This kind of split is one of the
  reasons why it is more convenient to work with Kraus representations
  of cp-maps instead of the cp-maps alone which would also be possible
  in principle.}

If there is now an operation performed which is independent of the
measured variable, the contribution to the Kraus set is 
appended to \emph{all} list components in this picture. Consider, 
for example, the application of some operator \(U\) to another quantum 
variable \(v\). The resulting four-tuple of Kraus aggregations then
looks like:
\begin{equation}
(\{\Gamma\}; \{P_{0}\}; \{U\}_{\#v}, \{\Gamma\}; \{P_{1}\}; \{U\}_{\#v}, 
 \{\Gamma\}; \{P_{2}\}; \{U\}_{\#v}, \{\Gamma\}; \{P_{3}\}; \{U\}_{\#v}).
\end{equation}

The first entry belongs to the case that \(x = 0\), the second to \(x
= 1\), and so on. Obviously, it is much simpler to write this in our
notation as
\begin{equation}
(\{\Gamma\}; \{M\}_{\#q}; \{U\}_{\#v})
\end{equation}
from which the tuple representation can be reconstructed.  This
also works if operations are considered that depend on the state of a
classical variable governed by a probability distribution. Consider,
for example, the case that an operator \(U\) is applied to some
quantum variable \(v\) if the value of \(x\) is \(2\) (the program
code for such an operation would be \texttt{if (x = 2) then 
v *= U;}). In our notation, the branching condition is
preserved in the probabilistic environment as shown in the
example given by Figure~\ref{seq_fgraph}. From this information,
the corresponding tuple representation
\begin{equation}
(\{\Gamma\}; \{P_{0}\}, \{\Gamma\}; \{P_{1}\}, 
 \{\Gamma\}; \{P_{2}\}; \{U\}_{\#v}, \{\Gamma\}; \{P_{3}\};).
\end{equation}
can be constructed. Note that \(U\) is only applied in the case \(x = 2\).

The transfer from our representation to tuples of Kraus aggregations is
easier when classical variables with defined values and no
associated uncertainty are considered, so we will not show an explicit
example for this.

As the forgoing considerations have demonstrated, the probabilistic
environment and the Kraus aggregation contained in the \((K,T,E)\)
tuple can be used to construct a tuple of Kraus aggregations where the
tuple contains one entry for every combination of values the classical
variables can be in. Fixed points of \((K,T,E)\) triples must
therefore be calculated separately for all possible Kraus aggregations
that can be constructed from the triple because each of them
represents another possible meaning of the program. This is possible
with the methods introduced before. After the fixed points have been
derived, the usual \((K,T,E)\) representation can be used again. Thus,
calculation of fixed points effectively only requires the properties
of Kraus sets as introduced before. In the following, we need thus
only make sure that the conditions for the existence of fixed points
of Kraus aggregations as given in Theorem~\ref{fixed_point_bounded_op}
are fulfilled to ensure the existence of fixed points for \((K,T,E)\)
tuples.

\subsubsection{Classical operations}
The classical subsystem of cQPL consists of the following parts:

\begin{itemize}
  \item Allocation and (implicit) deallocation of classical variables.
  \item If-then-else expressions.
  \item While-loops (this and the previous point require the evaluation
    of boolean conditions which must also be accounted for by the
    semantics).
  \item Sequential composition.
  \item Do-nothing-operation (skip).
  \item Sending and receiving of classical states.
  \item Assignment to classical variables.
  \item Calling procedures which manipulate classical data.
  \item Blocks.
\end{itemize}

Note that we do not cover sendig and receiving of classical data because
this has extensively been covered in the literature. Additionally, it is
in principle always possible to achieve the same effects with the transmission
of quantum mechanical information. In the following, we cover only
the valuation functions which are either absolutely indispensable or
are influenced by the quantum properties of our language. 

\paragraph{Sequential composition}
Modifications made to the typing context, the probabilistic environment
and the Kraus aggregation made by the first statement must be taken into
account when the semantics of the second statement is calculated:
\begin{equation}
  \EXP\sembrack{S_{1}; S_{2}} = \EXP\stdsembrack{S_{2}}
             (\underbrace{\EXP\sembrack{S_{1}}}_{= (K',T',E')})
\end{equation}
This is utilised many times in the denotational equations for quantum
communication.

\paragraph{Blocks}
Blocks are used to introduce multiple levels of scope into programs.
This can happen both implicitly (\eg, in loops) and explicitely (by 
syntactical specification of blocks), but there is no need to distinguish 
between these cases from a denotational point of view.

Superficially, a block looks just like a collection of multiple 
statements which are executed one after another; but some additional
points need to be taken into account:

\begin{itemize}
  \item New variables (both quantum and classical) may be declared inside
    blocks, but they cease to exist once the block's scope is left.
  \item New variables do overshade old ones if they share the identifier.
  \item Changed bindings of already existing variables are also 
    visible after the control flow has left the block's scope.
\end{itemize}

Thus, the probabilistic environment is partially affected by a 
block. The typing context before and after the block is identical 
and thus unaffected by the block's effect,\footnote{But note that
  the typing context \emph{within} the block may well be different
  than the one outside.} and the Kraus aggregation records everything that
has been done inside the block.

Since the mentioned problems appear in every programming language featuring
blocks, standard solutions are available in every 
textbook (as usual, cf. Refs.~\cite{mosses, reynolds, winskel}), so we
will not explictely present them here to save some formal overhead.

\paragraph{Conditionals and Operators}
Dyadic operators combine two subexpressions into one result, as, \eg,
all arithmetic operations do. Conditionals are operators which result
in a boolean variable, \ie, they evaluate to one of the values
\texttt{true} or \texttt{false} which are represented by \(1\) and
\(0\). In contrast to most classical languages, the result of both
types need not be fixed with certainty, but can be governed by a
probability distribution. Note that conditionals may not be used as
stand-alone expressions, but can only be part of conditional
statements.  This is why their semantic valuation does not result in
the usual \((K,T,E)\) tuple, but in a probability distribution for the
possible results which is, \eg, apt for inclusion into the
probabilistic environment. The basic valuation functions are given by:
\begin{align}
\OP\stdsembrack{\text{\(a\) DO \(b\)}}(T,E) &= 
               \DO\stdsembrack{DO}(\OP(a)(T,E) \otimes
                                           \OP(b)(T,E))\label{form:op_one}\\
\OP\stdsembrack{\text{MO \(a\)}}(T,E) &= \MO\stdsembrack{MO}(\OP(a)(T,E))%
                                                        \label{form:op_two}
\end{align}
where \(\text{DO} \in \Set{+,-,\wedge,\vee,\ldots}\) and 
\(\text{MO} \in \Set{\neg,-}\). The meaning of the operations is defined 
as usual, but the probability distribution nature of the arguments needs to 
be taken into account:
\begin{equation}
\DO\stdsembrack{DO}(a,b) = 
    \bigoplus_{v_{1}\in\range(a)}\bigoplus_{v_{2}\in\range(b)}
        \pi_{a}(v_{1})\pi_{b}(v_{2})\text{DO}(v_{1},v_{2}).
\end{equation}
This expression results in a new probability distribution that
can be used by the elements further up in the evaluation hierarchy.

Note that the denotation of a single variable is given by the
corresponding probability distribution which can be found in the
probabilistic environment:

\begin{equation}
\OP\stdsembrack{v}(T,E) = E(v)
\end{equation}

This definition ensures that chains of expressions using dyadic and monadic
operators (\eg, \(42+23+4\)) are covered by the semantics because \(a\) and 
\(b\) in Eqns.~\ref{form:op_one},~\ref{form:op_two} can either be values 
or other operator expressions.

Also note that the eventual action of the respective 
operators (\(+\), \(\wedge\), \ldots) can be seen intuitively, so we 
abstain from further formalisation and rely on the reader to use his common 
mathematical sense.

\begin{remark}
Note that we do neither consider any problems of numerical accuracy nor
of limited ranges of representable numbers for the specific data types.
Consequently, we also do not care for the problem of division by zero. 
We are aware that such pitfalls exist, but are not interested in their 
solution in this context since their nature is purely classical.
\end{remark}

\paragraph{If-Statements}
The semantic description of the if-statement is simplified by introducing
the following helper function:
\begin{align}
f((K_{0}, &T_{0}, E_{0}), (K_{1}, T_{1}, E_{1}), \pi,\nu, (K,T,E)) =\nonumber\\
    &\big(p(\nu=0)\cdot K\circ (K_{0}-K) + p(\nu=1)\cdot K\circ (K_{1}-K),\\
    &T \oplus \nu:\bit,
    p(\nu = 0)\cdot E\oplus\nu:\pi\oplus E_{0}
     + p(\nu=1)\cdot E\oplus \nu:\pi\oplus E_{1}
    \big)\nonumber
\end{align}
which eases selection of components of \((K,T,E)\) tuples gained by
other evaluations and additionally circumvents repeated semantic
evaluations of some components. The tuple \((K_{0}, T_{0},E_{0})\) is
the result of the evaluation of the \texttt{if}-branch, while
\((K_{1}, T_{1}, E_{1})\) is for the \texttt{then} branch. \(\pi\) is
the probability distribution governing the branch, and \(\nu\) is the
identifier which is used to represent this distribution in the
probabilistic environment. \((K_{i} - K)\) represents the Kraus aggregation
that contains only the elements that were appended to \(K_{i}\) in 
comparision to \(K\); this ensures that only new contributions introduced
in the branches are added to the Kraus aggregation finally.
The denotational description for the
if-statement then reads as
\begin{align}
\sembrack{
  \text{\texttt{if} }&\text{\(c\) \texttt{then} \(C_{0}\) 
        \texttt{else} \(C_{1}\)}} =\nonumber\\
    &f(\underbrace{\EXP\sembrack{C_{0}}}_{(K_{0}, T_{0}, E_{0})}, 
       \underbrace{\EXP\sembrack{C_{1}}}_{(K_{1}, T_{1}, E_{1})}, 
       \underbrace{\OP\stdsembrack{c}(T,E)}_{\pi}, 
       \underbrace{\text{uid}}_{\nu}, (K,T,E))
\end{align}
where \(\text{uid}\) is a unique identifier for the branch which
can be chosen at will, but must not be identical with other identifiers
already in use. Such a choice is simple for non-communicating 
programs. The extension to communicating systems is possible if
every identifier is given a unique prefix for each partner as
described before.

Note that although dyadic operators might syntactically be used
to describe arithmetic operations and not necessary conditionals,
this source of mistake is ruled out by the type system which only
allows boolean typed expressions for \(c\).

\paragraph{While-Statements}
While statements can be solved using the fixed-point theorem given in
Eqn.~\ref{math:fixed_point_theorem}. For this, note that the
denotation of the while function can be rewritten using the
previously considered if-function together with an explicit block
(\(c\) denotes a boolean condition and \(S\) a statement):
\begin{align}
\sembrack{w} &\equiv \sembrack{\text{\texttt{while} \(c\) \texttt{do}
    \(S\)}}\\
\sembrack{w} &= \sembrack{\text{\texttt{if} \(c\) then \(\{c; w\}\)
                                \text{else skip}}}\label{form:while_recursion}
\end{align}
Eqn.~\ref{form:while_recursion} is a recursive equation (\(c\) appears
both on the left hand and the right hand side) whose solution is a
fixed point. The agreement is that the least fixed point is taken to
be the denotational solution, and this is exactly what the fixed point
theorem delivers. To write this formally is now a standard exercise of
denotational semantics \cite{reynolds,mosses}, but we show it
nevertheless because it is an instructive example for the technique of
solving recursive equations which will be necessary for the denotation
of quantum communication. Consider the function \(F\) given by
\begin{equation}
F f\ (K,T,E) = \text{\texttt{if} \sembrack{c} \texttt{then}
                         \(f(\sembrack{S})\) \text{else} \((K,T,E)\)}.
\end{equation}
Then the fixed point solution can be formally written as
\begin{equation}
\stdsembrack{\text{\texttt{while} \(c\) \texttt{do} \(S\)}} = 
                                    Y_{\Sigma\rightarrow \Sigma_{\bot}} F,
\end{equation}
where \(\Sigma\) is any \((K,T,E)\) tuple as usual; assume that \(f\) 
is defined like
\begin{equation}
f(x) = \begin{cases}
  \bot & \text{if}\ x = \bot\\
  f(x) & \text{otherwise}
\end{cases}
\end{equation}
since we have to account for the case that the argument of \(f\) is
\(\bot\) because of the recursion (note that we would have to use two
different symbols for \(f\) to write this \"uberproperly).

\paragraph{Assignments}
Assignments in cQPL can be seen as a convenience mechanism which
extends the simple binding of identifiers to values; it is not
necessary to introduce the concept of stateful variables to the
language to be able to formalise assignments. The description can be
simplified if some syntactical transformations are applied. For this,
first consider the following program fragment:

\begin{verbatim} 
new int a := 10;
// Do something using a (part 1)
a := a + 4;
// Do something using a (part 2)
\end{verbatim}

\noindent The assignment is equivalent to introducing a new 
identifier \texttt{a'}:

\begin{verbatim} 
new int a := 10;
// Do something using a (part 1)
new int a' := a + 4;
// Do something using a' (and replace all a by a') (part 2)
\end{verbatim}

\noindent This strategy also works when blocks are taken into consideration:

\begin{verbatim} 
new int a := 10;
if (...) {
  new int a := 5;
  a := a + 1;
  ...
}
else { 
  new int a := 1;
  a := a + 1; 
  ...
}
\end{verbatim}

Note that not possible to employ a static renaming scheme in this case
because the identifiers \texttt{a} in both subsequent blocks would
then end up with identical names which leads to problems.  We thus
have to postulate that new identifiers are always chosen such that
they do not overlap with previous identifiers and must, of course,
also not overlap with identifiers which can be assigned by the user.
This is possible if the set of identifier strings is denoted by
\(\Sigma_{1}^{*}\), we introduce a second set of strings
\(\Sigma_{2}^{*}\) with \(\Sigma_{1} \cap \Sigma_{2} = \varnothing\);
each time an identifier is overshaded, it is replaced in its complete
scope with a new one in \(\Sigma_{2}^{*}\) that has \emph{not} been
used before. While this policy is hard to implement in
practice,\footnote{This is exactly the reason why mechanisms like
  call-by-name disappeared as a curiosity some thirty years ago.}  it
does not present any problem in theory (even cases which require an
infinite number of replacements are no problem because there are
infinitely many unique identifiers).

Most important, the approach is also valid when loops are introduced
because these can be rewritten using a (possibly infinite) chain of 
appropriate if-then constructions as is done in the denotation of them.

In conclusion, we do not need to take care for the obstacles introduced
by overshading, but can simply ignore the problem in the denotation
of assignments (\(\nu\) denotes some identifier):
\begin{equation}
  \sembrack{\text{\(\nu\) \texttt{:=} \(\epsilon\)}} = 
  (K,T,E \oplus \nu:\EQN\sembrack{\epsilon})
\end{equation}
where \(\epsilon\) is some arbitrary arithmetic expression (which
includes single identifiers); the denotation of this is obvious and
will not be considered further. Note that this valuation function
does \emph{not} cover the case that the result of a quantum variable
measurement is stored in a classical variable; this will be covered later
when we describe the denotation of the \texttt{measure} funtion
on Page~\pageref{form:measurements}.

\paragraph{Allocating and destroying variables}
As we have noted in the previous remarks, we do not need to take the
problem of overshading into account when dealing with assignments;
this obviously also applies to allocations. We refrain from a more
detailled description of this topic because everything necessary for
the solution can be readily found in the literature, \eg,
\cite{winskel, reynolds, mosses}. Note that allocating new variables
does not present any problems for the boundedness of the Kraus aggregation
because the number of qbits is limited by \(Q\), an arbitrary, but
finite quantity.

\paragraph{Procedure handling}
Procedures in cQPL follow the standard scheme of classical languages
for the non-quantum part. The denotation of such can therefore be
directly taken from the usual textbooks
\cite{reynolds,mosses,winskel}, so we will not repeat this here. The
interesting problem is given by recursive procedures, especially when
they act on quantum parameters. For simplicity, we consider directly
recursive procedures with quantum parameters; the case of indirect
recursion is in princple identical, but necessitates more formal
effort, so we skip it here. The solution is an adaption of the method
presented in~\cite[Sections 5.5 and 6.5]{selinger_qpl} for our
purposes.

\begin{figure}[htb]
  \centering\includegraphics[width=\textwidth-1cm]{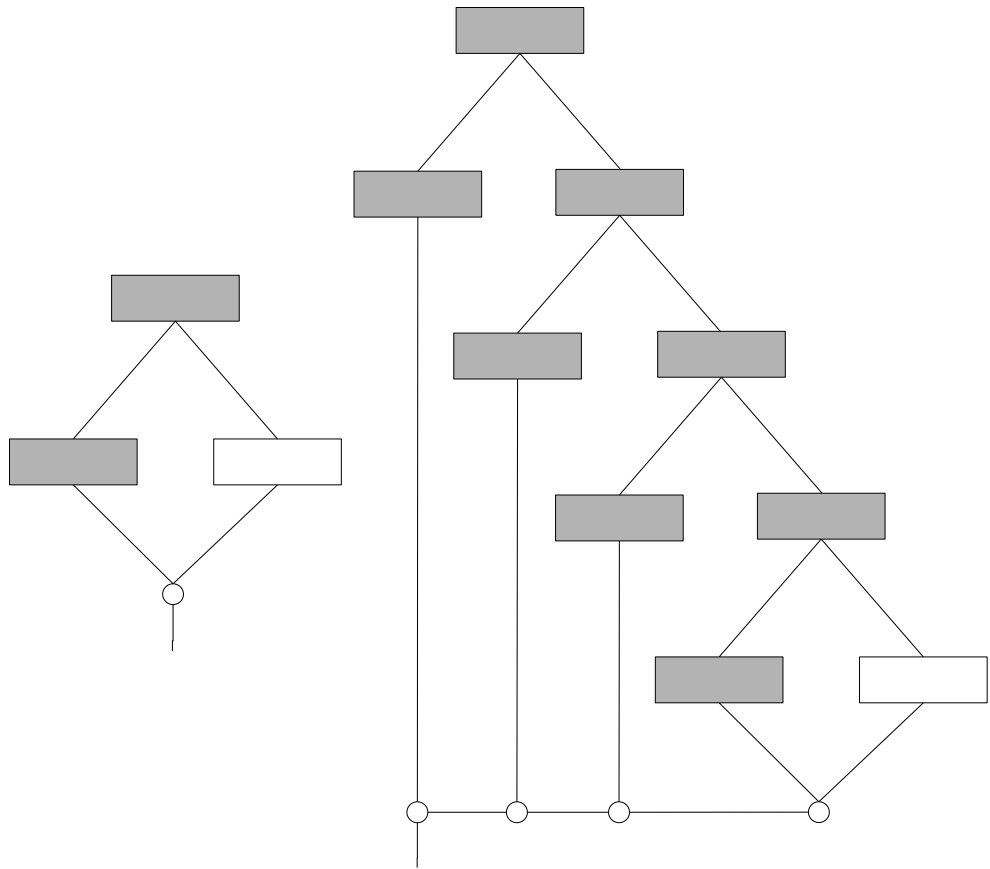}
  \caption{Unwinding a recursive flow chart is done by placing the flow
  chart with a hole (given by the white rectangle) into the hole and 
  repeating the process again and again.
  The limit of this sequence is given by a fixed point as explained
  in the text.}\label{form:rec_proc}
\end{figure}

Consider a procedure \(Y\) which depends on itself, \eg, 
\begin{equation}
Y = X(Y)
\end{equation}
If \(Y\) is given as a flow chart, this can be interpreted as shown
in Figure~\ref{form:rec_proc}: A ``hole'' in the representation of
\(X\) is replaced by \(Y\) with another hole, this is again replaced
by the same, \dots

\begin{verbatim}
proc rec: test:qbit {
  A
  if (cond) call rec(test');
  else { ... }
  B
}

\end{verbatim}

\noindent Formally, we can thus define an approximation relation given by
\begin{equation}
Y_{i+1} = X(Y_{i}).\label{form:rec_approx_def}
\end{equation}
where \(Y_{i}\) are approximations of \(Y\) which get better with increasing
\(i\). \(Y_{0} = \bot\) is the crudest approximation which simply does not
terminate. The solution of Equation~\ref{form:rec_approx_def} is another
case for the fixed-point theorem. Once a solution for \(Y\) has been found,
it can be applied to any \((K,T,E)\) tuple to calculate the required
state transformation. Note that although the procedure depends only
on quantum variables, there nevertheless needs to be a termination condition.
In the example, this condition is given by the if-then-else statement in
which the recursive call is wrapped; since the procedure only takes
a quantum parameter, the condition is either trivial or depends on a
measurement of quantum states. In the first case, the condition depends only
on classical variables which were allocated within the procedure, the condition
could thus be determined at compile-time and the recursion unrolled because 
the recursion depth is already known. In the second case, we really need
to unwind the procedure.

\subsubsection{Quantum operations}
The quantum mechanically relevant operations of cQPL are :

\begin{itemize}
  \item Application of (unitary) operators.
  \item Calling procedures which manipulate quantum data.
  \item Sending and receiving quantum states.
  \item Measurements.
  \item Creating new and destroying existing quantum states, where the
    last operation is not explicitely, but only implicitly possible
    when quantum variables drop out of the present frame. Sending quantum
    variables makes them disappear from the scope so that they can not be 
    accessed any more, but does not destroy respectively deallocate them.
\end{itemize}

We will give formal denotations for these constructs in the remaining part
of this section. Obviously, the description of communication is
our foremost concern, so we elaborate this in most detail.

\paragraph{Unitary operators}
Unitary operators induce isometries, so the corresponding Kraus set is
obviously bounded and filfills the requirement for fixed points.
Every unitary transformation can be expressed by a Kraus set with only
one element. The semantics of a unitary transformation acting on the
qbits \(q_{1}, \ldots, q_{k}\) is given by
\begin{equation}
  \EXP\sembrack{\text{(\(q_{1}\),...,\(q_{k}\)) \text{\texttt{*=}} U}} = 
      (K \circ \{U\}_{(\#q_{1}, \ldots, \#q_{k})},T,E).
\end{equation}
Note that the type system ensures that the variable/operator dimension on
both sides of the expression matches as required, \ie, the 
classical variable has the proper size to hold all potential measurement
outcomes. 

\paragraph{Measurements}\label{form:measurements}
Measurements are described by the following semantic equation:
\begin{equation}
  \EXP\sembrack{\text{\texttt{a := measure q}}} = 
      \left(K \circ \{M\}_{\#q}, T, E \oplus \mapvalmeas{a}{q}\right).
\end{equation}
Hereby, \(\PROJ(q,T)\) denotes the function to generate the set of
projectors for the type of \(q\) which can be resolved from the typing
context \(T\). The required basis for the projection (which is
actually nothing else than the basis for
\(\mathbbm{F}_{2}^{t_{q}(\chi(q))}\) in Dirac notation, also named the
standard basis) is given by
\begin{equation}
  \mathcal{B} = \left\{\bigotimes_{n=0}^{t_{q}(\chi(q))} \ket{i_{n}}\ 
         \forall i_{n}\in\{0,1\}\right\}.
\end{equation}
The corresponding Kraus elements are obvious. Accordingly, 
\(\pi_{a}:\PROJ(q,T)\) denotes the probability distribution which connects
the possible values of \(a\) with a probability distribution induced by
the projectors. Since the measurement operators work on discrete states in
a finite-dimensional Hilbert space, they are obviously bounded.

\begin{remark}
  Note that although we restrict measurements to projections onto the
  standard basis, projective measurements in arbitrary bases can be
  realised by applying appropriate unitary transformations prior to
  the measurement.  
\end{remark}

\paragraph{Sending and receiving qbits}\label{form:send_receive}
To consider sending quantum variables, we first split commands which
send lists of quantum variables into a list of commands that send one
quantum variable each:
\begin{align}
  &\EXP\sembrack{\text{\texttt{send} \(q_{1}, \ldots, q_{n}\) 
                       \texttt{to} \textit{module}}} = \nonumber\\
  &\EXP\sembrack{\text{\texttt{send} \(q_{1}\) \texttt{to} \textit{module};
  \texttt{send} \(q_{2}\) \texttt{to} \textit{module}; 
      \ldots; \texttt{send} \(q_{n}\) \texttt{to} \textit{module}}}.
\end{align}
The denotation of a single send command is given by
\begin{align}
  &\EXP\sembrack{\text{\texttt{send} \(q\) \texttt{to} \textit{module}}} = 
  \big(K\circ \{S\}_{\#q},
    T \ominus q,
    E \ominus q\big).
\end{align}
\(\{S\}_{\#q}\) is not a real physical map as other cp-maps are, but
only a ``placeholder'' to note that qbits have been sent. This will
become important when the semantics of parallel execution is
considered further below. The interesting thing here is that the
sent qbit is removed from the typing context and from the 
probabilistic environment. Thus, it is not visible any more in
the remaining statements. Further access to it can be detected 
as erroneous at compile time; the typing context gives the formal
basis for this.

Receiving qbits is described in a similar manner. First, a
receive operation with multiple quantum variables is split into
a sequence of single-variable receive operations:
\begin{align}   
  &\EXP\sembrack{\text{\texttt{receive} \(q_{1}:\qtype_{1}, \ldots,
               q_{n}:\qtype_{n}\) \texttt{from} \textit{module}}} =\\
  &\EXP\sembrack{\text{\texttt{receive} \(q_{1}:\qtype_{1}\) \texttt{from}
               \textit{module}; \ldots; {receive} \(q_{n}:\qtype_{n}\) 
                 \texttt{from} \textit{module}}}.\nonumber
\end{align}
The denotation of a single-variable receive command is given by:
\begin{align}
  &\EXP\sembrack{\text{\texttt{receive} \(q:\qtype\) \texttt{from} 
      \textit{module}}} =\\
  &\big(K \circ \{R\}_{\#q},
   T \oplus q:\qtype,
   E \oplus q:\textit{module}\big).\nonumber
\end{align}
Again, \(\{R\}_{\#q}\) is a placeholder required to denote the semantics
of parallel composition.

\paragraph{Parallel composition}\label{form:comb_proc}
The compositionality principle of denotational semantics implies that
the formal denotation of sending and receiving qbits must be
independent of the conjugate action, \ie, sending must be independent
from reception and reception must be independent from sending. This is
fulfilled in the formalism of cQPL as we have shown on page
\pageref{form:send_receive}.  Nevertheless, the denotation of the
communication as a whole requires (at last) two communicating
partners. It needs thus make use of both of them to assign semantics
to communication.

Before we start with the formal details, we want to motivate why
parallel composition is necessary at all. For this, consider the case
of two processes where \(A\) sends a quantum bit (which we call
\(a_{1}\)) to \(B\) and, later on, receives a quantum bit from \(B\) (which
we call \(a_{2}\)). We must distinguish two different 
cases (note that for more difficult cases with an arbitrary number of
send and receive statements, the conditions become more complicated
because we need to account for more general schemes of mutual influence.
These conditions will be developed stepwise in the following):

\begin{itemize}
  \item The returned quantum bit was not identical with the sent
    one, \(a_{1} \ne a_{2}\). The inequality refers to the positions
    occupied by the qbits on the combined quantum heap, it is
    \emph{not} related with the names of the qbits.
  \item \(B\) returned the quantum bit it got from \(A\), \(a_{1} = a_{2}\).
\end{itemize}

Consider the consequences for the semantics of parallel composition
when the interaction of both processes is considered (both cases can
be treated identically if only the separated semantics is taken into
account): While in the first case (\(a_{1} \ne a_{2}\)), operations on
\(a_{2}\) and \(a_{1}\) end up on different physical locations, the
same operations must in the second case (\(a_{1} = a_{2}\)) be applied
to the \emph{same} physical location.  Therefore, we must be able to
construct enough information from the composed systems such that it is
possible to distinguish between both cases. 

\begin{figure}[htb]
  \centering\includegraphics{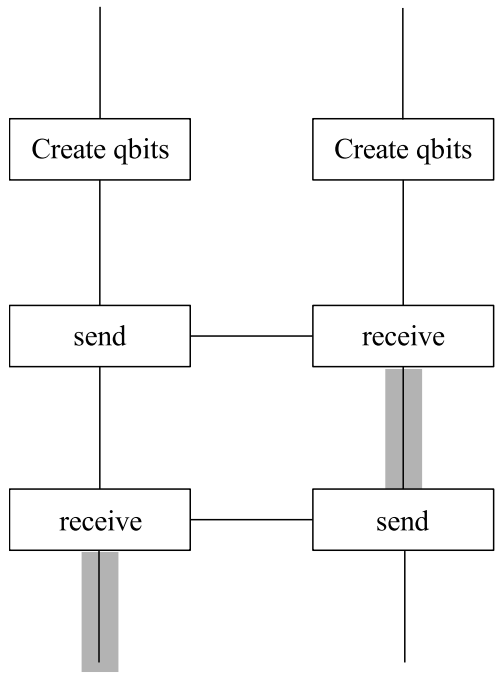}
  \caption{Flow graph for two communicating processes. \(A\) sends a qbit
    to \(B\) and \(B\) sends one to \(A\); if there are no operations
    on the same physical locations, the shaded regions commute.  This
    is obviously not the case if one or more identical quantum heap
    positions are modified in both paths: Since the shaded regions
    operate on the same physical qbit then, their actions need not
    necessarily be interchangeable because non-commuting
    operations may have been performed on the same physical location.}%
  \label{form:send_recv_comm_example}
\end{figure}

Additionally, the distinctness of qbits influences denotational
equivalence; many different orderings of the actions performed by
\(A\) and \(B\) lead to the same semantics (the exact conditions for
this will be formulated later on), but the class of possible
reorderings is usually bigger if there was no interaction on the same
physical location in communicating modules.
Figure~\ref{form:send_recv_comm_example} presents a visualisation of
this fact using a pseudo flow diagram.

There is no explicit statement in the syntax of cQPL which describes
the combination of two processes. As explained in Chapter~\ref{qprog},
communication is realised implicitly using \emph{modules} which interact among
each other. These modules must thus be given some semantical meaning;
this will be developed in the following. Note that we restrict the
description to two communicating processes which we call A and B
(this might stand as an abbreviation for the omnipresent Alice and
Bob) at first. Consider the following cQPL fragment:

\begin{verbatim}
module A {
...
};

module B {
...
};
\end{verbatim}

Any valid cQPL code (except the definition of new modules) may be contained 
in the bodies of \texttt{module A} and \texttt{module B}; both entities thus 
constitute regular cQPL programs whose meaning is made up by the meanings of 
all statements they consist of. Thus, we can write \(\PROG\stdsembrack{A}\) 
and \(\PROG\stdsembrack{B}\) for the denotation of the code as if the modules 
were regular, uncommunicating cQPL programs.

To consider the communicative interactions between both, we introduce
the operation
\begin{equation}
\PROG\stdsembrack{A || B} \equiv \COMM(\PROG\stdsembrack{A}, 
                       \PROG\stdsembrack{B})\kteinit\label{form:comm_simple}
\end{equation}
where \(\kteinit\) denotes the inital \((K,T,E)\) tuple without
content.  The valuation function \(\COMM\) is used to compute the
combined denotation of \(A\) and \(B\) which we also call
\emph{parallel execution}. It must obviously only depend on the
denotations \(\stdsembrack{A}\) and \(\stdsembrack{B}\), \ie,
\((K,T,E)\) tuples.  Since we do not want the semantics of
communication to depend on the order in which the communicating
partners are specified, the operator \(||\) must necessarily be
commutative: \(\PROG\stdsembrack{A || B} \stackrel{!}{=}
\PROG\stdsembrack{B || A}\).  This equivalence can easily be achieved:
It suffices to define the operator in such a way that the modules are
ordered lexicographically, then the order in which they are specified
does not influence the denotation; the commutation relation is thus
automatically fulfilled.

\(\PROG\stdsembrack{A}\) is used to evaluate a semantic equation with
empty initial context; this is obviously the case when the top-level
of a program is considered (as is the case for modules) where no definitions
can have been made yet. Thus, \(\PROG\stdsembrack{A} \equiv
\EXP\stdsembrack{A}\kteinit\).

Consider the evaluation of \(\stdsembrack{A}\) and \(\stdsembrack{B}\) which
both do as usual depend on \((K,T,E)\) tuples. \(\COMM\) results in the
creation of a valuation function which depends on the tensor product of
these tuples, \ie,
\begin{equation}
  \left.\begin{matrix}
    \VAL\stdsembrack{A}(K_{0}, T_{0}, E_{0})\\
    \VAL\stdsembrack{B}(K_{1}, T_{1}, E_{1})
  \end{matrix}\right\}\Rightarrow 
  \COMM(\VAL\stdsembrack{A}, \VAL\stdsembrack{B})\underbrace{(K,T,E)}_%
     {\makebox[1cm]{\small{\hss\((K_{0}\otimes K_{1}, T_{0}\otimes T_{1}, 
           E_{0}\otimes E_{1})\)\hss}}}.
\end{equation}
Both representations convey the same information. This is immediately
obvious if the expressions are written as direct \(\lambda\)-abstractions
as defined in Section~\ref{form:notat_remarks} together with an appropriate 
``untensoring'' function which separates the tensor product of the combined 
\((K,T,E)\) tuple into two components. We will not show this explicitely to 
avoid introducing even more symbols.

Note that termination of \(A\) and \(B\) alone does not imply
termination of \(\PROG\stdsembrack{A || B}\).\footnote{Here, the
  question arises how termination should be defined for communicating
  processes if only one part (A) of a total system is considered. We
  could, for instance, define a demonic partner which always supplies
  the required number of qbits the process needs and absorbs any
  number of qbits sent, but this will not necessary lead to
  termination of A. The exact solution of the problem depends on the
  context in which it is considered; we will not deal with it any
  further here, but define a process as terminating if there exists a
  demonic partner that behaves in such a way that the process
  terminates.}  
This can be seen by considering the following simple program:

\begin{verbatim}
module A {
  new qbit q1;
  send q1 to B;
};
\end{verbatim}

\begin{verbatim} 
module B {
  receive a1:qbit from A;
  receive a2:qbit from A;
}; 
\end{verbatim} 

While both processes represent valid cQPL programs, they will
obviously result in some non-terminating program (which should thus
denote \(\bot\)) because \(B\) will wait forever for the second qbit
(\(a_{2}\)) to be sent. Thus, \(\COMM(x,y) = \bot\) is possible
although both \(x\neq\bot\) and \(y\neq\bot\).

Extensions of the semantical components \((K,T,E)\) to the multi-party
case were defined in Section~\ref{form:multi_party_components}; they  
provide the basis for the denotation of parallel execution which will be 
developed stepwise by considering bipartite systems without communication, 
bipartite systems with single send/receive pairs, bipartite systems with
arbitrary send/receive statements and finally, arbitrary multipartite
systems. 

Thus, let \texttt{A} and \texttt{B} be two programs which do not
use any communication. The semantics of their parallel execution
can be readily denoted:
\begin{gather}
  \PROG\stdsembrack{A} = (K_{0},T_{0},E_{0}), 
         \PROG\stdsembrack{B} = (K_{1},T_{1},E_{1})\nonumber\\
  \Rightarrow\label{form:simple_composition}\\ \PROG\stdsembrack{A || B} = 
  (K_{0}\otimes K_{1}, T_{0}\otimes T_{1}, E_{0}\otimes E_{1}).\nonumber
\end{gather}
Note that the following equivalences hold if \(A_{i}, B_{i}\) are parts
of a program which do not contain any send/receive operations (we omit
the required \(\EXP\)s in the second line to simplify the notation):
\begin{align}
  \underbrace{\EXP\stdsembrack{A_{1} || B_{1};\ A_{2}||B_{2}}}_{} &= 
  \underbrace{\EXP\stdsembrack{A_{1}; A_{2} || B_{1}; B_{2}}}_{}\\
  \left(\COMM\left(\stdsembrack{A_{1}}, \stdsembrack{B_{1}}\right); 
   \COMM\left(\stdsembrack{A_{2}}, \stdsembrack{B_{2}}\right)\right) &= 
  \COMM(\stdsembrack{A_{1}; A_{2}}, \stdsembrack{B_{1}; B_{2}}).%
                                                   \label{form:comm_equiv}
\end{align}
This ensures that it does not make any
difference if we consider the sequential combination of two parallel
executions or the parallel execution of two sequential combinations as
shown in the formula; we will make use of this later on.

The situation gets more complicated if a single send/receive pair
is allowed, \ie, if a single quantum variable can be transferred from
A to B (the case B to A is nearly identical, so we restrict ourselfs
to the first case). Let \(\{S\}\) and \(\{R\}\) denote the Kraus 
pseudo-operations for sending and receiving. To see how these operations
influence the possible equivalent compositions of a program, observe the 
following two schematic Kraus aggregations:
\begin{align}
  \Lambda_{\text{A}} &= \{A_{1}\}, \ldots, \{A_{n}\}, \{S\}, 
                        \{A_{n+1}\}, \ldots, \{A_{m}\}\\
  \Lambda_{\text{B}} &= \{B_{1}\}, \ldots, \{B_{n'}\}, \{R\}, 
                        \{B_{n'+1}\}, \ldots, \{B_{m'}\}.
\end{align}
To consider how these aggregations can be rearranged, we define the
following function:
%
\begin{equation}
  \mathfrak{D}(A, B) = 
    \begin{cases}
      1 & \text{if \(A\) and \(B\) are operations on disjoint qbits}\\
      0 & \text{otherwise}
    \end{cases}.
\end{equation}

\begin{remark}
  The same effect could have been achieved with the standard
  commutator in principle, but we want the notation to emphasise
  additionally that not the simultaneous diagonalisability, but the
  fact that the operations work on disjoint bases is responsible for
  the exchangeability.  Additionally, using the standard commutator to
  show equivalence between permutations of communicating systems would
  have interferred with the standard permutation rules that would then
  have become more complicated.
\end{remark}

Since the two processes are necessarily totally uncorrelated before the 
transmission takes place, the parts given by the sub-aggregations
\begin{align}
  \Lambda_{\text{A}}^{1} &= \{A_{1}\}, \ldots, \{A_{n}\}%
                                       \label{form:simple_dis_start}\\
  \Lambda_{\text{B}}^{1} &= \{B_{1}\}, \ldots, \{B_{n'}\}
\end{align}
can be composed as in Eqn.~\ref{form:simple_composition} because
\(\mathfrak{D}(\Lambda_{\text{A}}^{1},\Lambda_{\text{B}}^{1}) = 1\).
Since the parts after sending respectively receiving the quantum
variable are uncorrelated as well (they both work on disjoint sets of
qbits; this property remains valid in the combined semantics because
the combination of the probabilistic environments creates an
appropriate combined quantum heap as described in
Section~\ref{form:multi_party_components}), the sub-aggregations
\begin{align}
  \Lambda_{\text{A}}^{2} &= \{A_{n+1}\}, \ldots, \{A_{m}\}\\
  \Lambda_{\text{B}}^{2} &= \{B_{n'+1}\}, \ldots, \{B_{m'}\}%
                                            \label{form:simple_dis_end}
\end{align}
can likewise be parallel composed as in Eqn.~\ref{form:simple_composition}.
The only thing which needs to be taken into account is that references
to the position of the received qbit must be replaced by the 
position of the qbit on the combined quantum heap in the combined 
probabilistic environment. This can formally be achieved by the following
function (here, \(S\) denotes a send and \(R\) a receive statement):
\begin{equation}\label{form:sr_simple_def}
  \SR(\COMM(\stdsembrack{S}, \stdsembrack{R}))(K^{\otimes},
  T^{\otimes}, E^{\otimes}) = (K^{\otimes}, T^{\otimes}, {E^{\otimes}}').
\end{equation}
Here, \(K^{\otimes}, T^{\otimes}, E^{\otimes}\) denotes the usual
\((K,T,E)\) parameter tuple with the additional requirement that it
must have the structure which is gained by combining two \((K,T,E)\)
tuples with a tensor product as given in
Section~\ref{form:multi_party_components}.  \(\SR\) itself is
responsible for two things: On the one hand, it applies the effect of
\(\COMM(\stdsembrack{S}, \stdsembrack{R})\) to the parameter tuple,
and on the other hand, it replaces the portion of \(E\) which contains
the information about the received quantum variable so that it now
points to the position of the \emph{sent} quantum variable on the
combined quantum heap; the resulting probabilistic environment is
denoted by \(E'\). This is obviously possible since \(\SR\) does have
access to the information provided both by send and receive.  To
illustrate the effect of \(\SR\), consider the following example: Let
the sent quantum variable be denoted by \texttt{q} and the received
one by \texttt{r}. The names of these variables will have thus been
changed to \(\LC(1)\text{\texttt{q}}\) and
\(\LC(2)\text{\texttt{r}}\). The receiving module does not have any
information about the received quantum variable except its type and
its local name; the position on the combined quantum heap is unknown.
This can now be changed by \(\SR\); for that, it inserts the
position of \(\LC(1)\text{\texttt{q}}\) on the combined quantum heap
into the combined probabilistic environment such that
\(\LC(2)\text{\texttt{r}}\) points to it. Nothing more is necessary
to identify the received quantum variable with the sent one.

Note that \(\RS\) defines the analogous function for the receive/send
case. This is necessary when bidirectional communication between processes
is considered. Except the inverted direction of data flow, the function
is completely equivalent to \(\SR\).

Thus, the semantics of parallel execution for processes with a 
single send/receive operation can be reformulated as 
\begin{equation}
  \PROG\stdsembrack{A || B} = \PROG\stdsembrack{A_{1} || B_{1};\ S || R;\ 
    A_{2} || B_{2}}\label{form:single_send_denot}
\end{equation}
where \(\stdsembrack{A_{i}}\) and \(\stdsembrack{B_{i}}\) denote the
parts of the program which induce the operations described by
\(\Lambda^{i}_{\text{A},\text{B}}\) as given by
Eqns.~\ref{form:simple_dis_start}--\ref{form:simple_dis_end}.  It is
already known how to compute the semantics of
\(\stdsembrack{A_{1}||B_{1}}\). To compute the denotation of the
complete statement, we first consider how to include the send/receive
pair into the description:
\begin{align}
\PROG\stdsembrack{A_{1}||B_{1};\ S||R} = 
    \SR&(\COMM(\stdsembrack{S}, \stdsembrack{R}))\nonumber\\
      &(\COMM(\PROG\stdsembrack{A_{1}},\PROG\stdsembrack{B_{1}})
         \kteinittensor).\label{form:simple_comm_part1}
\end{align}
Note that although \(\PROG\) with argument \(\stdsembrack{A_{1}},
\stdsembrack{B_{1}}\) appears both on the left and right hand side of
this equation, it is \emph{not} truly recursive, but can here be seen
as just a breakdown into simpler cases. A general recursive formula
will be derived in the following.

By using Eqn.~\ref{form:single_send_denot}, we can now give the denotation 
of the complete parallel composition as defined in Eqn.~\ref{form:comm_equiv};
for this, we define the righthand side of Eqn.~\ref{form:simple_comm_part1} 
to be denoted by \(\xi\) to increase clarity:
\begin{equation}
\PROG\stdsembrack{A_{1}||B_{1};\ S||R;\ A_{2}||B_{2}} = 
  \COMM(\PROG\stdsembrack{A_{2}}, \PROG\stdsembrack{B_{2}})(\xi).%
                                          \label{form:single_send_solution}
\end{equation}
Note that establishing the semantics of a given description is one thing
we need to do; finding equivalent descriptions for a given communication
is another task. For this, we need to consider all rearrangements that 
preserve semantics. This allows us to decide if two communicating programs 
are identical because we can check if they can be brought to the same form.

For the case of a single send/receive pair, the operations can be shifted in 
the Kraus aggregation if certain conditions hold:

\begin{itemize}
  \item The send operation can be postponed to the end of
    the aggregation (at least in the case where no more send/receive operations
    take place) or brought forward by \(k\) positions in the Kraus 
    aggregation if \(\mathfrak{D}(\{A_{n-l}\}, S) = 1 
      \ \forall l=0,\ldots,k-1\).
  \item The receive operation can be brought forward to the first
    position of the Kraus aggregation (again, this relies on the fact
    that only a single send/receive operations takes place) or
    postponed by \(k\) positions if \(\mathfrak{D}(\{B_{n+1+l}\}, R) = 1 
      \ \forall l=0,\ldots,k-1\).
\end{itemize}

Performing shifts characterised by these operations together with
reorderings as given by Eqn.~\ref{form:kraus_two_comm_equiv} generates the 
equivalence class of all programs with a single send/receive pair that
posses the same semantics. 

The next step is to include arbitrary send/receive operations into
the communication between \(A\) and \(B\). 
As in the case of a single send/receive combination, the send/receive 
statements act as synchronisation points; the denotation needs thus be 
aligned along them. The problem to solve is now given by
\begin{equation}
  \PROG\stdsembrack{A_{1}||B_{1};\ S_{1}||R_{1};\ A_{2}||B_{2};\ 
                    S_{3}||R_{3};\ \cdots;\ S_{n-1}||R_{n-1};\ A_{n}||B_{n}}.%
                                                     \label{form:multi_send}
\end{equation}
First, we will establish the semantics for the given ordering; 
semantics-preserving permutations will be considered afterwards.

Note that we only consider the denotation of data flow in one
direction for simplicity; the semantics of the case \(R||S\) which may
appear mixed with the other form is gained by replacing \(\SR\) with
\(\RS\) at the appropriate places). To find a solution for this
equation, define
\begin{equation}
A'||B' \equiv A_{2}||B_{2};\ S_{n}||R_{n};\ \cdots;\ 
S_{n-1}||R_{n-1};\ A_{n}||B_{n}.
\end{equation}
Eqn.~\ref{form:multi_send} then has the form
\begin{equation}
  \PROG\stdsembrack{A_{1}||B_{1};\ S_{1}||R_{1};\ A'||B'}.\label{form:temp1}
\end{equation}
According to Eqn.~\ref{form:single_send_solution}, the solution of
Eqn.~\ref{form:temp1} is given by
\begin{align}
  \EXP\stdsembrack{A' || B'}
     &(\SR(\COMM(\stdsembrack{S_{1}}, \stdsembrack{R_{1}}))\nonumber\\
     & (\COMM(\PROG\stdsembrack{A_{1}}, \PROG\stdsembrack{B_{1}}))
         \kteinittensor).
\end{align}
By introducing \(\xi_{1}\) as abbreviation for the part following 
\(\EXP\stdsembrack{A'||B'}\) and expanding \(A'||B'\) to
\(A_{2}||B_{2};\ S_{2}||R_{2};\ A''||B''\), the formula reads as
\begin{equation}
\EXP\stdsembrack{A_{2}||B_{2};\ S_{2}||R_{2};\ A''||B''}(\xi_{1}).
\end{equation}
This type of equation is already well-known; it can be further resolved to
\begin{align}
\EXP\stdsembrack{A''||B''}
     &(\SR(\COMM(\stdsembrack{S_{2}}, \stdsembrack{R_{2}}))\nonumber\\
     & (\COMM(\PROG\stdsembrack{A_{2}}, \PROG\stdsembrack{B_{2}}))
         (\xi_{1})).
\end{align}
By recursively defining
\begin{align}
\xi_{0} &= (K_{\bot}^{\otimes}, T_{\bot}^{\otimes}, E_{\bot}^{\otimes})\\
\xi_{i} &= (\SR(\COMM(\stdsembrack{S_{i}}, \stdsembrack{R_{i}}))
     (\COMM(\PROG\stdsembrack{A_{i}}, \PROG\stdsembrack{B_{i}}))
         (\xi_{i-1}))
\end{align}
we see that the final solution of Eqn.~\ref{form:multi_send} is given by
\begin{equation}
  \EXP\stdsembrack{A_{n}||B_{n}}(\xi_{n})
\end{equation}
where \(\xi_{n}\) needs to be expanded as defined above.

The whole process thus leads to a recursive valuation function given by
\begin{align}
&\EXP\stdsembrack{A||B;\ S||R;\ C||D}(K,T,E) = \nonumber\\
        &\EXP\stdsembrack{C||D}(\SR(\COMM(\stdsembrack{S}, \stdsembrack{R}))
             (\COMM(\EXP\stdsembrack{A}, \EXP\stdsembrack{B}))(K,T,E)%
                       \label{form:two_party_rec_sol}
\end{align}
whose solution can be found by resolving the recursion in the usual way.

Now, consider which alternative orderings of the Kraus aggregations
preserve semantics in the multi send/receive and receive/send case.
For this, observe the following symbolic representation of two Kraus 
aggregations (to save some notational effort and to increase lucidity, 
we represent the Kraus sets which are not concerned with communication by 
boxes. Although the boxes have identical widths, they do not need to contain
the same number of Kraus sets):
\begin{align}
&\boxvar{A_{1}}\ S_{1}^{\text{A}}\ \boxvar{A_{2}}\ R_{1}^{\text{A}}\ 
                                                             \boxvar{A_{3}}\\
&\boxvar{B_{1}}\ R_{1}^{\text{B}}\ \boxvar{B_{2}}\ S_{1}^{\text{B}}\ 
                                                             \boxvar{B_{3}}.
\end{align}

As in the case of a single send/receive pair, we know that the 
blocks \(A_{1}\) and \(B_{1}\) (considering again the compatible 
displacements of \(S_{1}^{\text{A}}\) and \(R_{1}^{\text{B}}\)) can 
be arbitrarily combined because they work on disjoint subsets of the 
quantum heap; the same holds for \(A_{2}, B_{2}\) and \(A_{3}, B_{3}\).
\emph{In addition} to the previously given rules, the following 
restrictions hold for shifting send and receive operations in multi 
send/receive scenarios:

\begin{itemize}
  \item Two consecutive send statements can only be interchanged if the
    corresponding receive statements are interchanged, and vice versa.
  \item If \(\mathfrak{D}(A_{i}, B_{i+1}) = \mathfrak{D}(B_{i},
    A_{i+1}) = 1\) the blocks \(A_{i}, A_{i+1}\) and \(B_{i},
    B_{i+1}\) can be taken like single blocks when possible orderings
    according to Eqn.~\ref{form:kraus_two_comm_equiv} are considered.
    Note that the validity of this condition 
    can only be detected if the combined quantum heap is considered. If
    it holds, then the sender does nothing to the sent and the receive
    does nothing to the received quantum bit; thus, the statements contained
    in the blocks can be executed in arbitrary order.
\end{itemize}

\noindent Two systems are identical if their denotations can be unified by
performing rearrangements according to these rules.

Another possibility that needs to be considered is the case where the
number of send/\hbox to 0mm{}re\-cei\-ve pairs in the parallel processes
does not match. As we have described before, this leads to a
non-terminating process because either the sending process wants to
ship a quantum variable, but cannot deliver it and thus blocks or the
receiver wants to get a variable, but blocks indefinitely because
there is no sender for one. The semantics should thus be given by
\(\bot\) for both cases.

For the case of two parties (where we do not need to consider the
case that a balanced number of send/receive statements is present, but
the distribution among the parties is unmatched), this is covered by
the following definition:
\begin{equation}
\EXP\stdsembrack{A||B} = \bot\ \text{if}\ \mathfrak{R}(\stdsembrack{A}) \neq 
                          \mathfrak{S}(\stdsembrack{A}) 
             \vee \mathfrak{S}(\stdsembrack{A}) \neq 
                  \mathfrak{R}(\stdsembrack{B})
\end{equation}
where \(\mathfrak{S}\) denotes the number of send and \(\mathfrak{R}\) 
the number of receive statements. The denotations \(\stdsembrack{A}\)
and \(\stdsembrack{B}\) are supposed to be those derived for the 
parallel composition. The extension to higher-dimensional systems is
obvious, so we omit it here.

Finally, we can describe the semantics for multi-party communication
with arbitrary send/receive statements which is the most general case
and therefore includes everything considered before. The problem 
which needs to be evaluated is given by
\begin{equation}
\PROG\stdsembrack{A_{1} || A_{2} || \cdots || A_{n}},
\end{equation}
where (note the different notation compared to before!) 
\(A_{i}\) represents all statements given in module \(i\); this may
contain any number of send/receive statements that are now denoted by
\(S_{k}^{i}\) and \(R_{k}^{i}\) where \(i\) is the receiver for \(S\) 
and destination for \(R\) and \(k\) the sequence number within the other 
send/receive statements of the communication channel the statement works 
in (if we consider for example three parties \(A_{1}\),\(A_{2}\) and 
\(A_{3}\), then there are the channels \(A_{1}\)--\(A_{2}\), 
\(A_{1}\)--\(A_{3}\), \(A_{2}\)--\(A_{3}\)).

The semantic context the evaluation is based on is given by the
tensor product of the semantic contexts of the subsystems, \ie,
\((K^{\otimes}, T^{\otimes}, E^{\otimes}) = (K_{1}\otimes\cdots\otimes K_{n},
T_{1}\otimes\cdots\otimes T_{n},E_{1}\otimes\cdots\otimes E_{n})\) and
equivalent for the initial context. The valuation function 
\(\COMM\) given by Eqn.~\ref{form:comm_simple} can be extended from the
two-party case to the \(n\)-party case without any problems, we
denote this by \(\COMM^{\otimes n}\). Since 
communication still takes place between two partners (although there are
now many choices for such two-partner subsystems), there is always a
pair of corresponding send/receive respectively receive/send statements. To 
take the \(n\)-dimensional semantical context into account, the definition
of \(\SR\) (the version for two parties is given in 
Eqn.~\ref{form:sr_simple_def}) needs to be adjusted as follows when
sending a quantum variable from system \(m\) to system \(m'\) is to
be covered:
\begin{equation}
  \SR^{m,m'}(\COMM^{\otimes n}(\stdsembrack{S^{m'}}, 
          \stdsembrack{R^{m}}))(K^{\otimes},
  T^{\otimes}, E^{\otimes}) = (K^{\otimes}, T^{\otimes}, {E^{\otimes}}').
\end{equation}

\(\SR\) is again responsible to apply the effect of 
\(\COMM^{\otimes n}(\stdsembrack{S}, \stdsembrack{R})\) to the parameter 
tuple; note that in this case, do-nothing-operations \(\mathbbm{1}\) are 
used for all systems except \(m\) and \(m'\) because these are not concerned
with the communication. This is to ensure that the dimensionality matches.

Additionally, \(\SR\) replaces the portion of \(E\) which contains
the information about the received quantum variable so that it now points
to the position of the \emph{sent} quantum variable on the combined quantum
heap. This is identical to the effect in the simplified case for two systems.
If in this case the sent quantum variable is denoted by \texttt{q}
in system \(m\) and the received one by \texttt{r} in system \(m'\), then
the names of these variables will have been changed to 
\(\LC(m)\text{\texttt{q}}\) and  \(\LC(m')\text{\texttt{r}}\). 
\(\SR\) simply inserts the position of \(\LC(m)\text{\texttt{q}}\) on 
the combined quantum heap into the combined probabilistic environment such 
that \(\LC(m')\text{\texttt{r}}\) points to it. 

With this, we can generalise the recursive definition of 
Eqn.~\ref{form:two_party_rec_sol} to the case with an arbitrary number
of participants:
\begin{align}
\EXP&\stdsembrack{A_{1}||\cdots||A_{n};\ S^{m}||R^{m'};\ 
                  B_{1}||\cdots||B_{n}}(K,T,E) = \nonumber\\
    &\EXP\stdsembrack{B_{1}||\cdots ||B_{n}}(\SR^{m,m'}(\COMM^{\otimes n}
        (\stdsembrack{S^{m}}, \stdsembrack{R^{m'}}))%
                           \label{form:many_party_rec_sol}\\
     &(\COMM(\EXP\stdsembrack{A_{1}}, \ldots, \EXP\stdsembrack{A_{n}}))%
                                                             (K,T,E).\nonumber
\end{align}

Finally, this is the solution to the most general case of communication
which can be expressed in cQPL.


\subsection{Explicit transformations of density matrices}
Although the abstract view on quantum operations which we have
presented in this work is quite suitable for reasoning about general
formal properties of quantum systems, the demands of practical work
are usually of a different nature: Here, one is interested in the
calculation of explicit states and probabilities which determine a
system and allow predictions about its past, present and future
behaviour. This goal is usually achieved by specifying the initial
state of the system, subjecting this to diverse transformations and
measuring the required properties which give rise to the desired
explicit probability distributions.

The semantics of a cQPL program can be used to generate exactly this
information: The abstract transformation given by the semantical
denotation of a program is a function which maps the density matrix of
the input state to the density matrix of the output state. Obviously,
the election of a certain density matrix as initial state implies loss
of generality, but in turn allows to infer real-world information, not
just abstract properties of generalised systems.


\subsection{The type system}\label{form:type_system}
\emph{Typing judgements} make statements about the connection
between expressions and their types; cf., \eg,\cite{cardelli} for an 
introduction. For our purposes, the following two building blocks are
necessary to describe the properties of cQPL:
\begin{align}
E \vdash e:T &\Leftrightarrow\ \text{Expression \(e\) has type \(T\) in E}\\
E \vdash F &\Leftrightarrow\ \text{\(F\) is well-typed in \(E\)}.
\end{align}
Proper typing is necessary to eliminate certain runtime errors by
applying appropriate compile time checks (cf.
Section~\ref{sem:runtime_errors}).  Additionally, it is the key to
showing that our formalism ensures that quantum bits can -- especially
in communicating systems -- be only manipulated by one party at a time
(a similar line of reasoning, albeit for a quite different formalism,
was used in \cite{gay_nagarajan}). The typing context provided
by \(T\) in the \((K,T,E)\) tuple is the basis for this.

Properties of the type system are customary expressed with 
judgements of the following general form:
\begin{equation}
\inferdots{P_{1}}{P_{n}}{C}
\end{equation}
where the \(P_{i}\) are called the \emph{premises} and \(C\) the
\emph{conclusion}. If all premises are true, the conclusion is
fulfilled. Such judgements can be used to deduce the type of a given
composite expression in an automated, formal manner.
The following elementary typing judgements hold for
cQPL:\footnote{Remember: \(c(x) = 1\) ensures that the data type of \(x\) is
  purely classical. }
\newlength\dummylength        
\setlength{\dummylength}{1mm} 
\begin{align}
\text{Scalars}\quad & 
          \infersimple{i \in \mathbbm{N}}{E \vdash i:\cint}\ 
                       \text{(analogous for bits, floats etc.)}\\[\dummylength]
\text{\code{new} t n \code{:= v}}\quad & \infersimple{E\vdash v:t}
                                         {E \vdash n:t}
\end{align}
\begin{align}
C_{1}; C_{2}\quad & \infer{E \vdash C_{1}}{E \vdash C_{2}}
                          {E \vdash (C_{1};C_{2}):\void}\ 
                  \text{(Composition preserves well-typedness)}\\[\dummylength]
                  \text{Conditionals}\quad & \infer{E \vdash
                    v_{1}:t_{1}\wedge E \vdash v_{2}:t_{2}} {c(t_{1})
                    = c(t_{2}) = 1}
                  {E \vdash \text{op}(v_{1}, v_{2}):\bit}\ 
                  (\text{op} \in \Set{<,>,=,\ldots})\\[\dummylength]
                  \text{Arithmetic}\quad & \infer{E\vdash
                    v_{1}:t_{1}\wedge E\vdash v_{2}:t_{2}} {c(t_{1}) =
                    c(t_{2}) = 1} {E \vdash \text{op}(v_{1},
                    v_{2}):\max(t_{1}, t_{2})}\ (\text{op} \in \Set{+,-,\cdot,:,
                    \ldots})
\end{align}

Again, we do not consider division by zero or overflows; up- and
downcasting of data types and procedure handling is also skipped.
An equivalence relation between two types was given by 
Eqn.~\ref{form:type_equiv} in Section~\ref{form:type_cont};
this can be immediately carried forward to typing judgements:
\begin{equation}
\sigma_{1} \cong \sigma_{2}
    \Rightarrow \infersimple{E \vdash x:\sigma_{1}}{E \vdash x:\sigma_{2}}.
\end{equation}
Note that we do not consider these equivalences explicitely in the following
to simplify the notation, all statements are automatically supposed to
hold for all equivalent types as well without further noting this.

Also note that subtyping (\ie, considering one type as a subtype of
another and allowing appropriate conversions) is not explicitely
taken into account because this is also a problem which is specific to the
classical data types of cQPL and thus not of too much interest here.

\subsubsection{Quantum variable tuples}
Tupling of variables must make sure that no component appears more than 
once in the list because this would allow to write programs which violate 
the no-cloning principle and thus lead to runtime errors. Formally, the
requirement is given by\footnote{Remember: \(q(k) = 1\) ensures that
  the data type does not contain any classical components,
  \(\#q\) denotes the positions in the quantum heap occupied by a
  quantum variable.} 
\begin{equation}
\infersimple{E \vdash x_{1}:\sigma_{1}\cdots 
                    E\vdash x_{n}:\sigma_{n}\hspace{4mm}\begin{array}[b]{l}
  \forall k=1,\ldots,n:q(\sigma_{k}) = 1\  \wedge \\ \#x_{k}\cap
  (\#x_{1} \cup \cdots \cup\#x_{k-1}\cup\#x_{k+1}\cup\cdots\cup \#x_{n}) = 
                                                                 \varnothing
  \end{array}}{E \vdash
  (x_{1}, \ldots, x_{n}):\sum_{i}\sigma_{i}}.
\end{equation}
The meaning of this is as follows: \(E \vdash x_{i}:\sigma_{i}\) formulates
the requirement that all variables are well-defined.  The condition
\(\forall k=1,\ldots,n:q(\sigma_{k}) = 1\) requires that all components are
quantum data types; the tuple thus has no classical components which
is justified by our abdication of mixed types. The condition
\(\#x_{k}\cap
(\#x_{1}\cup\cdots\cup\#x_{k-1}\cup\#x_{k+1}\cup\cdots\cup\#x_{n}) =
\varnothing\) is a formal version of the requirement that no variable
may appear more than once in the list of variables. The conclusion
which can be drawn from these premises is that \((x_{1}, \ldots,
x_{n})\) is a proper quantum variable tuple, \ie, a well-typed
expression in the current environment.

\subsubsection{Application of operators}
The application of unitary operators requires that the dimension
of the operator matches the dimension of the variable or variables 
it is applied to. Formally, this is written as
\begin{equation}
\infer{E \vdash (x_{1}, \ldots, x_{n}):q}
      {t^{q}(q) = \dim(U) \wedge U \in U(n)}
      {\forall k: E \vdash x_{k}:q_{k} \wedge 
       ((x_{1}, \ldots, x_{n}) \text{\texttt{*=}} U):\void}.
\end{equation}

The statement additionally ensures that the typing of the qbits
involved is not influenced by the operator application. Note that we
do not explicitely specify a formal condition for the unitarity of an
operator \(U\) given in terms of a function of its components. The
membership in \(U(n)\) is sufficient for our purposes.  The
distinctness of the destination variables for the transformation is
already ensured by the tupling requirements given above, so it does
not need to be checked explicitely.

\subsubsection{\textmd{\texttt{If}} conditionals and \textmd{\texttt{while}} 
loops}
The condition for this construction must have type \bit, whereas both
possible paths must be well-typed:
\begin{equation}
\infer{E \vdash c:\bit}{E \vdash P \wedge E \vdash Q}
  {E \vdash (\text{\texttt{if} \(c\) \texttt{then} \(P\) 
                   \texttt{else} \(Q\)}):\void}.
\end{equation}

\noindent A similar condition holds for the while loop:
\begin{equation}
\infer{E \vdash c:\bit}{E \vdash P}
      {E \vdash \text{(\texttt{while}(c) \text{do} \(P\))}:\void}.
\end{equation}

\subsubsection{Measurements}
The classical data type used to store the result of a measurement
must have the same number of bits as there are qbits in the
quantum variable. This is represented by the condition
\begin{equation}
\infer{E \vdash a:\sigma_{1}, c(\sigma_{1}) = 1 \wedge E \vdash b:\sigma_{2}, q(\sigma_{2}) = 1}{t^{q}(\sigma_{2}) = t^{c}(\sigma_{1})}
      {E \vdash \text{(a := measure b)}:\void}
\end{equation}

\subsubsection{Communication}
\paragraph{Sending qbits}
When qbits are sent, the type system has to make sure that no qbit is sent
twice because this would result in the same effects as if operators could
be applied to multiple copies of the same qbit; using a tuple to
combine the sent qbits automatically solves this problem:
\begin{equation}
\infersimple{E \vdash (x_{1}, \ldots, x_{n}):\sigma, q(\sigma) = 1}
            {E \vdash \text{(send \(q_{1}, \ldots, q_{n}\)):\void}}.
\end{equation}

Note that the type system is not concerned with the actual receiver of
the qbits; this information is only required for the denotation of the 
expression, but not to ensure well-typedness.

\paragraph{Receiving qbits}
When qbits are received, the type system must make sure that the
destination variables are not yet defined in the receiver's context,
\ie, they must not be well-typed expressions. Afterwards, the
variables used in the receive statement are well-defined in the typing
context and have the data type required by the statement.  This can be
formally written as
\begin{equation}
\infersimple{\forall i: \neg(E \vdash x_{i}), q(\sigma_{i}) = 1}
{E \vdash \text{(receive \(x_{1}:\sigma_{1}, \ldots, x_{n}:\sigma_{n}\)):\void}
   \wedge \forall i: E \vdash x_{i}:\sigma_{i}}.
\end{equation}
As in the case of sending, it is not interesting for the type system
from which communication partner the qbits originate.


\summaryodd{We have given the denotational semantics of all language
  components of cQPL excluding some standard cases that are
  readily available in the literature. Together with the definition of the
  type system (by intentional omission of all technical details), this
  completes the effort of assigning a precise meaning to quantum
  programs written in cQPL.}

\section{Avoidance of runtime errors}\label{sem:runtime_errors}
QPL is a functional language with a static type system which
guarantees the absence of runtime errors (note that functionality is
subject to a precise definition of the term; it is certain that
classical languages need to have additional properties -- most
important higher-order functions -- to be called fully functional. But
this is not really relevant from a physicist's point of view, 
as we have discussed before). It is very desirable that runtime
errors can be avoided as far as in principle possible, from a
physicists point of view, it does not matter how this is achieved.
This desire was brought forward into cQPL and manifests itself in two
points: Cloning is (as in QPL) prevented already at the syntactical
level, and communication does not allow different processes to access
qbits concurrently.

\subsection{Unique ownership of qbits}
\begin{observation}
No part of the quantum heap is accessible to two or more parties
at the same time during parallel execution of arbitrary cQPL programs.
\end{observation}

\begin{rationale}
When a module is considered stand-alone, it is obvious that all
qbits present in the system are owned by one party. Uniqueness of
the ownership is guaranteed by the quantum part of the probabilistic
environment which ensures (as described in 
Section~\ref{form:formal_definitions}) that it is impossible for two or 
more names to refer to overlapping sets of qbits. 

Parallel composition of systems is performed by always considering
pairs of send/receive statements; while sending removes the sent qbit
from the typing context of the originating system, receiving adds it
to the typing context of the destination. Since both commands are
always considered in pairs and denoted atomically,\footnote{This means
  that nothing can happen in between sending and receiving the quantum
  bit.} a quantum variable may not be in two typing contexts at the
same time. Access to quantum variables is only possible for a user when the
variable is present in his typing context, this ensures (together with
the fact that quantum heaps of communicating systems cannot overlap by
virtue of Definition~\ref{form:multi_party_components}) that it is
impossible for two or more modules to access identical qbits at a
time.

It is possible to prove this statement formally based on the
observations in Section~\ref{form:type_system}. Since this is on the one hand
a general problem of semantics theory and on the other hand
burdened with many technical difficulties, we omit a precise proof here,
but refer to \cite{wright_felleisen} where the exact details can be
found. \cite{gay_nagarajan} is one source where the proofs of the
aforementioned reference have been adapted to a quantum system which
fulfils exactly the same properties as ours (basically, not too much except
notational details needs to be changed).
\end{rationale}

\subsection{Prevention of cloning and unphysical situations}
One of the fundamental consequences of quantum mechanics is that it
is impossible to define a unitary operator that can duplicate
arbitrary quantum states with perfect fidelity; a straightforward
calculation shown in nearly every text on quantum mechanics 
(\eg, \cite{nielsen_chuang,preskill}) proves this. Obviously, quantum
programming languages must make sure that cloning is forbidden
because otherwise, processes contrary to the laws of physics 
could be simulated. Since most other approaches to quantum programming
(\eg, \cite{oemer_msc,betelli,knill,sanders_zuliani}) require the 
possibility to address quantum bits via references or pointers, 
they cannot ensure at compile-time that two distinct variables 
do not share the same quantum bit; they must provide appropriate checks at 
runtime which ensure this condition. Aside from efficiency considerations,
this is unsatisfying because especially for long-running programs,
termination with an error which was caused by a programming mistake
is undesirable.

The static typing of QPL allows together with some syntactical checks
to ensure that once a program was approved to be correct by the 
static syntactical and semantical analysis of the compiler,
no runtime errors caused by unphysical cloning of quantum states can
happen. A similar statement can be observed for cQPL:

\begin{observation}
cQPL programs which do not use communication primitives can be
guaranteed to execute without runtime errors if the syntactic and
semantic analysis deems them correct.
\end{observation}

\begin{rationale}
No quantum bit can be referred to by multiple identifiers in cQPL,
as was shown in the previous section. Thus, the distinctness of
the quantum components of a list of identifiers (which is used to
specify the list of qbits an operator works on or given as parameter
to a procedure) can be guaranteed by ensuring that the same identifier 
does not appear multiple times in the list. Therefore, the same line
of reasoning for the impossibility of cloning or generating unphysical
situations as in \cite[Section 4.8]{selinger_qpl} applies.
\end{rationale}

\begin{remark}
Note that non-termination is something different than a runtime error.
\end{remark}

\begin{remark}
Note that static typing can also prevent the possibility for
some runtime errors which originate from the classical parts of
the language; this is well-known in programming language theory
(cf., \eg, Refs.~\cite{rechenberg_pomberger,wilhelm_maurer,appel} for details) 
so that we will not dwell into this any further here.
\end{remark}

\subsection{Unavoidable non-termination conditions}\label{form:unavoid_rt_error}
Albeit cQPL tries to prevent runtime errors as good as possible, the
introduction of communication opens the possibility of writing
programs that cannot be checked at compile time if they will terminate
at runtime although nothing would hinder the separate modules to
terminate. Nevertheless, by restricting the code to a certain subset
of cQPL,\footnote{Which can, in principle, solve all problems that
  might arise in quantum programming, but is not a very practical.} it
is still possible to produce programs which will execute guaranteed
without termination problems and without runtime errors. Note that
non-termination is not considered as a runtime error. If a program of
the form \texttt{while (1) do skip} is provided, then executing the
\texttt{skip} command forever (and thus doing nothing forever) is
exactly the intention of the program and therefore the correct
behaviour which should be reflected by the denotation.

In Section~\ref{form:comb_proc}, we have already considered an example of
a non-terminating program. The culprit here was the different number of sent 
versus received qbits, but since the number of sent and received
qbits is fixed at compile time on both sides, this error can
obviously be detected by the semantic analysis; the
program can be rejected. Unfortunately, this possibility is
not always the case because the exact number of how many qbits will
be sent and how many will be received can not be decided in general. 
Consider the following example:

\begin{verbatim}
module A {
  new qword nq;
  nq *= H(8);
  new word n := measure nq;
  while (n >= 0) {
     new qbit q;
     send q to B;
     n := n-1;
  }
}; 

module B {
   receive q1:qbit, q2;qbit, q3:qbit; 
}; 
\end{verbatim}

Since \texttt{n} in module \texttt{A} may contain (with equal probability)
any value in \([0,2^{8}-1]\), the number of sent qbits cannot be determined
with certainty, but is governed by the probability distribution 
of \texttt{n}. It may be the case that the program terminates (namely,
if exactly three qbits are sent by A), but it may also be the case that 
less or more than three qbits are sent. This results in either 
a blocking process \texttt{A} which cannot find a receiver for the qbits
it wants to transmit, or in a blocking process \texttt{B} which
is not satisfied with a proper number of qbits and blocks to wait for the
missing ones.

Fortunately, there are only three commands in cQPL that allow to
execute a sequence of communication commands for which it is not
possible to determine at runtime how many there will be, so we can
make the following observation:

\begin{observation}
cQPL programs using communication can be guaranteed to execute without
runtime errors if the syntactic and semantic analysis deems them correct
and the following possibilities of the language are not used:

\begin{itemize}
  \item While-loops with a termination condition that contains a
    probabilistic variable.
  \item If-conditionals that are based on a probabilistic variable.
  \item Recursive procedures whose recursion depth cannot be 
    determined at compile time.
\end{itemize}
\end{observation}

\begin{rationale}
All send and receive statements which are given as a sequence of commands
(which may include the use of blocks) can be counted at compile time;
their order is obviously also known. If the if-statement is used with
a condition that can be computed at compile time, one path can be eliminated.
Thus, the statement is nothing else than a regular contribution to the list
of statements. While-loops with a compile-time computable number of
iterations can be replaced by inlining the loop body the appropriate
number of times, so they also become only a regular contribution to a sequence
of commands. If the recursion depth of a procedure can be calculated,
it be converted to an iteration where the number of steps and thus
the number of communication commands are known. Therefore, it is also
just a regular contribution to a sequence of commands.
\end{rationale}

\begin{remark}
Note that the checks required to determine the number of loop iterations
etc.~at compile-time are based on well-understood analysis techniques 
in computer science; nevertheless, we did not actually implement these
checks in the cQPL compiler because it is nothing else than a routine
task with little benefit and no gain of any valuable insight, but just
a technical problem.
\end{remark}

\newpage\thispagestyle{plain}
\chapter{Prospects}{I ain't no physicist, but I know what matters.}%
{\hfill Popeye the sailor}\label{prospect} 

\section{Outlook}
The field of quantum programming languages is -- as everything
connected with quantum information -- still a young one, and many
things that are standard in classical programming languages still need
to be adapted for these. Some ideas which were tried to be realised
during the work on this thesis, but did not reach fruition are:

\begin{itemize}
  \item Integration of higher-order functions. Everything we tried ended
    up in requiring closures for an implementation, but this is (to our
    knowledge) impossible to achieve because of the no-cloning theorem.
    Having them would be quite desirable for many applications.
  \item The ability to describe the complete loss of quantum bits
    caused by imperfect channels or eavesdroppers. This is obviously 
    hard to integrate into a programming language,\footnote{Just think
      about the situation that would arise if variables in classical
      programming languages could randomly disappear\dots}
    but should be possible by heading for a protocol specification
    variant of cQPL.
  \item Consideration of more general eavesdropping models where the
    strategy needs not be fixed, but can be one of multiple independent 
    alternatives. The work provided in \cite{hartog} would possibly
    provide a suitable basis with demonic choices.
  \item Most texts about quantum programming languages extensively use
    categories to describe the underlying structures. We found that
    this does not really add any substantial points, but merely more
    notation and nomenclature, so we did not follow this style
    although some effort was made in the beginning to become familiar
    with the field.
\end{itemize}

Nevertheless, the material provided here could serve as starting point
for the following possible extensions:

\begin{itemize}
  \item Quantum instead of classical control, \ie, allowing conditions
    to be based on quantum and not classical logic. It would be possible
    to simulate this with QCL, but the benefit is questionable because
    no known quantum algorithm makes use of such a feature.
  \item The method presented here could provide a basis to formulate
    quantum process algebras as, \eg, presented in
    \cite{gay_nagarajan, adao_mateus}.
  \item An extension from discrete to continuous systems would allow
    the simulation of general quantum systems and could thus be useful
    for a much wider range of quantum information applications.
  \item Faulty hardware models could be integrated at the simulation
    layer, but this would presumably be very challenging at the
    semantic level.
\end{itemize}

Initially, it was planned to also investigate the possibility of
integrating the semantic framework into a theorem prover which could
possibly facilitate automated analysis techniques for quantum
protocols.  Some preliminary experiments were performed by describing
the BB84 protocol in a classical protocol simulator, but this has only
shown that the gap between the requirements for such an automatisation
and what is currently available is still very wide for all approaches
to quantum programming.

\section{Latest developments}
After this thesis was finished, another QPL compiler written by D.
Williams was presented in a joint work by Nagarajan, Papanikolaou and
Williams~\cite{nagarajan_compiler}. Since both efforts work on closely
related fields, it seems apt to sketch similarities and differences of them
(to distinguish it from our implementation, we call William's compiler
sqrQPL\footnote{Because their compiler targets a virtual machine 
  which is termed \emph{sequential quantum random access machine}.} in the
following):

\begin{itemize}
  \item Both compilers use the quantum computer model defined by
    Knill~\cite{knill} as basic architecture.
  \item sqrQPL provides an own quantum simulator which is called
    \emph{sequential quantum random access machine}. Code generated 
    for this architecture resembles machine language quite closely.
  \item sqrQPL provides support for a smaller subset of QPL
    than cQPL.
  \item sqrQPL has the ability to automatically decompose arbitrary
    unitary matrices into a set of standard gates. As a result (and in
    addition to theoretical elegance) of this, the simulator needs
    only provide support for very few different elementary gates.
  \item The semantics of sqrQPL is fully covered by the one given for
    QPL, while cQPL needs to provide additional semantics for the
    added features.
  \item cQPL already includes support for communication and concurrency, 
    whereas work to bring these abilities to sqrQPL will be started in the
    future according to~\cite{nagarajan_compiler}.
\end{itemize}

It would be interesting (and should be possible without too much
effort) to provide an SQRAM-backend for cQPL; since
Ref.~\cite{nagarajan_compiler} states that support for communication
and concurrency is (at least in preliminary form) already present in
their simulator, no major obstacle does seem to exist to hinder such an
endeavour. In summary (and, obviously, seen from the author's
subjective point of view) sqrQPL is a straight implementation of QPL
where the ability to decompose complicated gates into a set of simpler
ones is the essential feature. The focus of cQPL is mainly on the
semantics of (quantum) communication; although the cQPL compiler seems
(at the time of writing) to provide a bigger and more versatile
language core than sqrQPL, it is more or less a by-product of the
actual work.

\begin{appendix}
\chapter{List of symbols}{\dots\ und Lasse sagte, die Sprache der %
Jungen sei sowieso die einzig wahre.}{Astrid Lindgren, Wir Kinder %
aus Bullerb\"u}\label{symbols}
The following presents a list of symbols used in this work. Note
that the meanings given here are not necessarily the only ones with which
they were used. 
\newcommand{\symentry}[2]{\noindent\begin{tabular}{ll}%
\parbox[b]{2.3cm}{\noindent#1\ \dotfill\ \ }%
\parbox[t]{0.5\textwidth-3cm}{\noindent\raggedright#2}%
\end{tabular}\par\vspace{0.5ex}}
\addtolength\columnsep{2mm}
\begin{multicols}{2}
\symentry{\(\#\)number}{Unique node id}
\symentry{\(\#\)string}{Position(s) occupied by quantum variable variable on 
                        the quantum heap}
\symentry{\(\stdsembrack{}\)}{Separate syntax and semantics}
\symentry{\(||\)}{Parallel composition}
\symentry{\(\$\)}{Superoperator on \(\BH\)}
\symentry{\(\bigsqcup\)}{Least upper bound}
\symentry{\(\bot\)}{Least element of a partial order}
\symentry{\(\cong\)}{Equivalence, reflexive and transitive}
\symentry{\(\sqsubseteq\)}{Binary partial order}
\symentry{\(\kteinit\)}{Initial \((K,T,E)\) tuple}
\symentry{\(\Gamma\)}{List of Kraus sets}
\symentry{\(\Lambda\)}{Completely positive map}
\symentry{\(\chi(v)\)}{Type associated with a variable \(v\)}
\symentry{\(\omega\)}{Increasing chain of natural numbers}
\symentry{\(\varphi\)}{A permutation}
\symentry{\(\pi:\mathcal{B}\)}{Probability distribution obtained by applying a
  projective measurement defined by the basis \(\mathcal{B}\)}
\symentry{\(\varrho\)}{A density operator}
\symentry{\(\Sigma\)}{State in form of a \((K,T,E)\) tuple}
\symentry{\(\sigma\)}{Signature for types}
\symentry{\(\mathcal{A}\)}{Set of all possible Kraus aggregations}
\symentry{\(\mathcal{A}\)}{Observable algebra}
\symentry{\(A\)}{Finite ordered set}
\symentry{\(A\#b\)}{The \(b^{\text{th}}\) element of the ordered set \(A\)}
\symentry{\(\BH\)}{Set of all bounded operators on Hilbert 
          space \(\mathcal{H}\)}
\symentry{\(\COMM\)}{Valuation function for parallel execution}
\symentry{\(\CX\)}{Complex-valued functions \(X \rightarrow \mathbbm{C}\)}
\symentry{\(c(\sigma)\)}{Check if a given data type is purely classical}
\symentry{\(\card(X)\)}{Cardinality of \(X\)}
\symentry{\(\mathcal{D}_{n}\)}{Set of all density operators of dimension \(n\)}
\symentry{\(\mathcal{S}(k)\)}{Set of all decompositions 
                              of \(k \in \mathbbm{N}\)}
\symentry{\(\mathfrak{D}(A,B)\)}{Determine if \(A\) and \(B\) operate on
  disjoint qbits}
\symentry{\(\DO\)}{Valuation function for dyadic operators}
\symentry{\(E\)}{Environment}
\symentry{\(E \vdash x:T\)}{\(x\) has type \(T\) is valid in environment \(E\)}
\symentry{\(E \vdash F\)}{\(F\) is well-typed in environment \(E\)}
\symentry{\(\EA\)}{Effects of \(\mathcal{A}\)}
\symentry{\(\EQN\)}{Valuation function for arithmetic expression}
\symentry{\(\EXP\)}{Valuation function for expressions}
\symentry{\(\mathcal{F}\)}{Set of all fixed points of a permutation}
\symentry{\(\mathbbm{F}_{2}\)}{Binary group/ring/field}
\symentry{\(\operatorname{fix}\)}{Fixed point}
\symentry{\(\inj\)}{Injection}
\symentry{\(I(M)\)}{Set of all intervals in \(M\)}
\symentry{\(K\)}{Kraus aggregation}
\symentry{\(\mathcal{K}\)}{Set of all unparametrised Kraus agregations}
\symentry{\(\mathfrak{L}_{c}\) }{Set of labels for communication partners}
\symentry{\(M\)}{Finite set}
\symentry{\(\MO\)}{Valuation function for monadic operators}
\symentry{\(\OP\)}{Valuation function for operators}
\symentry{\(\mathcal{P}(M)\)}{Powerset of \(M\)}
\symentry{\(n^{c}\)}{Classical data type with \(n\) bits}
\symentry{\(n^{q}\)}{Quantum data type with \(n\) qbits}
\symentry{\(\pos(x, L)\)}{Position of \(x\) in the list \(L\)}
\symentry{\(\PROG\)}{Valuation function for programs}
\symentry{\(\PROJ\)}{Generate Kraus set with projection operators for a
  quantum type}
\symentry{\(Q\)}{Size of the quantum heap (global constant!)}
\symentry{\(\mathcal{Q}\)}{Set of all quantum variables in a typing context}
\symentry{\(q(n^{\tau})\)}{Distinguish between classical and quantum 
                           components of a data type}
\symentry{\(q(\sigma)\)}{Check if a given data type is purely quantum}
\symentry{\(\qtype\)}{Arbitrary quantum data type}
\symentry{\(\mathfrak{R}\)}{Number of receive statements in a Kraus 
                            aggregation}
\symentry{\(\RS\)}{Valuation function for a receive/send pair}
\symentry{\(\mathfrak{S}\)}{Number of send statements in a Kraus aggregation}
\symentry{\(\SA\)}{States of \(\mathcal{A}\)}
\symentry{\(\smashp\)}{Smash product (with a single bottom element)}
\symentry{\(\SR\)}{Valuation function for a send/receive pair}
\symentry{\(\sym(M)\)}{Symmetric group over \(M\)}
\symentry{\(\mathcal{T}(\sigma)\)}{Set of data types equivalent to \(\sigma\)}
\symentry{\(T\)}{Typing context}
\symentry{\(t_{q}(\sigma)\)}{Number of quantum bits contained in a 
                             data type \(\sigma\)}
\symentry{\(t_{c}(\sigma)\)}{Number of bits contained in a data 
                             type \(\sigma\)}
\symentry{\(U(n)\)}{Unitary group of degree \(n\)}
\symentry{\(\VAL\)}{Arbitrary valuation function}
\symentry{\(X\)}{Finite set}
\symentry{\(x:t\)}{Variable \(x\) with type \(t\)}
\symentry{\(\Ycomb_{D}\)}{Fixed point combinator on cpo D}
\end{multicols}

\forgechapter{2}
\chapter{Glossary}{Die Bedeutung eines Wortes ist das, was die Erkl{\"a}rung 
der Bedeutung erkl{\"a}rt.}{\hfill\parbox{5.7cm}%
{\raggedright Ludwig Witt\-gen\-stein, Phi\-lo\-so\-phi\-sche Gram\-ma\-tik}}%
\label{glossary}
Some terms used in this work are not too commonplace in physics, so we
collected the most important definitions to remind the reader of their 
meaning if it cannot be immediately recollected. 
Some of the definitions were inspired by~\cite{wikipedia}.

\begin{multicols}{2}
\begin{description}
  \item[Abstract syntax] Grammar used to specify the possible shapes
    of the parse tree.
  \item[Backus-Naur Form] Metasyntax with a standardised set of 
    symbols and notations which is used to express context-free grammars.
  \item[Compiler-Compiler] A program used to generate a parser which
    can perform syntax analysis on programs that follow a given grammar.
  \item[Compile time] refers to all actions which are performed by
    the compiler before the program is executed, \eg, syntactical and
    semantical analysis, scoping rule enforcement, type analysis, 
    optimisation, code generation etc.
  \item[Concrete syntax] Syntax in which textual representations of
    programs must be specified.
  \item[Context free grammar] Formal grammar in which every production
    rule needs to be of the form \(V\rightarrow w\) where \(V\)
    is a non-terminal and \(w\) a list of terminal and non-terminal symbols.
  \item[Data type] A data type is a name or label for a set of values and 
    some operations which can be performed on that set of values.
  \item[EBNF] Extended Backus-Naur Form
  \item[Environment] Structure which provides a mapping between identifiers
    of variables and the values associated with them.
  \item[FIFO] Queue with first-in, first-out behaviour, \ie, the output
    of the queue is in the same order as the input.
  \item[Functional languages] do not work on 
    explicit states, but use transformations that map input 
    to output parameters (\(\rightarrow\) referential 
    transparency). Assignment to variables is not possible 
    since they represent immutable bindings for values. In
    our notation, functionality is exploited as far as it is 
    necessary for the ability to guarantee freedom from runtime 
    errors. Classical examples of functional languages include 
    Lisp, ML, and Haskell.
  \item[Identifier] Name of a variable in a program.
  \item[Imperative languages] work on a global state that is modified during 
    runtime. The most widespread languages (C, C++, Pascal etc.) follow
    this approach. It is normally impossible to decide if a program
    written in an imperative language will terminate or produce errors
    without executing it.
  \item[Lexer] A lexer is the part of a compiler which takes the source 
    code of a program (in textual form) and disseminates it into
    a stream of \emph{token}s which is fed to the \emph{parser}.
  \item[Lexicographic order] Two strings \(x, y\in \Sigma^{*}\) 
    can be ordered such that \(x > y\) if \(x\#i - y\#i > 0\) for the 
    first \(i\in\mathbbm{N}\) for which \(x\#i \neq y\#i\).
  \item[LALR(1)] Certain class of context-free grammars that needs to be
    specified subject to some constraints on its form, but can be
    handled by Yacc-style parsers.
  \item[Mutex] Mutual exclusion. A technique realised with the aid
    of special variables which ensures that only one component of 
    a parallel program can be in the region protected by the mutex 
    at a time.
  \item[Non-terminal symbol] A symbol that is composed of terminal symbols
    and possibly other non-terminal symbols.
  \item[Parser] The parser is the part of a compiler which analyses
    the grammatical structure of a program (which is fed to him in
    the form of \emph{token}s produced by the \emph{lexer}). This
    process is also known as \emph{syntactical analysis}.
  \item[Parse tree] Representation of a program which is generated by
    the parser. Since tree-based data structures are used to 
    represent the information, the abstract syntax of the language
    (which is easier to analyse) can be utilised.
  \item[Runtime] refers to the time when a program is executed and
    the compiler has no more influence on what happens. Alternatively,
    it may denote a library with helper functions supplied by
    the compiler which are necessary for the generated code to
    work (this may be also referred to as \emph{runtime library}).
  \item[Scope] Rules used to determine what, if any, entity a given 
    occurrence of an identifier in a program refers to.
  \item[Semantic analysis] is the part of a compiler that adds semantic
    information to the parse tree (for example, the required space in memory 
    for variables) and performs sanity checks which may detect errors
    in the code before it is executed.
  \item[Static typing] means that once a type has been assigned to an
    object, it cannot be changed any more.
  \item[Strong typing] means that not only values, but also identifiers
    are typed.
  \item[Terminal symbol] A symbol of a grammar that represents a constant.
  \item[Token] Tokens are the smallest elementary parts of a program
    form the parser's point of view. While \texttt{int} is
    a three-letter word respectively a string of characters in the source 
    code, the parser regards it as a single entity which describes
    the data type of integers. 
  \item[Type] Also called \emph{data type}. It is a label for a set of 
    values together with some operations that can be performed on the set.
  \item[Type system] Set of rules that determines which type a given
    object has, how types can be combined etc.
  \item[Type checking] is the pass of a compiler which ensures
    that all operations of a program are applied to variables
    of proper type; it can, \eg, ensure that string concatenation
    is not tried to be performed on integers.
  \item[Yacc] Yet another compiler compiler. One of the early approaches
    to automated parser generation. Most modern parser generators follow
    the concept of this program.
\end{description}
\end{multicols}
\newcommand{\tp}[1]{\textrm{\textsl{#1}}}
\chapter{Formal syntax}{Der Satz ist der sprachliche Ausdruck 
daf\"ur, dass sich die Ver\-bin\-dung mehrerer Vor\-stel- lungen  
in der Seele des Spre\-chen\-den voll\-zo\-gen 
hat, und das Mit\-tel da\-zu, die n\"am\-li\-che Ver\-bin\-dung der 
n\"am\-li\-chen Vor\-stel\-lun\-gen in der 
Seele des H\"o\-ren\-den zu er\-zeu\-gen.}{H.~Paul, Prin\-zi\-pi\-en der
Sprach\-ge\-schich\-te}\label{comp:formal_syntax}
The formal syntax for cQPL is defined by the following rules which are
used to generate the parser. Words in \texttt{typewriter} face denote
tokens recognised by the lexer, whereas \tp{slanted} text is used for
productions. \tp{identifier}s are given by a letter followed by an
arbitrary number of letters, digits and underscores. The empty
production is denoted by \(\epsilon\).

\begin{alltt}
\tp{program}: \tp{stmt\_list} EOF
         | \tp{module\_list} EOF
\tp{stmt\_list}: \tp{statement};
         | \tp{stmt\_list} \tp{statement};
\tp{module\_list}: \tp{module\_def};
           | \tp{module\_list} \tp{module\_def};
\tp{module\_def}: module \tp{identifier} \verb/{/ \tp{stmt\_list} \verb/}/
\tp{proc\_decl}: proc \tp{identifier}:\tp{context} -> \tp{context} \tp{block} in \tp{statement}
         | proc \tp{identifier}:\tp{context} \tp{block} in \tp{statement}
\tp{context}: \tp{identifier}:\tp{var\_type} \tp{more\_context} | \(\epsilon\)
\tp{nonempty\_context}: \tp{identifier}:\tp{var\_type} \tp{more\_context}
\tp{more\_context}: , \tp{identifier}:\tp{var\_type} \tp{more\_context} | \(\epsilon\)
\tp{block}: \verb/{/ \tp{stmt\_list} \verb/}/
\tp{var\_type}: bit | qbit | qint | int | float
\tp{send\_stmt}: send \tp{args} to \tp{identifier}
\tp{receive\_stmt}: receive \tp{nonempty\_context} from \tp{identifier}
\tp{allocate\_stmt}: new \tp{var\_type} \tp{identifier} := \tp{arith\_expr}
\tp{arith\_expr}: \tp{int\_value}
          | \tp{float\_value}
          | true
          | false
          | \tp{identifier}
          | (\tp{arith\_expr})
          | \tp{arith\_expr} + \tp{arith\_expr}
          | \tp{arith\_expr} - \tp{arith\_expr}
          | \tp{arith\_expr} * \tp{arith\_expr}
          | \tp{arith\_expr} / \tp{arith\_expr}
          | \tp{arith\_expr} < \tp{arith\_expr}
          | \tp{arith\_expr} > \tp{arith\_expr}
          | \tp{arith\_expr} <= \tp{arith\_expr}
          | \tp{arith\_expr} >= \tp{arith\_expr}
          | \tp{arith\_expr} == \tp{arith\_expr}
          | \tp{arith\_expr} != \tp{arith\_expr}
          | \tp{arith\_expr} \& \tp{arith\_expr}
          | \tp{arith\_expr} | \tp{arith\_expr}
          | - \tp{arith\_expr}
          | ! \tp{arith\_expr}
\tp{proc\_call}: call \tp{identifier} (\tp{args})
         | (\tp{var\_list}) := call \tp{identifier} (\tp{args})
\tp{args}: \tp{identifier} \tp{more\_args} | \(\epsilon\)
\tp{more\_args}: , \tp{identifier} \tp{more\_args} | \(\epsilon\) 
\tp{if\_stmt}: if \tp{arith\_expr} then \tp{statement}
        | if \tp{arith\_expr} then \tp{statement} else \tp{statement}
\tp{measure\_stmt}: measure \tp{identifier} then \tp{statement} else \tp{statement}
\tp{assign\_stmt}: \tp{identifier} := \tp{arith\_expr}
\tp{assign\_measure\_stmt}: \tp{identifier} := measure \tp{identifier}
\tp{while\_stmt}: while \tp{arith\_expr} do \tp{statement}
\tp{gate\_stmt}: \tp{var\_list} *= \tp{gate}
\tp{gate}: H | CNot | Not | Phase \tp{float\_value}
     | FT (\tp{int\_value})
     | [[ \tp{number\_list} ]]
\tp{number\_list}: \tp{sign} \tp{float\_value}
            | \tp{sign} \tp{int\_value}
            | \tp{sign} \tp{float\_value} PLUS \tp{sign} \tp{imaginary\_value} 
            | \tp{sign} \tp{int\_value} PLUS \tp{sign} \tp{imaginary\_value} 
            | \tp{sign} \tp{imaginary\_value} 
            | \tp{number\_list}, \tp{sign} \tp{float\_value} 
            | \tp{number\_list}, \tp{sign} \tp{int\_value} 
            | \tp{number\_list}, \tp{sign} \tp{imaginary\_value} 
            | \tp{number\_list}, \tp{sign} \tp{float\_value} + \tp{sign} \tp{imaginary\_value} 
            | \tp{number\_list}, \tp{sign} \tp{int\_value} + \tp{sign} \tp{imaginary\_value}
\tp{sign}: - | + | \(\epsilon\)
\tp{var\_list}: \tp{identifier} | \tp{var\_list}, \tp{identifier} 
\tp{skip\_stmt}: skip
\tp{print\_stmt}: print "\tp{string}" 
          | print \tp{arith\_expr} 
          | dump \tp{var\_list}
\tp{statement}: \tp{proc\_call} 
          | \tp{proc\_decl}
          | \tp{while\_stmt}
          | \tp{allocate\_stmt}
          | \tp{if\_stmt}      
          | \tp{print\_stmt}   
          | \tp{assign\_stmt}  
          | \tp{assign\_measure\_stmt}
          | \tp{measure\_stmt}
          | \tp{skip\_stmt}   
          | \tp{block}       
          | \tp{gate\_stmt}
          | \tp{send\_stmt}
          | \tp{receive\_stmt}
\end{alltt}

\fancyhead[LE]{\nouppercase{\leftmark}}
\fancyhead[RO]{\nouppercase{\rightmark}}
\fancyhead[RE]{}
\fancyhead[LO]{}
\cleardoublepage
\addcontentsline{toc}{chapter}{\numberline{}Bibliography}
\bibliographystyle{alpha}
\nocite{*}
\bibliography{da}\cleardoublepage

\thispagestyle{plain}
\chapter*{Thanks}

\begin{itemize}
  \item To PD Dr.~Norbert L\"utkenhaus for taking the peril of a journey
    into the strange and unaccustomed, his support by discussions and
    suggestions and for giving me complete freedom in deciding what to
    work on.
  \item To Prof.~Dr.~Dr.~Volker Strehl for many pointers into the
    right direction and for sacrificing time for a student from
    another faculty.
  \item To Tobias Moroder, Hans Loehr, Martin Trini, Johannes Rigas,
    Volkher Scholz and Markus Diefenthaler for proofreading 
    and many valuable corrections and suggestions.
  \item To the guys in my office (Tobi Moroder, Johannes Rigas and 
    Dr.~Matthias Jakob) for providing a pleasant environment to work in, 
    many inspiring level eights and our shared pleasure of working
    under illumination provided by the moon.
  \item To an anonymous reviewer for encouraging comments.
  \item To \url{dict.leo.org} for countless suggestions on english
    vocabulary; quick answers to many questions were given by 
    \url{www.wikipedia.net}.
  \item To Dr.~Peter Selinger for detailled explanations regarding
    his work.
  \item To the red and the green forrest fairy because they are way too 
    mythical to not be thanked; without any doubt, the same
    holds for HM Queen Elizabeth II.
  \item To all members of the QIT group (Tobi, Johannes, Philippe,
    Matthias 1, Matthias 2, Joe, Geir Ove, Marcos, Ivan) for their
    help, valuable discussions and the pleasant working environment,
    not to forget the shared fun among some of us in chasing little
    bouncing objects on diverse courts.
  \item To my parents and my family for their overall and ubiquitous 
    love and support in any aspect of life.
\end{itemize}
\newpage\thispagestyle{plain}
\end{appendix}
\end{document}